%% file: main.tex
\documentclass[10pt,journal,compsoc,dvipsnames]{IEEEtran}

\pdfoutput=1
\IEEEoverridecommandlockouts
\usepackage{cite}
\usepackage[export]{adjustbox}
\usepackage{makecell}
\usepackage{csquotes}
\usepackage{amsmath,amssymb,amsfonts}
\usepackage{algorithmic,algorithm}
\usepackage[dvipsnames]{xcolor}
\usepackage{subfigure}
\usepackage{enumitem}
\usepackage{comment}
\usepackage{xspace}
\usepackage{balance}
\usepackage{fancyhdr}
\usepackage{multirow}
\usepackage{listings}

\newcommand{\mycomment}[1]{}
\newcommand{\oran}{O-RAN\xspace}
\newcommand{\pandora}{PandORA\xspace}
\newcommand{\ai}{\gls{ai}\xspace}
\newcommand{\ml}{\gls{ml}\xspace}
\newcommand{\drl}{\gls{drl}\xspace}
\newcommand{\ran}{\gls{ran}\xspace}
\usepackage{siunitx}
\usepackage{soul}
\usepackage{multirow}
\usepackage{booktabs}
\usepackage{tabularx}
\usepackage{tikz}
\usepackage[normalem]{ulem}
\usepackage{pgfplots}
\usepackage{adjustbox}
\pgfplotsset{compat=newest}
\pgfplotsset{plot coordinates/math parser=false}
\newlength\fheight
\newlength\fwidth
\usetikzlibrary{plotmarks,patterns,decorations.pathreplacing,backgrounds,calc,arrows,arrows.meta,spy,matrix,scopes}
\usepgfplotslibrary{patchplots,groupplots}
\usepackage{tikzscale}
\usepackage[draft]{hyperref}
\usepackage{cleveref}
\newif\ifexttikz
\exttikzfalse

\ifexttikz
	\usetikzlibrary{external}
	\usepackage{fontspec}
\fi

\usepackage[acronyms,nonumberlist,nopostdot,nomain,nogroupskip,acronymlists={hidden}]{glossaries}
\newglossary[algh]{hidden}{acrh}{acnh}{Hidden Acronyms}
\input{acronyms.tex}
\glsdisablehyper

\definecolor{codegray}{rgb}{0.25,0.25,0.25} 
\definecolor{codepurple}{rgb}{0.58,0,0.82}
\definecolor{desireRed}{RGB}{230,57,60}%
\definecolor{darkPurple}{RGB}{59,31,43}%
\definecolor{springGreen}{RGB}{37,223,145}%
\definecolor{queenBlue}{RGB}{69,123,157}%
\definecolor{spaceCadet}{RGB}{29,53,87}%

\colorlet{punct}{red!60!black}
\definecolor{delim}{RGB}{20,105,176}
\colorlet{numb}{magenta!60!black}

\lstdefinelanguage{json}{
    basicstyle=\color{Periwinkle}\ttfamily\scriptsize,
      breakatwhitespace=true,         
      breaklines=true,                 
      captionpos=b,
      frame=tb,
      keepspaces=true,                 
      numbers=left,                    
      numbersep=5pt,                  
      showspaces=false,                
      showstringspaces=false,
      showtabs=false,                  
      tabsize=2,
      xleftmargin=10pt,
      belowskip=-10pt,
    literate=
     *{0}{{{\color{numb}0}}}{1}
      {1}{{{\color{numb}1}}}{1}
      {2}{{{\color{numb}2}}}{1}
      {3}{{{\color{numb}3}}}{1}
      {4}{{{\color{numb}4}}}{1}
      {5}{{{\color{numb}5}}}{1}
      {6}{{{\color{numb}6}}}{1}
      {7}{{{\color{numb}7}}}{1}
      {8}{{{\color{numb}8}}}{1}
      {9}{{{\color{numb}9}}}{1}
      {:}{{{\color{punct}{:}}}}{1}
      {,}{{{\color{punct}{,}}}}{1}
      {\{}{{{\color{delim}{\{}}}}{1}
      {\}}{{{\color{delim}{\}}}}}{1}
      {[}{{{\color{delim}{[}}}}{1}
      {]}{{{\color{delim}{]}}}}{1},
}
\definecolor{red}{rgb}{0, 0, 0}
\definecolor{blue}{rgb}{0.0, 0.0, 0.0}

\usepackage{dblfloatfix} 

\ifexttikz
\else
\usepackage{tikzpagenodes,etoolbox}
\usetikzlibrary{calc}
\usepackage[contents={}]{background}
\AddEverypageHook{%
\ifnumequal{\thepage}{1}{%
 \tikz[remember picture,overlay]{%
     \node[draw,
     text width=0.95\textwidth,
     font=\footnotesize
     ]
     at ($(current page header area) - (0,5pt)$)
     {%
     This paper has been accepted for publication on \textit{IEEE Transactions on Mobile Computing}.\\
\copyright~2024 IEEE. Personal use of this material is permitted. Permission from IEEE must be obtained for all other uses, in any current or future media, including reprinting/republishing this material for advertising or promotional purposes, creating new collective works, for resale or redistribution to servers or lists, or reuse of any copyrighted component of this work in other works.
     };
 }%
}{}
}
\fi

\begin{document}





\title{PandORA: Automated Design and\\Comprehensive Evaluation of Deep Reinforcement Learning Agents for Open RAN\vspace{-.15cm}}


\author{\IEEEauthorblockN{Maria Tsampazi\IEEEauthorrefmark{1}, Salvatore D'Oro\IEEEauthorrefmark{1}, Michele Polese\IEEEauthorrefmark{1}, Leonardo Bonati\IEEEauthorrefmark{1},\\Gwenael Poitau\IEEEauthorrefmark{4}, Michael Healy\IEEEauthorrefmark{4}, Mohammad Alavirad\IEEEauthorrefmark{4}, Tommaso Melodia\IEEEauthorrefmark{1}}\\
\IEEEauthorblockA{\IEEEauthorrefmark{1}Institute for the Wireless Internet of Things, Northeastern University, Boston, MA, U.S.A.\\E-mail: \{tsampazi.m, s.doro, m.polese, l.bonati, t.melodia\}@northeastern.edu\\\IEEEauthorrefmark{4}Dell Technologies, P\&O OCTO – Advanced Wireless Technology\\E-mail: \{gwenael.poitau, mike.healy, mohammad.alavirad\}@dell.com\vspace{-0.8cm}} 

\thanks{This article is based upon work partially supported by Dell Technologies and by the U.S.\ National Science Foundation under grants CNS-1925601, CNS-2112471, CNS-1923789 and CNS-2120447.}
}

\makeatletter
\patchcmd{\@maketitle}
  {\addvspace{0.5\baselineskip}\egroup} 
  {\addvspace{-2.8\baselineskip}\egroup} 
  {}
  {}

\makeatother
\maketitle

\glsunset{usrp}

\begin{abstract}
The highly heterogeneous ecosystem of \gls{nextg} wireless communication systems calls for novel networking paradigms where functionalities and operations can be dynamically and optimally reconfigured in real time to adapt to changing traffic conditions and satisfy stringent and diverse \gls{qos} demands.
Open \gls{ran} technologies, and specifically those being standardized by the \oran Alliance, make it possible to integrate network intelligence into the once monolithic \gls{ran} via intelligent applications, namely, xApps and rApps.
These applications enable flexible control of the network resources and functionalities, network management, and orchestration
through data-driven intelligent control loops. 
Recent work has showed how \gls{drl} is effective in dynamically controlling \oran systems. However, how to
design these solutions in a way that manages heterogeneous optimization goals and prevents unfair resource allocation is still an open challenge, with the logic within \gls{drl} agents often considered as a black box.
\textcolor{red}{In this paper, we introduce \pandora, a framework to automatically design and train \gls{drl} agents for Open RAN applications, package them as xApps and evaluate them in the Colosseum wireless network emulator.}
\textcolor{red}{We benchmark $23$ xApps that embed \gls{drl} agents trained using different architectures, reward design, action spaces, and decision-making timescales, and with the ability to hierarchically control different network parameters. We test these agents on the Colosseum testbed under diverse traffic and channel conditions, in static and mobile setups.}
\textcolor{red}{Our experimental results indicate how suitable fine-tuning of the \gls{ran} control timers, as well as proper selection of reward designs and \gls{drl} architectures can boost network performance according to the network conditions and demand. Notably, finer decision-making granularities can improve \gls{mmtc}'s performance by $\sim56\%$ and even increase \gls{embb} Throughput by $\sim99\%$.
}
\end{abstract}

\glsresetall
\glsunset{usrp}

\vspace{-.35cm}
\begin{IEEEkeywords}
Open RAN, O-RAN, Resource Allocation, Network Intelligence, Deep Reinforcement Learning.
\end{IEEEkeywords}

\vspace{-.55cm}
\section{Introduction} \label{Section I}
Programmable, virtualized, and disaggregated architectures are seen as key enablers of \gls{nextg} cellular networks. Indeed, the flexibility offered through softwarization, virtualization, and open standardized interfaces provides new self-optimization capabilities through \gls{ai}. These concepts are at the foundation of the Open \gls{ran} paradigm, which is being specified by the \oran Alliance. Thanks to the \glspl{ric} proposed by \oran (i.e., the near- and non-real-time \glspl{ric}), intelligence can be embedded into the network and leveraged for on-demand closed-loop control of its resources and functionalities~\cite{polese2023understanding}. This is achieved via intelligent applications, called xApps and rApps, which execute on the near- or non-real-time \glspl{ric}, respectively.
Through the \glspl{ric}, these applications interface with the network nodes and implement data-driven closed-loop control based on real-time statistics received from the \gls{ran}, thus realizing the vision of resilient, reconfigurable and autonomous networks.
Since they do not require prior knowledge of the underlying network dynamics~\cite{sutton2018reinforcement}, \gls{drl} techniques are usually preferred in the design of such control solutions for the Open \gls{ran}~\cite{wang2022self,polese2022colo,d2022orchestran}.
\vspace{-.35cm}

\subsection{Related Work} \label{Section RL}
Intelligent control in \oran through xApps has widely attracted the interest of the research community. For example, \cite{johnson2022nexran} proposes the NexRAN xApp to control
and balance the throughput of different \gls{ran} slices. \textcolor{blue}{In \cite{kak2023hexran}, an \oran-compliant \gls{ran} platform is introduced for \gls{mrat} environments, enabling network slicing and slice-specific scheduling. The FlexSlice framework in \cite{EURECOM+7416} addresses \gls{ran} slicing and scheduling with finer control loop granularity for real-time efficiency.} \textcolor{red}{In \cite{wiebusch2023towards}, the authors discuss an \oran-based framework for predictive \gls{ul} network slicing, leveraging a Deep Learning-based xApp for dynamic reconfiguration of \gls{ran} scheduling for \gls{urllc} services. Similarly, in~\cite{yeh2023deep}, the authors present a Deep Learning-based approach to develop a \gls{ran} slicing xApp.}
The authors of~\cite{kouchaki2022actor} develop a \gls{rl} xApp to assign resource blocks to certain users according to their \gls{csi} and with the goal of maximizing the aggregated data rate of the network.
A deep Q-learning-based xApp for controlling slicing policies to minimize latency for \gls{urllc} slices is presented in \cite{filali2023communication}, while \textcolor{red}{a cooperative multi-agent \gls{rl} algorithm for Open RAN slicing and radio resource management, taking into account various \gls{sla} constraints, is discussed in~\cite{10329913}.}
The authors of~\cite{polese2022colo} experimentally evaluate and demonstrate three \gls{drl}-based xApps under a variety of traffic and channel conditions, and investigate how different action space configurations impact the network performance. 
Finally, other research efforts focus on coordinating multiple xApps to control different parameters via a combination of federated learning and team learning\textcolor{red}{~\cite{iturria2022multi,zhang2022team,zhang2022federated,9933014}}.

\textcolor{blue}{\gls{drl} has been widely used for resource allocation~\cite{luong2019applications,alwarafy2021deep} in the wireless communications and networking field. In~\cite{suh2022deep}, the authors propose a \gls{qos}-oriented resource allocation scheme for network slicing and throughput maximization in \gls{5g} and beyond networks. DeepSlicing in~\cite{liu2020deepslicing} relies on a \gls{drl} approach
to determine the number of resources  required by the \glspl{ue} in each slice to ensure their \gls{qos} demands are met.} \textcolor{red}{In~\cite{sun2019dynamic}, authors present a \gls{drl}-based framework for \gls{nextg} \gls{ran} slicing to improve resource utilization and meet the slices' \gls{qos} requirements by leveraging feedback generated by the \gls{dqn} agents.}
In~\cite{yan2023deep}, the authors discuss a two-level scheduling framework to jointly
maximize the \gls{qoe} and \gls{se}. Specifically, the higher layer controller manages bandwidth allocation, while the lower level controller is responsible for scheduling the \gls{prb} and power allocation at a smaller timescale, utilizing \gls{mimo} antenna technology. In~\cite{naderializadeh2021resource} a \gls{marl} approach is introduced for joint optimization of \gls{ue} scheduling and power control in a wireless environment characterized by the coexistence of multiple \glspl{tx} with multiple associated \glspl{ue}. In~\cite{azimi2021energy}, an energy-efficient \gls{ran} slicing scheme utilizing \gls{drl}-assisted learning for resource allocation in \gls{5g} networks is introduced. Specifically, a combination of Deep Learning and \gls{drl} is employed for resource allocation on large and small time-scales, respectively. In~\cite{rahimi2022novel} a hierarchical \gls{drl}-based approach is employed for resource allocation, coupled with a load balancing strategy to enhance energy efficiency while ensuring user fairness from a \gls{qos} perspective.

\begin{figure*}[t!]
  \centering 
  \includegraphics[width=.95\textwidth]{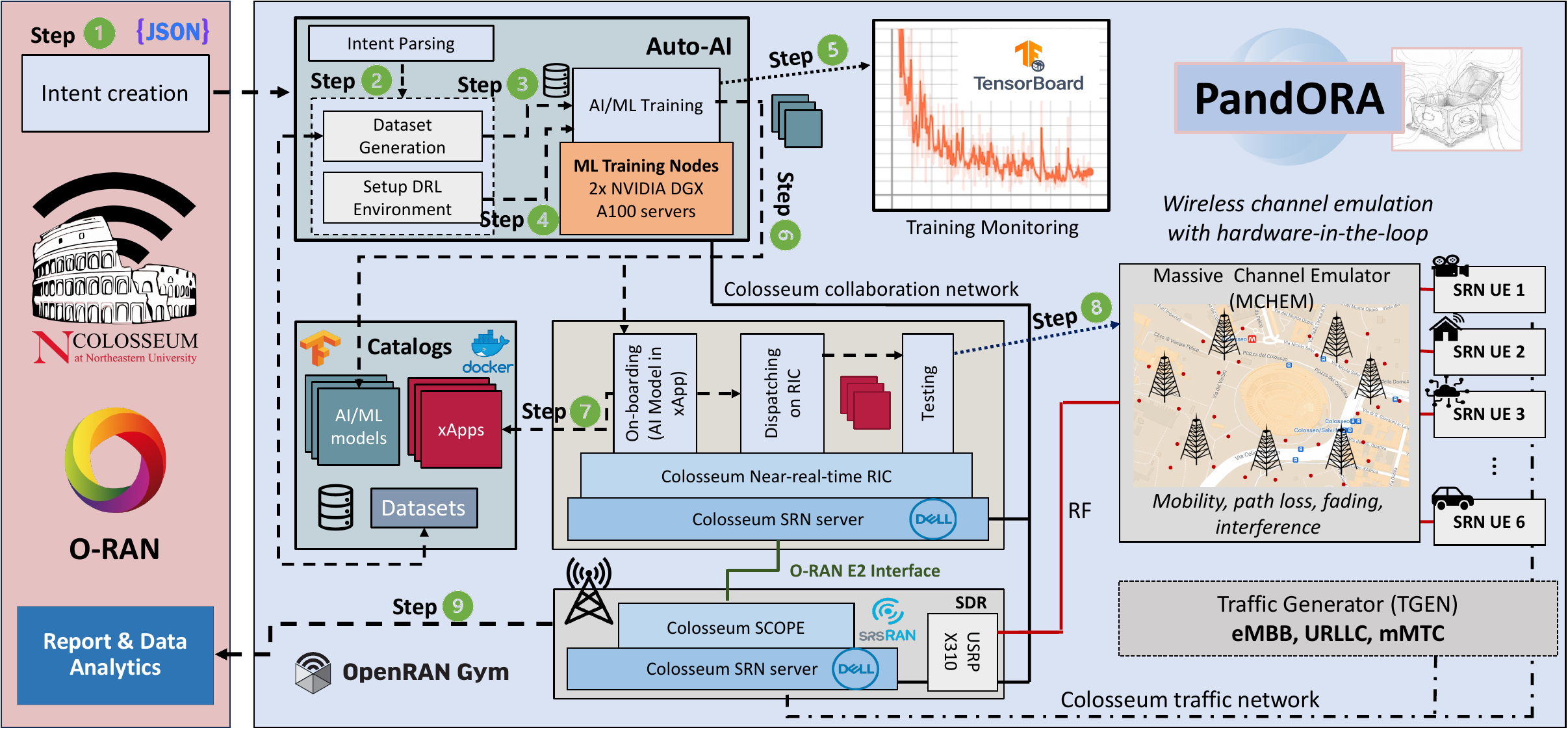} 
   \setlength\abovecaptionskip{-.1cm}
  \caption{\textcolor{blue}{PandORA framework for intent-driven DRL training, xApp on-boarding, and testing with Open RAN in Colosseum.}}
  \label{fig:toolbox}
   \vspace{-0.55cm}
\end{figure*}

\vspace{-0.35cm}
\subsection{Contributions and Outline} \label{Section IB}

The above works clearly show that \gls{drl} and \gls{ai}
are catalysts in the design and development of intelligent control solutions for the Open \gls{ran}. However, despite early results showing their success and effectiveness, designing \gls{drl} agents which are effective at controlling and optimizing complex Open \gls{ran} scenarios---characterized by the coexistence of diverse traffic profiles and potentially conflicting \gls{qos} demands---is still an open challenge that, as we describe below, we aim at addressing in this paper. Specifically, our goal is to go beyond merely using \gls{ai}, and specifically \gls{drl}, in a black-box manner. Instead, we try to address some fundamental design questions that are key for the success of intelligence in Open \gls{ran} systems.

We consider an Open \gls{ran} delivering services to \gls{urllc}, \gls{mmtc} and \gls{embb} network slices.
Specifically, we use OpenRAN Gym~\cite{bonati2023openran}---an open-source framework for \gls{ml} experimentation in \oran---to deploy such Open \gls{ran} on the Colosseum wireless network emulator~\cite{bonati2021colosseum}, and control it through xApps using \textcolor{red}{$23$} different \gls{drl} agent designs.
These xApps have been trained to perform slice-based resource allocation (i.e., scheduling profile selection and \gls{ran} slicing control) and to meet the diverse requirements of each slice.
We investigate the trade-off between long-term and short-term rewards, we discuss and compare different design choices of action set space and \textcolor{blue}{\gls{drl} architecture}, hierarchical decision-making policies and action-taking timescales.
Finally, we show how these choices greatly impact network performance and affect each slice differently.  

\textcolor{blue}{To the best of our knowledge, in~\cite{tsampazi2023comparative}, we conducted the first experimental study that comprehensively evaluated the design choices for \gls{drl}-based xApps to provide insights on the design of xApps for \gls{nextg} Open \glspl{ran}.
In this paper, we extend our previous work~\cite{tsampazi2023comparative} and
present \pandora,
a large-scale evaluation and profiling of \gls{drl} agents for Open RAN, leveraging a framework to automate the training of \gls{drl} agents and their on-boarding as xApps to be executed in the near-real-time \gls{ric}. This all-in-one solution, spanning from the automated end-to-end streamlining of \gls{drl} model training, to testing on an Open \gls{ran} network, is enabled by the Colosseum testbed. Moreover, we extend the analysis in \cite{tsampazi2023comparative} by considering two new directions that affect \gls{drl}-based xApp design. Specifically, (i) we investigate the trade-offs between using a global \gls{drl} agent that takes decisions for all slices against the case of training and deploying per-slice dedicated xApps; and (ii) we analyze the effect of \gls{ran} control timing (i.e., the temporal granularity we use to compute and enforce new control actions). We explore the aforementioned directions by training a variety of \gls{drl} agents, which we then onboard on xApps. We test these agents both under network conditions similar to those encountered in the training dataset (i.e., in-sample experimental evaluation), as well as under previously unseen network conditions not encountered during the training process (i.e., out-of-sample experimental evaluation).}

The remainder of this paper is organized as follows.
\textcolor{blue}{Section~\ref{Section II} describes the \pandora System Model and Evaluation Framework.}
Section~\ref{Section III} presents the different \gls{drl} optimization strategies considered in this work, while Section~\ref{Section IV} details our experimental setup and training methodology, through \textcolor{blue}{\pandora.} 
Experimental results are discussed in Sections~\ref{sec:experimental-evaluation},~\ref{sec:experimental-evaluation2} and\textcolor{red}{~\ref{sec:generalization-pandora}}.
Finally, Section~\ref{conclusion} draws our conclusions.

\vspace{-.35cm}
\section{The \pandora System Model and Evaluation Framework} \label{Section II}

Before providing thorough implementation details of the \pandora system design and its operational procedures (Section~\ref{sec:pandora}), we first briefly introduce the \oran system model considered in this paper (Section~\ref{sec:system_model}).\\

\vspace{-0.65cm}

\subsection{System Model and Reference Use-Case Scenario} \label{sec:system_model}


In this work, we consider an Open \gls{ran} multi-slice scenario where \glspl{ue} generate traffic with diverse profiles and \gls{qos} demands. Without loss of generality, we assume that traffic generated by \glspl{ue} can be classified into \gls{embb}, \gls{urllc}, or \gls{mmtc} slices.  

We focus on the case where a set of xApps execute in the near-real-time \gls{ric} to intelligently control the resource allocation process in a way that satisfies the diverse \gls{qos} demands required by each slice. These xApps take these control decisions by leveraging \gls{ai}/\gls{ml} algorithms that can control \gls{bs} parameters and functionalities such as \gls{ran} slicing (i.e., the portion of available \glspl{prb} that are allocated to each slice at any given time) and \gls{mac} layer scheduling policies (i.e., how such \glspl{prb} are internally allocated to users belonging to each slice). \textcolor{red}{It is worth mentioning that xApps can assign a certain scheduler profile (to be selected among \gls{rr}, \gls{wf} and \gls{pf}) to each slice, but do not take \gls{tti}-level scheduling decisions, which is instead left to the \gls{gnb}.} xApps make decisions based on \gls{ue} traffic demand, load, performance and network conditions that are given by \glspl{kpm} periodically reported by the \gls{ran}. 

\vspace{-0.45cm}

{\color{blue}
\subsection{\pandora Overview and Procedures}\label{sec:pandora}

The architecture and building blocks of \pandora are illustrated in Fig.~\ref{fig:toolbox}. These are key to our \textcolor{red}{extensive} evaluation \textcolor{red}{campaign}, as they streamline the \gls{drl} design and xApp evaluation. 
\pandora leverages automated pipelines that cover the end-to-end lifecycle of \gls{ai} development for \oran systems. Specifically, it embeds the following components and functionalities:
\begin{itemize}
    \item \textbf{Catalogs}. These include \gls{ai}/\gls{ml} models trained offline, xApps, and datasets that can be used to train and test data-driven solutions.
    \item \textbf{Intent-driven Auto-\ai Engine.} This is a fully-customizable software component that parses training intents defined in JSON format and automatically extracts the necessary data from datasets in the catalog. Once the training process is instantiated, \ai solutions are trained based on the intent, and the \ai training is configured to match the desired inputs, outputs, and goals. It also offers web-based dashboards (e.g., TensorFlow's TensorBoard~\cite{tensorflow2015-whitepaper}) to monitor the training phase.
    \item \textbf{xApp On-boarding Module.} Templates and pipelines are provided to convert trained \gls{ai} solutions into xApps that are subsequently published to the catalog.
    These xApps are composed of two interconnected units, namely a service model connector and a data-driven logic unit. The former is responsible for handling the communication with the near-real-time \gls{ric} (e.g., extracts live \glspl{kpi} from the \gls{ran} nodes or  sends control actions, via the E2 termination), while the latter is tasked with embedding trained \ai solutions~\cite{bonati2023openran}.
    
    \item \textbf{xApp Dispatcher Module.} This module automates the instantiation of xApps in order to facilitate their \emph{testing} process. This is done through scripts that automatically create Docker containers of the xApps to dispatch and deploy them on the near-real-time \gls{ric}.
    \item \textbf{Integration with OpenRAN Gym and Colosseum.} \pandora is seamlessly integrated with OpenRAN Gym~\cite{bonati2023openran} and Colosseum~\cite{bonati2021colosseum} to enable end-to-end \oran-compliant testing of xApps in a reproducible RF environment. Via OpenRAN Gym, 
    \pandora also offers data collection and analytics functionalities that are useful to validate xApp behavior and evaluate their performance under diverse conditions and deployments.
\end{itemize}
}

\vspace{-0.15cm}
\textcolor{blue}{
In a nutshell, \pandora users (e.g., telco operators, xApp developers) can generate an intent in JSON format (Step~$1$ in Fig.~\ref{fig:toolbox}) by specifying the control objective (e.g., maximizing a certain set of \glspl{kpm} for each slice), desired control parameters and observable inputs (e.g., throughput measurements, \gls{cqi}, to name a few). \pandora parses the intent (Step $2$) and: (i) produces a dataset to be used to train the \ai algorithms (Step $3$); (ii) configures the \ai algorithm to be trained, e.g., a \drl agent based on the \gls{ppo} architecture controlling slicing policies to maximize a certain reward function (Step $4$); and (iii) trains the \ai solution while offering a dashboard to monitor the training phase (Step $5$). Once the training phase is complete, the \gls{ai}/\gls{ml} models are published to the Catalog, and \pandora integrates with OpenRAN Gym~\cite{bonati2023openran} to convert the trained \ai solution into an xApp (Step $6$), which then gets published to the respective xApp Catalog (Step $7$). Once the publishing is complete, the xApp can be tested on Colosseum under one or more Colosseum scenarios (Step $8$), and testing data is collected and stored to enable performance evaluation, xApp validation and data analytics (Step $9$). 
}

{\color{blue}
\pandora is developed in Python using Tensorflow libraries~\cite{tensorflow2015-whitepaper}. Although it supports the training of any type of \ai/\ml model, for the scope of this work, we primarily focus on training \drl agents which are the state-of-the-art for intelligent decision making in \oran~\cite{wang2023resource, filali2023communication,fiandrino2023explora,brik2022deep}. 

Following \oran specifications and requirements~\cite{polese2023understanding}, \pandora  trains \ai solutions offline using an extended and customized version of the \texttt{tf\_agents} library~\cite{TFAgents} where the configuration of the training environment, the \drl agents and the datasets are generated at run-time starting from a JSON-based intent file.
}

{\color{blue} 
 A simplified example of a minimal intent document is shown in Listing~\ref{lst:intent}. The example shows how it is possible to \textcolor{red}{specify the slices} to be controlled by the agent, the actions that can be controlled, the \glspl{kpi} for each slice 
 that constitute the observation of the state, and 
 which should be considered to compute the reward. 
 \pandora \emph{intents} use a modular approach where \pandora users can specify slice-specific rewards and set a global reward to combine them. 
 In Listing~\ref{lst:intent}, we illustrate a case where a \drl agent is trained to maximize both the average throughput and  number of transmitted packets in the \gls{dl} for the \gls{embb} slice (Lines $4$ - $9$), \textcolor{red}{while maximizing the average number of \gls{dl} transmitted packets for the \gls{mmtc} slice (Lines $10$ - $15$)} and minimizing the maximum \gls{dl} buffer occupancy for the \gls{urllc} slice (Lines $16$ - $21$).\footnote{Note that the weight associated to the \gls{urllc} slice is configured in Line $25$ and has negative value. Therefore, the $\texttt{MaxElemReward}$ function in Line $18$ effectively results in minimizing the maximum buffer occupancy. \textcolor{red}{It is worth mentioning that the \gls{gnb} does not have direct access to end-to-end latency measurements, therefore we use the buffer occupancy as a proxy for latency~\cite{tsampazi2023comparative}.}} 
 
 These three slice-specific rewards are then combined using a global reward that aims at maximizing the weighted sum of the individual reward contributions for each slice (weights are defined at Line $25$).

 Users can specify a general control action space (e.g., controlling \ran slicing policies and scheduling profile as shown in Line $23$) without the need to specify the explicit values of these control actions. At run-time, \pandora parses the JSON-formatted intent and generates a dataset that fulfills the intent. This dataset includes all necessary \glspl{kpm} required to compute observations and rewards. Additionally, it generates the proper action space from the available datasets. For instance, \pandora processes these datasets, automatically determining the available scheduling profiles (e.g., \gls{rr}, \gls{wf}, and \gls{pf} as described in Section~\ref{sec:system_model}), along with the \ran slicing policies suitable for the given slices. More advanced \pandora users can also specify the subset of actions that must be considered by the agent. For example, they can force the agent to only consider PF and RR schedulers and exclude \gls{wf}. 

 Apart from offering modules to generate custom reward functions, \pandora provides a set of pre-defined reward functions that can be selected and combined to generate a global reward. This includes rewards aimed at maximizing or minimizing specific metrics through average, maximum, minimum, and median values. It enables the prioritization of certain \glspl{kpm} by configuring weights, as demonstrated in Line $25$. 
\pandora also offers pre-configured slice configurations with tailored rewards and observable \glspl{kpm} that users can utilize. These configurations serve as default values for each slice, and they can be overrided by the user.
}

It is also worth noting that intents, slice configurations, rewards, actions, and observation configurations are modular and can be combined to generate new intents. In this way, new intents specific to each slice are created, which are later combined, providing further flexibility to \pandora and its users.

 Via \texttt{tf\_agents}~\cite{TFAgents}, \pandora provides access to a range of \drl agents, including \gls{dqn}, \gls{ppo}, \gls{ddpg}, \gls{td3}, and various others. Users can utilize \pandora to select the type of agents they want to use, along with specifying hyperparameters. Once the selection is complete, \pandora configures the agents at run-time, adjusting action and observation spaces as well as rewards based on the specified intent. Similarly to intents, users have the flexibility to customize their \drl architecture and hyperparameters or use default values pre-configured in \pandora.

\begin{lstlisting}[float=t,floatplacement=b,language=json,
caption={{\color{blue}A simplified example of a JSON intent for \pandora.}},
label={lst:intent}]
{
  "intent": {
    "slices": [
      {
        "name": "embb",
        "reward": "MaxAverageReward",
        "reward_KPIs": ["dl_brate", "dl_tx_pkts"],
        "observation_KPIs": ["dl_buffer", "dl_tx_pkts"]
      },
      {
        "name": "mmtc",
        "reward": "MaxAverageReward",
        "reward_KPIs": ["dl_tx_pkts"],
        "observation_KPIs": ["dl_brate", "dl_tx_pkts"]
      },
      {
        "name": "urllc",
        "reward": "MaxElemReward",
        "reward_KPIs": ["dl_buffer"],
        "observation_KPIs": ["dl_buffer", "dl_brate"]
      }
    ],
    "actions": ["scheduling", "ran_slicing"],
    "global_reward_type": "NestedSumWeightedReward",
    "global_reward_weights": [0.5, 0.25, -1]
  }
}
\end{lstlisting}

\vspace{-0.25cm}
\subsection{DRL Agent Architectures tested in this work}\label{Section IIA}

\textcolor{blue}{To control \gls{ran} slicing, and scheduling, we focus on data-driven approaches that rely on \gls{rl}. In \gls{rl}, an \emph{agent} iteratively interacts with the \emph{environment} by performing \emph{observations}, in order to learn the optimal control \emph{policy} that maximizes the desired cumulative \emph{reward}. More specifically, the \emph{agent} explores the \emph{environment} and takes \emph{actions} in several environmental \emph{states}, without knowing a priori which actions are more beneficial, and eventually learns the best policy through experience. The \emph{reward} is a metric that defines the effectiveness of an \emph{action}, while a sequence of \emph{states}, \emph{actions}, and \emph{rewards} that result in a terminal \emph{state} is called an \emph{episode}. \emph{State Space $S$}, represents all the possible states of the environment $s \in S$, while \emph{Action Space $A$} defines all the feasible actions $a \in A$ that can be taken by the \emph{agent}. Finally, in this work, we focus on \gls{drl} agents due to their ability to learn directly from experience, without relying on pre-existing models or explicit knowledge of the wireless environment~\cite{mnih2015human,kaloxylos2021ai,yang2019application} and therefore of its complex and dynamically changing conditions. Moreover, we focus on discrete actions and, for this reason, we limit our analysis to \gls{ppo}~\cite{schulman2017proximal} and \gls{dqn}~\cite{mnih2013playing} architectures which are two state-of-the-art algorithms respectively for on-policy and off-policy \gls{drl} for discrete action spaces.}

\textcolor{blue}{\textbf{\gls{ppo}} has been demonstrated several times to outperform other architectures~\cite{polese2022colo, kouchaki2022actor}. It is based on an actor-critic network architectural approach,} where the actor and critic network \blockquote{work} cooperatively to learn a policy that selects actions that deliver the highest reward possible for each state. While the actor's task is to take actions based on current network states, the critic's target is to evaluate actions taken by the actor network and provide feedback that reflects how effective the action taken by the actor is. In this way, the critic helps the actor in taking actions that lead to the highest rewards for each given state. \textcolor{blue}{The Clipped Surrogate Objective function used by \gls{ppo} is defined in Eq.~\eqref{eq:clip-surog-fun}}

\textcolor{blue}{
{\small
\begin{equation}\label{eq:clip-surog-fun}
\begin{aligned}
   L^{\text{CLIP}}(\theta)=\hat{\mathbb{E}}_t\left[\min \left(q_t(\theta) \hat{A}_t, \operatorname{clip}\left(q_t(\theta), 1-\epsilon, 1+\epsilon\right) \hat{A}_t\right)\right],
\end{aligned}
\end{equation}%
}
}

\noindent
\textcolor{blue}{where $\hat{\mathbb{E}}_t$ represents the empirical average; $\hat{A}_t$ is the estimator of the advantage function denoted as $A_t$, which
assesses how well an action performed compared to the expected performance under the current policy; and $q_t(\theta)=\frac{\pi_\theta\left(a_t \mid s_t\right)}{\pi_{\theta_{\text {old }}}\left(a_t \mid s_t\right)}$. This ratio represents the probability of taking action $a_t$ at state $s_t$ following the current policy $\pi_{\theta}$, divided by the respective probability when following the previous policy (i.e., $\pi_{\theta_{\text {old }}}$). 
When multiplied by the estimator of the advantage function at time-step $t$ (i.e., $\hat{A}_t$), the unclipped part
Eq.~\eqref{eq:clip-surog-fun} is obtained, as shown in Eq.~\eqref{eq:unclipped-ratio}, where \gls{cpi} stands for the technique leveraged to avoid large
policy updates~\cite{kakade2002approximately}.}

\textcolor{blue}{
{\small
\begin{equation}\label{eq:unclipped-ratio}
\begin{aligned}
   L^{\text{CPI}}(\theta)=\hat{\mathbb{E}}_t\left[\frac{\pi_\theta\left(a_t \mid s_t\right)}{\pi_{\theta_{\text {old }}}\left(a_t \mid s_t\right)} \hat{A}_t\right]=\hat{\mathbb{E}}_t\left[q_t(\theta) \hat{A}_t\right].
\end{aligned}
\end{equation}%
}
}

\noindent
\textcolor{blue}{
Since maximizing Eq.~\eqref{eq:unclipped-ratio} directly would result in large policy updates, by \emph{clipping} the probability ratio, we constrain the surrogate objective function (i.e., Eq.~\eqref{eq:clip-surog-fun}) and remove the incentive to move the probability ratio (i.e., $q_t(\theta)$) outside the interval $[1-\epsilon, 1+\epsilon]$. $\epsilon$ is a hyperparameter that determines the clip range and is chosen in a way to ensure that the policy update will not be too large, indicating a significant divergence between the old and the current policy. Lastly, we take the minimum of the clipped and non-clipped objectives. Therefore, the final objective serves as a lower pessimistic bound of the unclipped objective. This indicates that we select either the clipped or the non-clipped objective based on $q_t(\theta)$ and the advantage. The final form of the Clipped Surrogate Objective Loss for the actor-critic implementation of \gls{ppo} is given in Eq.~\eqref{eq:final-ppo-obj}. It is a combination of the Clipped Surrogate Objective function (i.e., \small$L_t^{\text{CLIP}}$), the squared-error value loss function (i.e., {\small$L_t^{\text{VF}}(\theta)=\left(V_\theta\left(s_t\right)-V_t^{\operatorname{targ}}\right)^2$)} which is the difference between predicted and target cumulative reward estimations, and the entropy bonus (i.e., \small$S"$), with the latter ensuring sufficient exploration, while \textcolor{blue}{$c_1$,$c_2$ are control parameters.}}

\textcolor{blue}{
{\small
\begin{equation}\label{eq:final-ppo-obj}
\begin{aligned}
  L_t^{\text{CLIP+VF+S}}(\theta)=\hat{\mathbb{E}}_t\left[L_t^{\text{CLIP}}(\theta)-c_1 L_t^{\text{VF}}(\theta)+c_2S"\left[\pi_\theta\right]\left(s_t\right)\right]
\end{aligned}
\end{equation}%
}
}

\noindent
\textcolor{red}{In addition, based on a comparative analysis between models with diverse number of layers and neurons, the actor and critic networks we select for this work both consist of fully-connected neural networks with 3 layers of 30 neurons each.} The hyperbolic tangent serves as the activation function while the learning rate is set to $10^{-3}$.

\textcolor{blue}{The \textbf{\gls{dqn}} algorithm leverages Q-learning principles to estimate the Q-function. The Q-function represents the expected discounted cumulative reward obtained by taking action $a$ in state $s$ and then following the policy $\pi$. The optimal state-action pair is computed using the Bellman Equation, defined in Eq.~\eqref{eq:bellm-eq}}

\textcolor{blue}{
{\small
\begin{equation}\label{eq:bellm-eq}
   Q^*(s, a)=\left[r+\gamma \max _{a^{\prime}} Q^*\left(s^{\prime}, a^{\prime}\right)\right].
\end{equation}%
}
}

\noindent 
\textcolor{blue}{This equation is approximated by a \gls{dqn} that calculates the cumulative future reward, denoted as $r_t$ at time $t$, and discounted by a factor $\gamma \in[0,1]$, while $s'$ and $a'$ represent the state and the action taken in the next time-step. In addition, the algorithm uses a replay buffer to store experience and to cope with instabilities caused by correlation between consecutive episodes. The set of experiences is denoted as $D_t=\left\{e_1, \ldots, e_t\right\}$, where each $e=\left(s, a, r, s'\right)$ is a tuple that represents the state, action, reward, and the next state. The system transitions to the next state $s'$ after the agent takes an action $a$ at each time-step $t$ during the training phase, while the experience vector is stored in the replay buffer. In order to reduce the correlation between the Q-function value and the optimal $Q^*$ value, a second Q-Network is employed, namely the target Q-Network. Both the main and target Q-Network share the same structure, while the weights of the latter are periodically updated to match those of the former. During the training phase, the replay buffer is used to randomly select batches of previous experiences. These experiences are then utilized to calculate the \gls{dqn} weights, denoted as $\theta_t$ in Eq.~\eqref{eq:dqn-loss}. This calculation involves minimizing the loss function using \gls{sgd} method. Concludingly, the loss equation leveraged during the Q-Learning update at iteration $t$ is given by Eq.~\eqref{eq:dqn-loss}}

\textcolor{blue}{
{\small
\begin{equation}\label{eq:dqn-loss}
\begin{aligned}
L_t\left(\theta_t\right)=\mathbb{E}_{s, a, r, s^{\prime}}\left[\left(r+\gamma \max _{a^{\prime}} Q\left(s^{\prime}, a^{\prime};\theta_t^{-}\right)-Q\left(s,a; \theta_t\right)\right)^2\right],
\end{aligned}
\end{equation}%
}
}

\noindent
\textcolor{blue}{where $\theta_t^{-}$ are the weights of the target Q-Network, and $\gamma$ is the discount factor that prioritizes  instantaneous rewards against long-term rewards. Additionally, the $\epsilon_{\text{DQN}}$-greedy is employed, where the \gls{drl} agent's choice of action depends on the parameter $\epsilon_{\text{DQN}}\in[0,1]$. In detail, with a probability of $1-\epsilon_{\text{DQN}}$, an action computed by the \gls{dqn} is selected. Conversely, with a probability of $\epsilon_{\text{DQN}}$, the agent selects a random action from the action space $A$. This strategy effectively balances the exploration of new actions, randomly selected, and the exploitation of the agent's learned knowledge, based on actions selected by the \gls{dqn}. In our analysis, the latter is implemented as a feed-forward multi-layer \gls{nn}, consisting of $5$~hidden layers of $50$~neurons each. The learning rate is set to $10^{-3}$, while the discount factor is $\gamma=0.95$. Finally, the replay buffer memory size is set to $10,000$, while $\epsilon_{\text{DQN}}=0.1$.}

We follow the same approach as in \cite{polese2022colo}, where the input (e.g., \glspl{kpm}) of the \gls{drl} agent is first processed by the encoding part of an autoencoder for dimensionality reduction. This also synthetically reduces the state space and makes training more efficient in terms of time and generalization. In detail, the autoencoder converts an input matrix of $K=10$ individual measurements of $M=3$ \gls{kpm} metrics (i.e., \gls{dl} throughput, buffer occupancy, and number of transmitted packets) into a single $M$-dimensional vector. The \gls{relu} activation function and four fully-connected layers of $256$, $128$, $32$ and $3$ neurons are also used in the encoder.
\textcolor{red}{Although the channel state information (e.g., \gls{cqi}, \gls{mcs}) is not directly fed to the AI/ML agents, it is worth mentioning that its effect on performance is indirectly captured by the \glspl{kpi} used to feed the \gls{drl} agents (e.g., Throughput, Buffer Occupancy, Transmitted Packets), which has been shown to be sufficient to implicitly capturing channel effects on network performance~\cite{polese2022colo}.}

\begin{figure}[t!]
  \centering
  \includegraphics[width=3.45in]{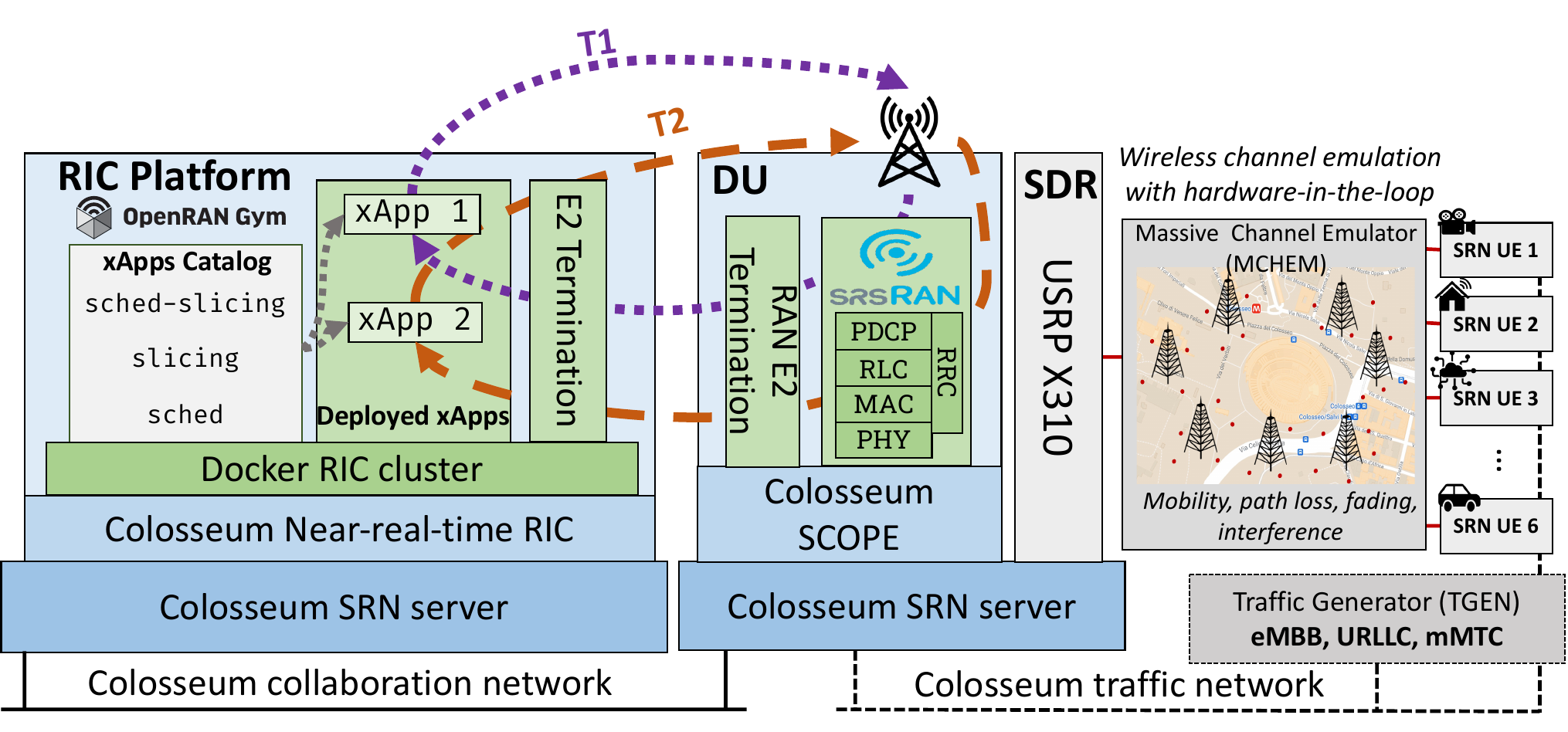} 
  \setlength\abovecaptionskip{-.1cm}
  \caption{Reference \oran testing architecture with focus on the case of two xApps operating at different time scales, $T_{i}$, as described in Section~\ref{Section IV-B}.}
  \label{fig:ext-arch}
  \vspace{-0.5cm}
\end{figure}

\textcolor{blue}{In the case of the joint-slice optimization,} the cumulative average reward function of the \gls{drl} agent is designed to jointly satisfy the \gls{qos} demand of the three slices with respect to their \gls{kpm} requirements. For instance, \gls{embb} users aim to maximize throughput, while \gls{mmtc} users aim at maximizing the number of transmitted packets. Finally, the goal of \gls{urllc} users is to deliver packets with minimum latency. Since the base station cannot measure end-to-end application-layer latency (which is instead measured at the receiver side), we measure latency in terms of number of bytes in the transmission buffer, the smaller the buffer, the smaller the latency. The reward is formulated as the weighted sum in Eq.~\eqref{eq:weighted_reward}

{\small
\begin{equation}\label{eq:weighted_reward}
   R = \sum_{t=0}^{\infty} \gamma^t \left( \sum_{j=1}^{N} w_{j} \cdot r_{j,t} \right), 
\end{equation}%
}%

\noindent
where $t$ represents the training step, and $N=3$ is the total number of slices, $w_{j}$ represents the weight associated to slice $j$, considered for reward maximization in the three corresponding slices. Finally, $\gamma$ is the discount factor and $r_{j,t}$ describes the slice-specific reward obtained at each training step $t$. In our case, $r_{j,t}$ represents the average value of the \gls{kpm} measured by all users of slice $j$ at time $t$ (e.g., throughput for the \gls{embb} slice). Note that the weight $w_{j}$ for the \gls{urllc} slice is negative to model the minimization of the buffer occupancy. The models that we have designed and trained are deployed as xApps on the near-real-time \gls{ric}, as illustrated in Fig.~\ref{fig:ext-arch}.

In the case of the per-slice optimization, the respective weighted average reward function of the \gls{drl} agent is given as follows in Eq.~\eqref{eq:per_slice_reward}

 {\small
 \begin{equation}\label{eq:per_slice_reward}
    R = \sum_{t=0}^{\infty} \gamma^t \left( w_{j} \cdot r_{j,t} \right), 
 \end{equation}%
 }%

\label{Section IIIC}
\begin{figure}[t!]
  \centering
  \includegraphics[width=3in]{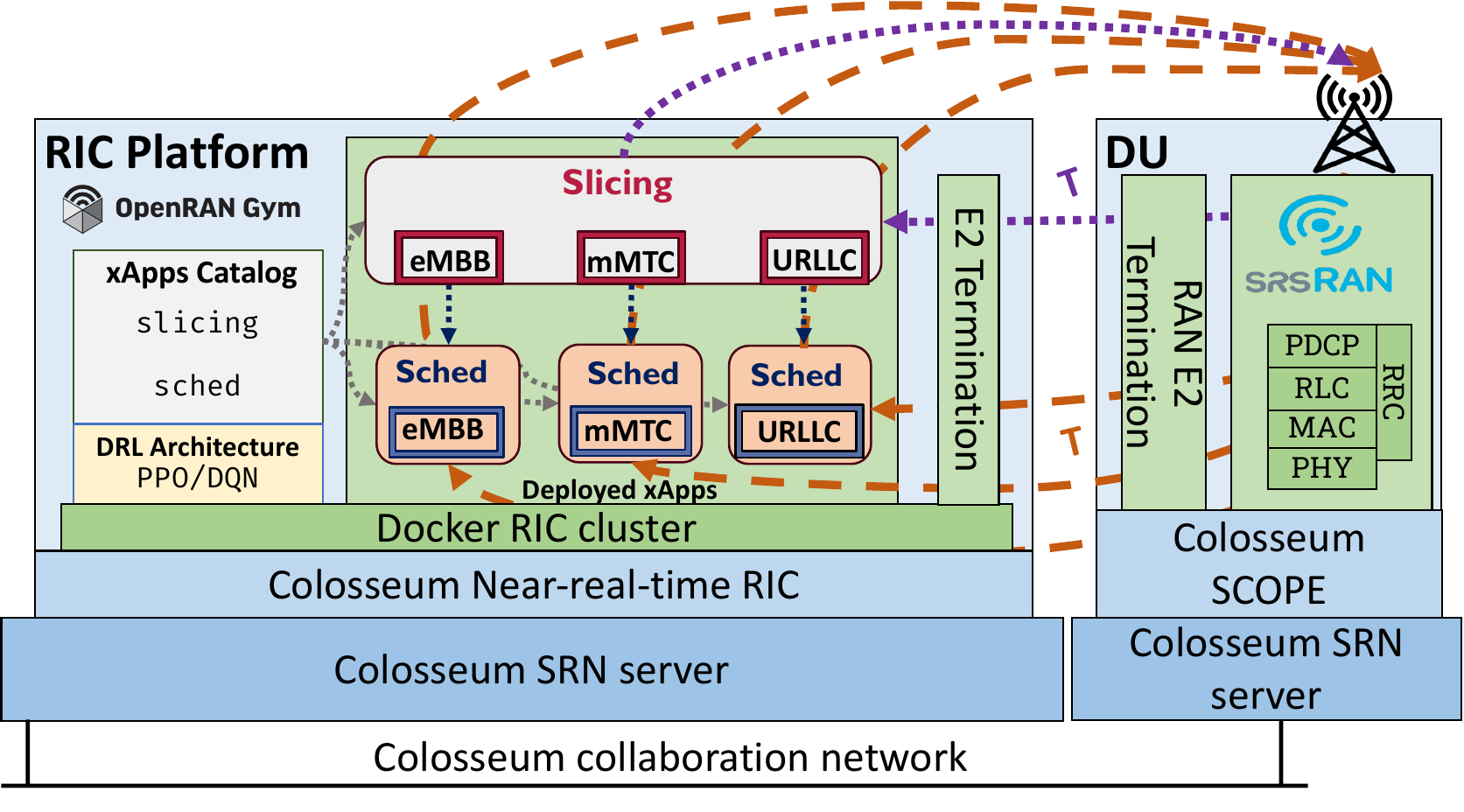}
  \setlength\abovecaptionskip{-.1cm}
  \caption{\textcolor{blue}{Reference \oran testing architecture with focus on the case of four xApps operating at time scale, $T$, as described in Section~\ref{Section IV-C}.}}
  \label{fig:4xapps-arch}
  \vspace{-0.5cm}
\end{figure}

\noindent
\textcolor{blue}{
The models that we have designed and trained are deployed as xApps on the near-real-time \gls{ric}, as illustrated in Fig.~\ref{fig:4xapps-arch}.}

\begin{figure*}[t!]
\centering
\subfigure[\gls{embb} Throughput]{\includegraphics[height=2.85cm]{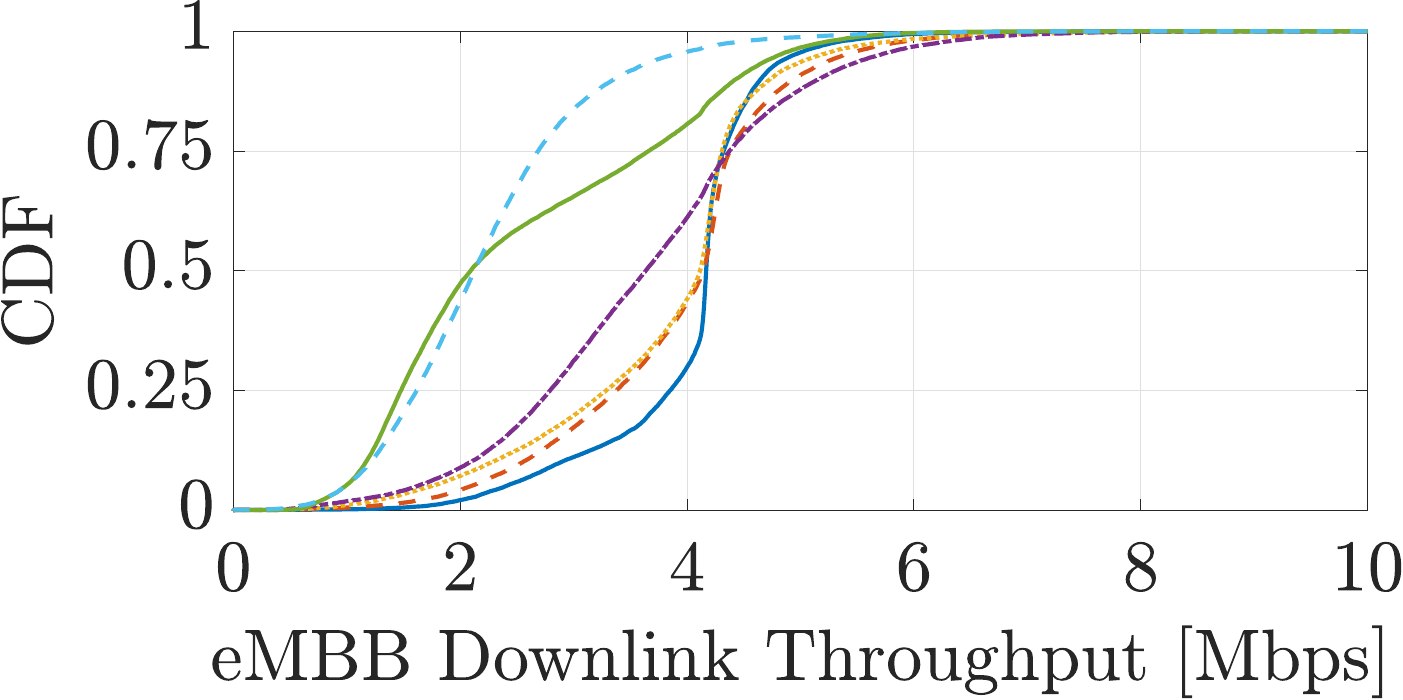}
\label{fig:Figure4a}}
\hfil
\subfigure[\gls{mmtc} Packets]{\includegraphics[height=2.85cm]{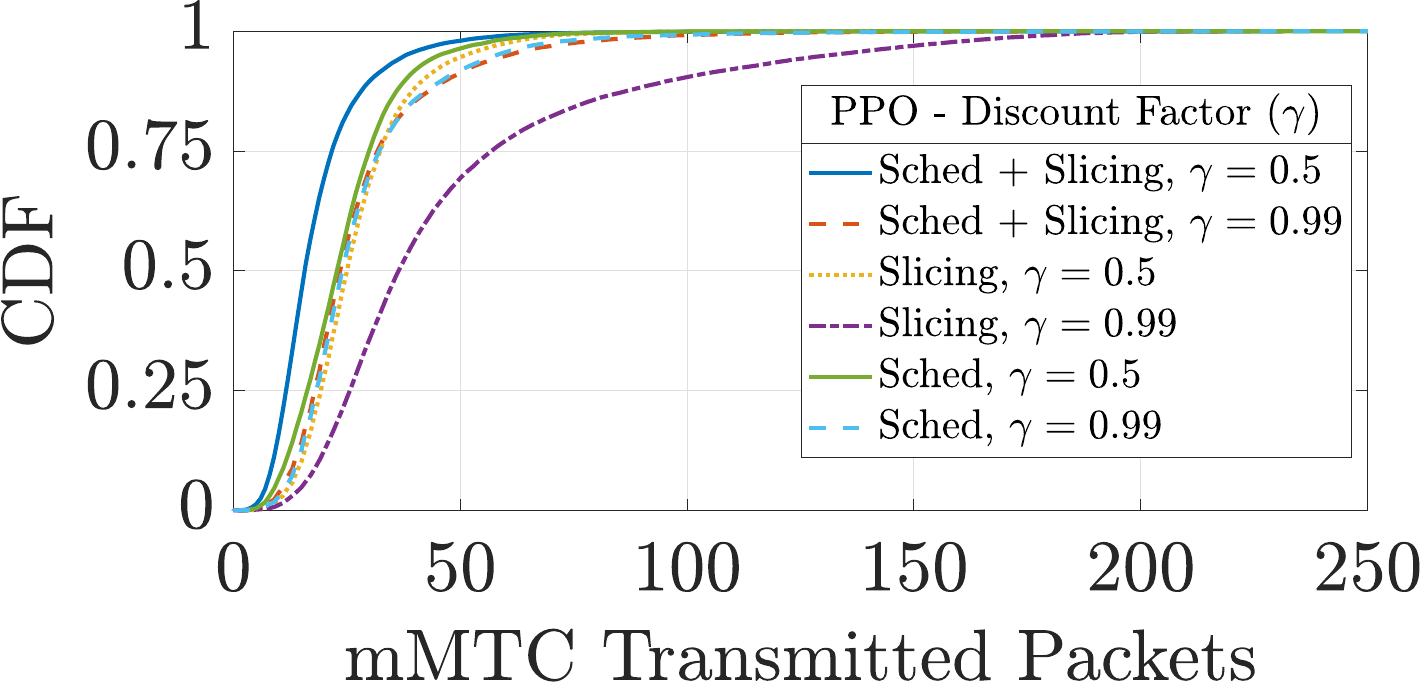}
\label{fig:Figure4b}}
\hfil
\subfigure[\gls{urllc} Buffer Occupancy]{\includegraphics[height=2.85cm]{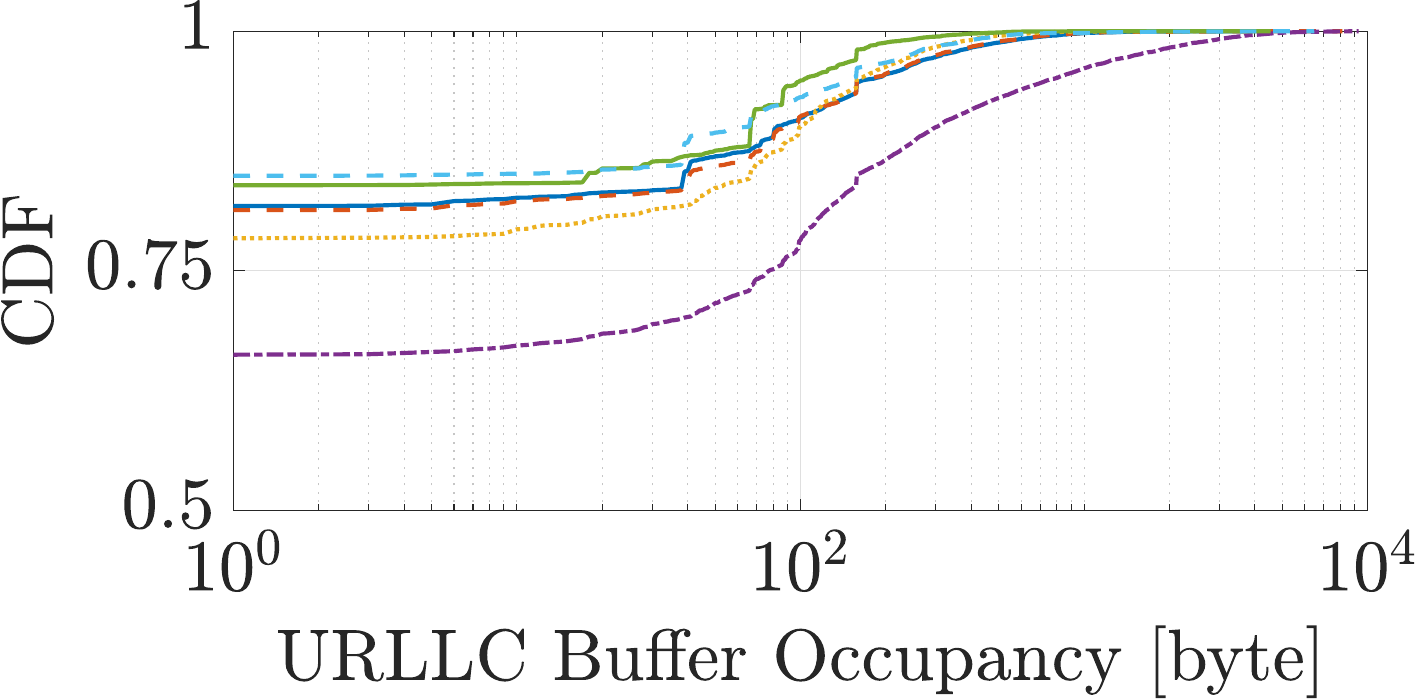}
\label{fig:Figure4c}}
\setlength\abovecaptionskip{-.02cm}
\caption{\textcolor{blue}{Performance evaluation under different action spaces and values of the $\gamma$ parameter with the PPO DRL Architecture.}}
\label{Figure4-1a}
\vspace{-0.55cm}
\end{figure*}

\begin{figure}[t!]
\centering
\subfigure[\gls{embb} DL Throughput]{\includegraphics[width=1.6in]{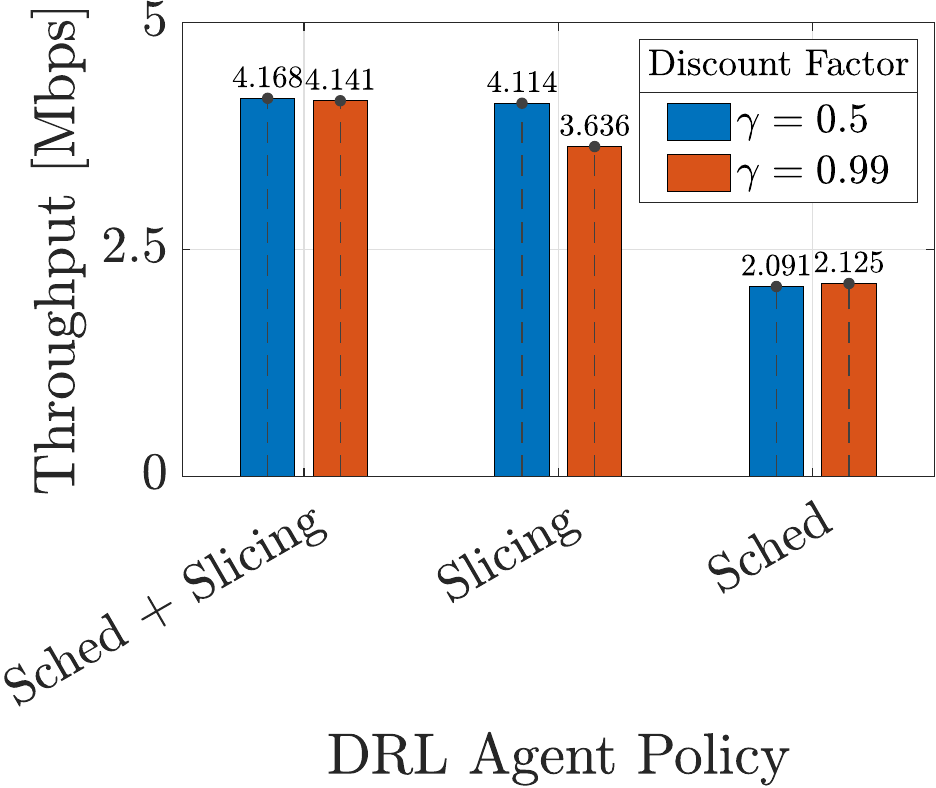}
\label{fig:Figure4d}}
\hfil
\subfigure[\gls{mmtc}  Packets]{\includegraphics[width=1.6in]{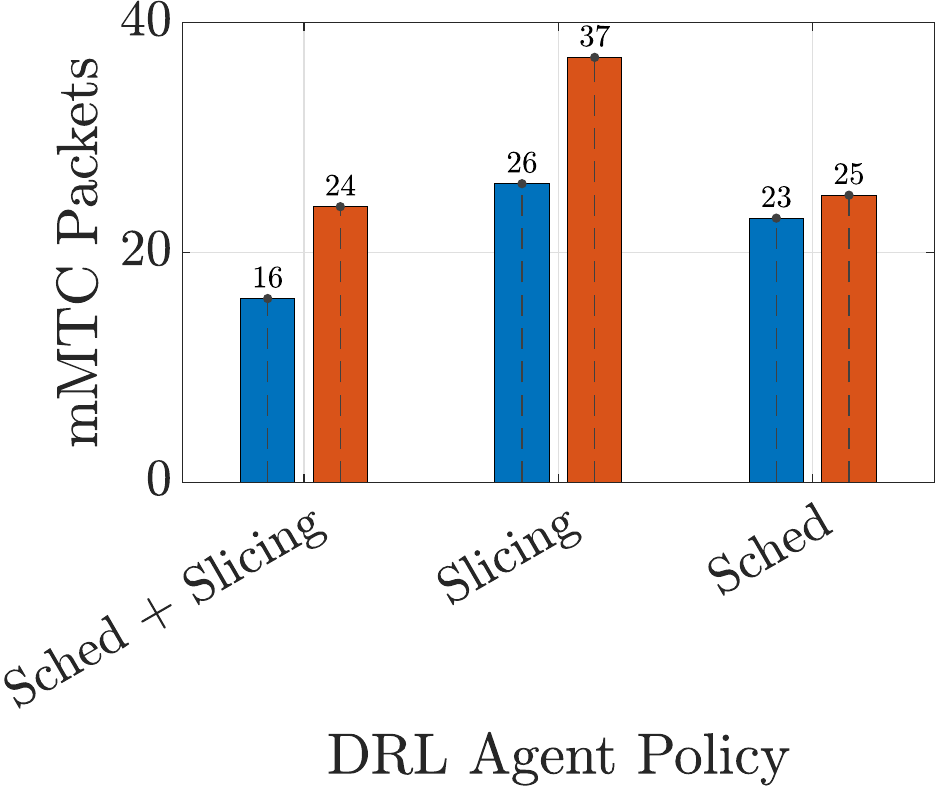}
\label{fig:Figure4e}}
\setlength\abovecaptionskip{-.02cm}
\caption{\textcolor{blue}{Median values under different action spaces and values of $\gamma$ with the PPO DRL Architecture.}}
\label{Figure4-1b}
\vspace{-0.55cm}
\end{figure}

\vspace{-0.35cm}
\section{DRL Optimization Strategies} \label{Section III}

Our goal is to investigate how different design choices affect the effectiveness and decision-making of the \gls{drl}-based xApps. For this reason, we consider the following design choices, which resulted in the generation and testing of a total of \textcolor{red}{$23$} xApps.

\textbf{Short-term vs. Long-term Rewards.}\label{Section IIIA}
We train \gls{drl} agents with different values of the discount factor $\gamma$. The \gls{ppo} discount factor weights instantaneous rewards against long-term rewards. A higher value prioritizes long-term rewards, while a lower $\gamma$ prioritizes short-term rewards. Results of this exploration are provided in Section \ref{Section IV-A}.

\textbf{Hierarchical Decision-Making.} \label{Section IIIB}
We investigate the case of two xApps configuring different parameters in parallel but at different timescales. In this way, we investigate how multiple xApps with different optimization goals and operating timescales impact the overall network performance. The findings of this investigation are provided in Section \ref{Section IV-B}, and
a practical example is illustrated in Fig.~\ref{fig:ext-arch}.

\textbf{Per-Slice Scheduling Profile Selection.} 
\textcolor{blue}{We explore a hierarchical decision-making configuration of four xApps operating simultaneously at the same time granularity and reconfiguring the \gls{bs}'s control parameters. In detail, one slicing xApp simultaneously allocates \glspl{prb} to all slices. The remaining three xApps are exclusively assigned one slice each and select a dedicated scheduling profile for each slice.
In this way, we aim to examine the effects of multiple xApps operating at the same timescale, 
 and to compare a  widely used state-of-the-art off-policy architecture  (i.e., the \gls{dqn}), with its on-policy counterpart (i.e., the \gls{ppo} algorithm). 
A practical example is depicted in Fig.~\ref{fig:4xapps-arch}, while the results of this investigation are presented in Section~\ref{Section IV-C}.}

\textbf{Impact of Reward's Weights.} \label{Section IIID}
We test different values for the weights $w_i$ of the slices in Eq.~\eqref{eq:weighted_reward}. A different weight configuration affects how \gls{drl} agents prioritize each slice. The results of this analysis are reported in Section \ref{Section IV-D}, where we show how weights significantly impact the overall performance and can result in inefficient control strategies.

\textbf{Effect of \gls{ran} Control Timers.} \label{Section IIIE}
\textcolor{blue}{Finally, we aim at understanding how different \gls{ran} control timers (i.e., different periodicities for \gls{ran} telemetry, report and control) will have an impact on the decision-making process and how different control sets will prioritize one slice over the other. Specifically, we look into three control timers namely the \gls{kpm} log time, the \gls{du} Report Timer, and the action update time, i.e., how frequently the actions sent back by the \gls{ric} get updated and enforced by the \gls{bs}. The reported findings of this strategy are provided in Section~\ref{Section IV-E}.
}

\vspace{-0.35cm}

\textcolor{red}{It is reminded that the action set $A$ consists of both scheduling and \gls{ran} slicing policies. In our analysis, we consider the cases where an agent can control either scheduling or slicing decisions individually, or control both jointly. The state $S$ is represented by the output of the autoencoder which is used to compress input \glspl{kpi} collected over the E2 interface and convert them into latent representations. Finally, although in our analysis we evaluate diverse reward designs that consider long-term and short-term goals, and combine diverse target \glspl{kpi} for each slice, the general form of the reward $R$ considered in all of our \gls{drl} agents is defined in Eq.~\eqref{eq:weighted_reward}
}.
\vspace{-.5cm}

\begin{table}[bt]
\centering
\small
\setlength\abovecaptionskip{-.1cm}
\caption{Traffic Profiles}
\begin{adjustbox}{width=1\linewidth}
\begin{tabular}{@{}>{\centering\arraybackslash}p{2cm}@{\hspace{0.1mm}}>{\centering\arraybackslash}p{2cm}@{\hspace{0.1mm}}>{\centering\arraybackslash}p{2cm}@{\hspace{0.1mm}}>{\centering\arraybackslash}p{2cm}@{}}
\toprule
\multicolumn{1}{c}{\textbf{Profile Name}} & \multicolumn{1}{c}{\textbf{eMBB [Mbps]}} & \multicolumn{1}{c}{\textbf{mMTC [kbps]}}  & \multicolumn{1}{c}{\textbf{URLLC [kbps]}} \\ \midrule 
\centering Profile 1 & $1$ & $30$ & $10$ \\
\centering Profile 2 & $4$ & $44.6$ & $89.3$ \\
\bottomrule
\end{tabular}
\end{adjustbox}
\label{table:traffic-profiles}
\vspace{-.55cm}
\end{table}

\section{Experimental Setup and DRL Training}\label{Section IV}

To experimentally evaluate the \gls{drl} agents, we leverage the capabilities of OpenRAN Gym \cite{bonati2023openran}, an open-source experimental toolbox for end-to-end design, implementation, and testing of \gls{ai}/\gls{ml} applications in \oran. It features:
\begin{itemize}
   \item End-to-end \gls{ran} and core network deployments though the srsRAN~\cite{gomez2016srslte} softwarized open-source protocol stack; 

    \item Large-scale data collection, testing and fine-tuning of \gls{ran} functionalities through the SCOPE framework~\cite{bonati2021scope}, which adds open \glspl{api} to srsRAN for the control of slicing and scheduling functionalities, as well as for \glspl{kpm} collection;
    
    \item An \oran-compliant control architecture to execute \gls{ai}/\gls{ml}-based xApps via the ColO-RAN near-real-time \gls{ric}~\cite{polese2022colo}. 
    The E2 interface between \gls{ran} and the \gls{ric} and its \glspl{sm}~\cite{polese2023understanding} manage streaming of \glspl{kpm} from the RAN and control actions from the xApps. 
\end{itemize}

\begin{table}[bt]
\centering
\small
\setlength\abovecaptionskip{-.1cm}
\caption{Weight Configurations}
\begin{adjustbox}{width=0.75\linewidth}
\begin{tabular}{@{}l@{\hspace{3mm}}l@{\hspace{3mm}}l@{\hspace{3mm}}l@{}}
\toprule
\multicolumn{1}{c}{\textbf{Weights}} & \multicolumn{1}{c}{\textbf{eMBB}} & \multicolumn{1}{c}{\textbf{mMTC}}  & \multicolumn{1}{c}{\textbf{URLLC}} \\ \midrule 
\texttt{Default} & $72.0440333$ & $0.229357798$ & $0.00005$ \\
\texttt{Alternative} & $72.0440333$ & $1.5$ & $0.00005$ \\
\bottomrule
\end{tabular}
\end{adjustbox}
\label{table:weight-confs-list}
\vspace{-0.55cm}
\end{table}


\begin{figure*}[t]
\centering
\subfigure[\gls{embb} Throughput]{\includegraphics[height=2.85cm]{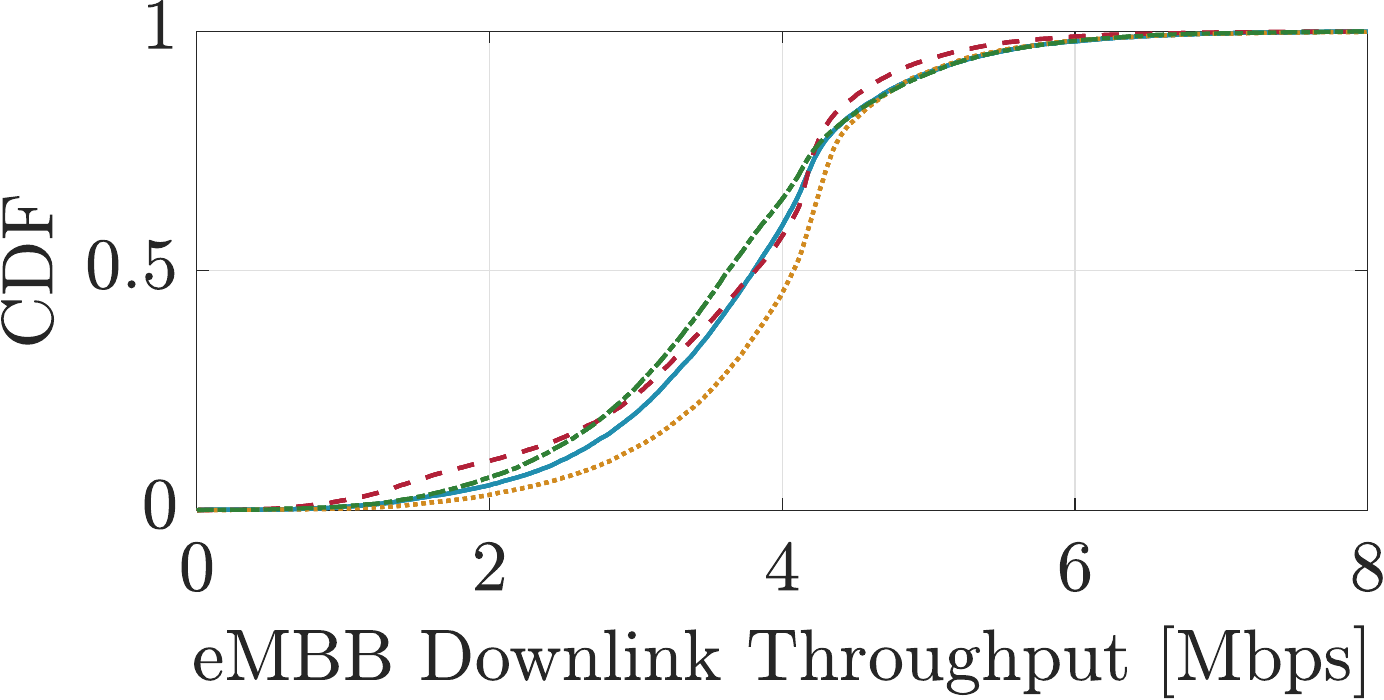}
\label{Figure4a2}}
\hfil
\subfigure[\gls{mmtc} Packets]{\includegraphics[height=2.85cm]{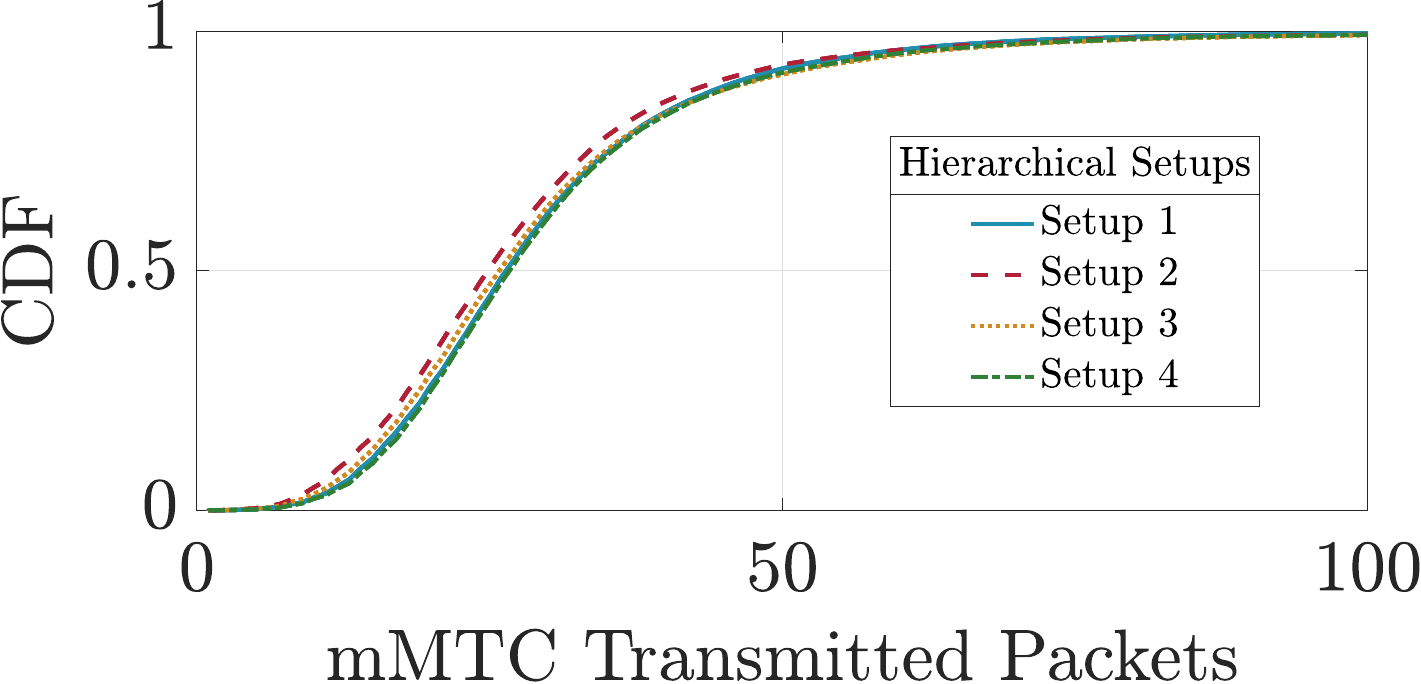}
\label{Figure4b2}}
\hfil
\subfigure[\gls{urllc} Buffer Occupancy]{\includegraphics[height=2.85cm]{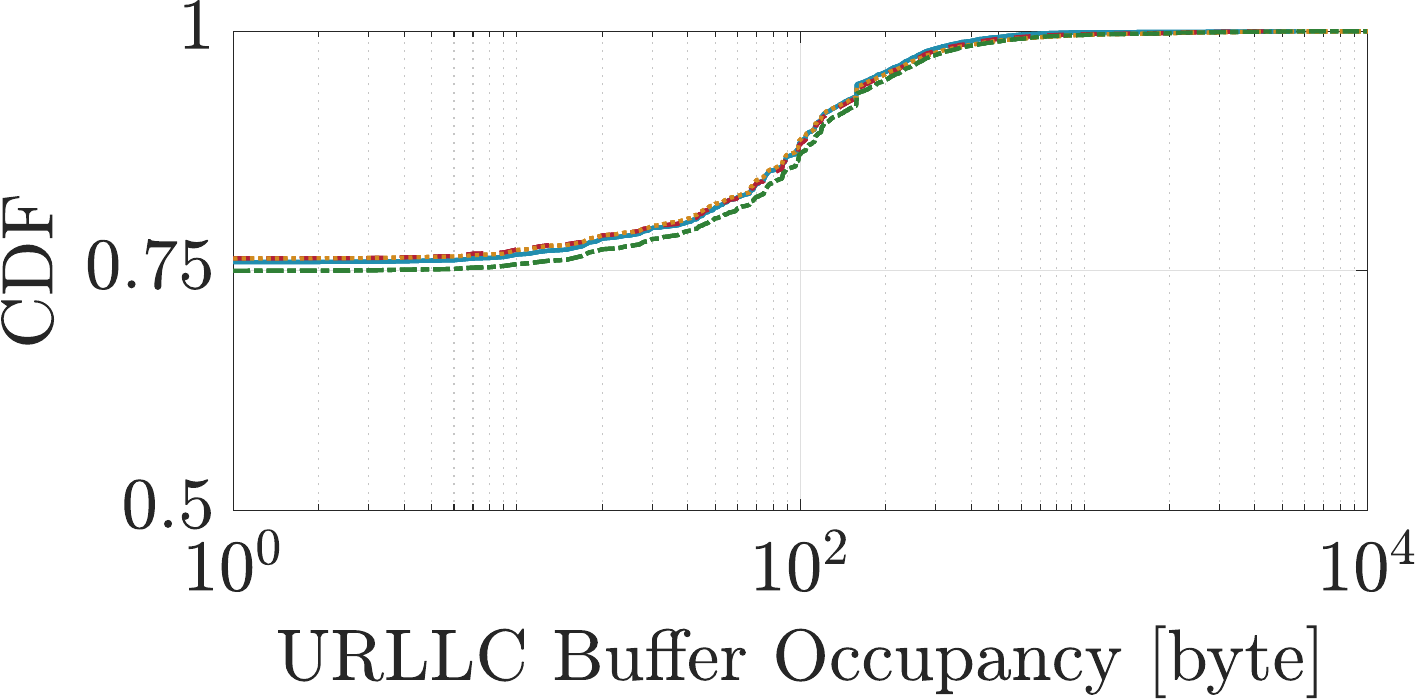}
\label{Figure4c2}}
\setlength\abovecaptionskip{-.02cm}
\caption{Performance evaluation under different hierarchical configurations \textcolor{blue}{with the PPO DRL Architecture.}}
\label{Figure4-2a}
\vspace{-0.55cm}
\end{figure*}

\begin{figure}[t!]
\centering
\subfigure[\gls{embb} DL Throughput]{\includegraphics[width=1.6in]{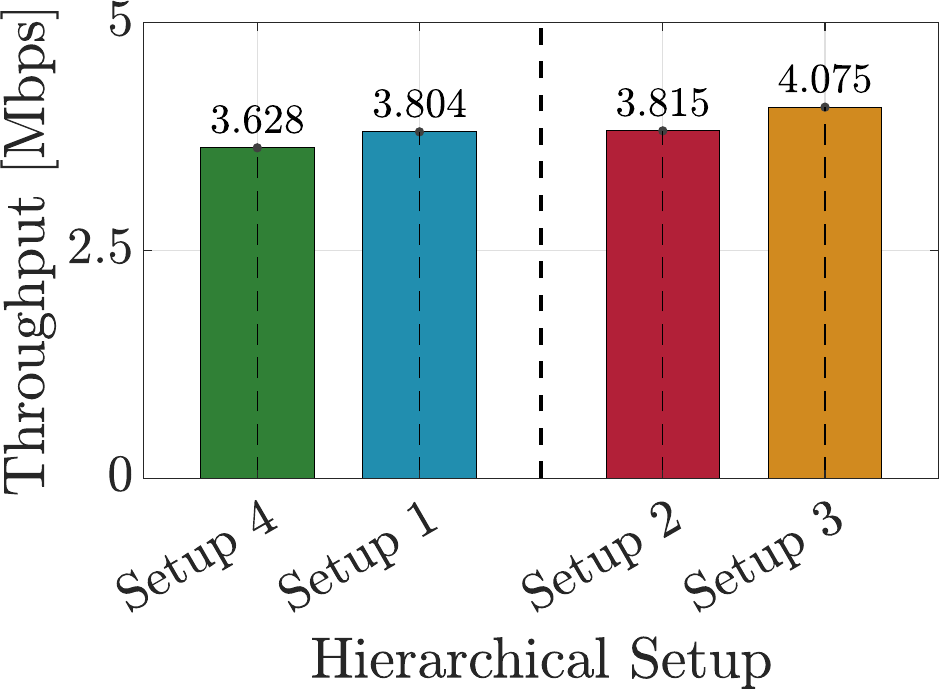}
\label{Figure4d2}}
\hfil
\subfigure[\gls{mmtc}  Packets]{\includegraphics[width=1.6in]{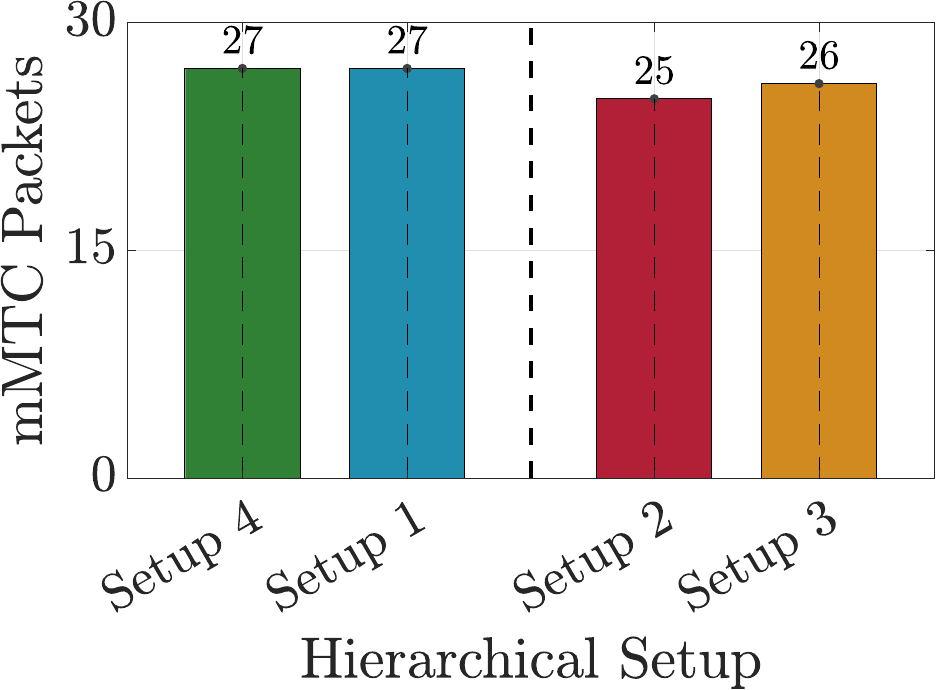}
\label{Figure4e2}}
\setlength\abovecaptionskip{-.02cm}
\caption{Median values under different hierarchical configurations \textcolor{blue}{with the PPO DRL Architecture.}}
\label{Figure4-2b}
\vspace{-0.55cm}
\end{figure}

\textcolor{red}{
We deploy OpenRAN Gym on Colosseum~\cite{bonati2021colosseum}, a publicly available testbed with $128$~\glspl{srn}, i.e., pairs of Dell PowerEdge R730 servers and NI \gls{usrp} X310 \glspl{sdr}.
Colosseum enables large-scale experimentation in diverse \gls{rf} environments and network deployments.
The channel emulation is done through the \gls{mchem} component, which leverages \gls{fpga}-based \gls{fir} filters to reproduce different conditions of the wireless environment (e.g., path loss, fading, attenuation, interference of signals) modeled a priori through ray-tracing software, analytical models, or real-world measurements.
Similarly, the Colosseum \gls{tgen}, built on top of the \gls{mgen} TCP/UDP traffic generator~\cite{mgen}, emulates different traffic types, demand, and distributions (e.g., Poisson, periodic).}

We deploy a 3GPP-compliant cellular network with one base station and $6$~\glspl{ue} uniformly distributed across $3$~different slices.
These are: (i)~\gls{embb} that concerns high traffic modeling of high-quality multimedia content and streaming applications; (ii)~\gls{urllc} for time-critical applications, such as autonomous driving in \gls{v2x} scenarios; and (iii)~\gls{mmtc} for \gls{iot} devices with low data rate requirements
but with high need for consistent information exchange. \textcolor{blue}{In terms of physical deployment, we consider two different topologies both located in the urban environment of Rome, Italy~\cite{bonati2021scope}. Specifically, \textbf{Location 1} concerns \glspl{ue} uniformly distributed within a $50$\:m radius from the \gls{bs}, while in \textbf{Location 2}, the corresponding \glspl{ue} are placed within a $20$ m radius from the \gls{bs}.}

The bandwidth of the \gls{bs} is set to $10$\:MHz (i.e., $50$ \glspl{prb}) and is divided among the $3$ slices, with $2$ users statically assigned to each slice. \textcolor{red}{Slice-based traffic is generated following the specifications reported in Table~\ref{table:traffic-profiles}. Specifically, we consider two different traffic profile configurations (i.e., Profile $1$ and Profile $2$) with different source bitrates. However, for both profiles, we consider a constant bitrate traffic for \gls{embb} users, while \gls{urllc} and \gls{mmtc} \glspl{ue} generate traffic based on a Poisson distribution.}

\begin{table}[b]
\centering
\small
\setlength\abovecaptionskip{-.1cm}
\caption{Per-Slice Top performing xApps under the Default Weight Configuration}
\begin{adjustbox}{width=0.85\linewidth}
\begin{tabular}{@{}l@{\hspace{0.25mm}}ll@{\hspace{0.25mm}}l@{}}
\toprule
\multicolumn{1}{c}{\textbf{}} & \multicolumn{1}{c}{\textbf{eMBB}} & \multicolumn{1}{c}{\textbf{mMTC}} \\ \midrule
\texttt{1)} & \texttt{Sched \& Slicing 0.5} & \texttt{Slicing 0.99}  \\
\texttt{2)} & \texttt{Sched \& Slicing 0.99} & \texttt{Slicing 0.5}  \\
\texttt{3)} & \texttt{Slicing 0.5} & \texttt{Sched 0.99} \\
\texttt{4)} & \texttt{Slicing 0.99} & \texttt{Sched \& Slicing 0.99}  \\
\texttt{5)} & \texttt{Sched 0.99} & \texttt{Sched 0.5}  \\
\texttt{6)} & \texttt{Sched 0.5} & \texttt{Sched \& Slicing 0.5}  \\
\bottomrule
\end{tabular}
\end{adjustbox}
\label{table:default-comp-analysis}
\vspace{-.55cm}
\end{table}

To train the \gls{drl} agents, we used the publicly available dataset described in~\cite{polese2022colo}.
This dataset contains about $8$\:GB of \glspl{kpm} collected by using OpenRAN Gym and the Colosseum network emulator over $89$~hours of experiments, and concerns
setups with up to 7~base stations and 42~\glspl{ue} belonging to different \gls{qos} classes, and served with heterogeneous scheduling policies.
Each \gls{drl} model evaluated in the following sections takes as input \gls{ran} \glspl{kpm} such as throughput, buffer occupancy, number of \glspl{prb}, and outputs resource allocation policies (e.g., RAN slicing and/or scheduling) for \gls{ran} control.

Abiding by the \oran specifications\textcolor{red}{~\cite{alliance2021ran}, which do not permit the
deployment of untrained data-driven solutions}, we train our \gls{ml} models \emph{offline} on Colosseum's GPU-accelerated environment, which includes two NVIDIA DGX A100 servers with $8$~GPUs each. \textcolor{red}{Notably, \gls{ai}/\gls{ml} solutions should be trained offline to avoid actions that can potentially
lead to performance degradation in the network~\cite{polese2023understanding}. Then, the} trained \gls{drl}-agents are onboarded on xApps inside softwarized containers implemented via Docker and deployed on the ColO-RAN near-real-time \gls{ric}. 

\begin{table}[ht]
\vspace{-0.1cm}
\centering
\small
\setlength\abovecaptionskip{-.1cm}
\caption{Hierarchical Reporting Setup}
\begin{adjustbox}{width=0.65\linewidth}
\begin{tabular}{@{}c@{}c@{}c@{}}
\toprule
\multicolumn{1}{c}{\textbf{Setup ID}} & \multicolumn{1}{c}{\textbf{\texttt{Slicing 0.5}}} & \multicolumn{1}{c}{\textbf{\texttt{Sched 0.99}}} \\ \midrule
\textbf{1} & $1$\:s & $10$\:s  \\
\textbf{2} & $1$\:s & $5$\:s  \\
\textbf{3} & $10$\:s & $1$\:s \\
\textbf{4} & $5$\:s & $1$ \:s \\
\bottomrule
\end{tabular}
\end{adjustbox}
\label{table:hier-setups}
\vspace{-0.2cm}
\end{table}

\vspace{-0.35cm}
\section{In-Sample Experimental Evaluation }\label{sec:experimental-evaluation}

In this section, we present the results of an extensive performance evaluation campaign, with more than \textcolor{red}{$38$~hours} of experiments \textcolor{blue}{in Colosseum}, to profile the impact of the strategies discussed in Section~\ref{Section III}. These results were produced by taking the median as the most representative statistical value of a dataset, and averaged over multiple repetitions of experiments in the \gls{dl} direction of the communication system. \textcolor{blue}{Last, all the experiments of the in-sample experimental evaluation campaign concern \textbf{Location 1} which is the configuration used to collect training data, and all the \glspl{ue} are assumed static. Out-of-sample evaluation, i.e., testing the \drl agents against data collected in a entirely different network deployment, will be the focus of Section~\ref{sec:experimental-evaluation2}.}

\vspace{-0.35cm}
\subsection{Impact of Discount Factor on the Action Space}\label{Section IV-A}

We explore how \gls{ran} slicing, \gls{mac} scheduling, and joint slicing and scheduling control are affected by training procedures that favor short-term against long-term rewards. \textcolor{blue}{All the reported results were obtained with $\gamma\in\{0.5,0.99\}$, while
the reward's weight configuration is shown in Table \ref{table:weight-confs-list} and identified as \texttt{Default}.}

In Fig.~\ref{Figure4-1a}, we report the \gls{cdf} of individual \glspl{kpm} for each slice and for different xApps trained to control different sets of actions and using different values of $\gamma$. The median of such measurements for the \gls{embb} and \gls{mmtc} slices is instead reported in Fig.~\ref{Figure4-1b}.
The median for the \gls{urllc} slice is not reported, as this value is zero in all configurations.
The best performing configurations for the \gls{embb} and \gls{mmtc} slices are instead listed in numerical order in Table~\ref{table:default-comp-analysis} from best to worst performing.

\begin{figure*}[t!]
\centering
\subfigure[\gls{embb} Throughput]{\includegraphics[height=2.85cm]{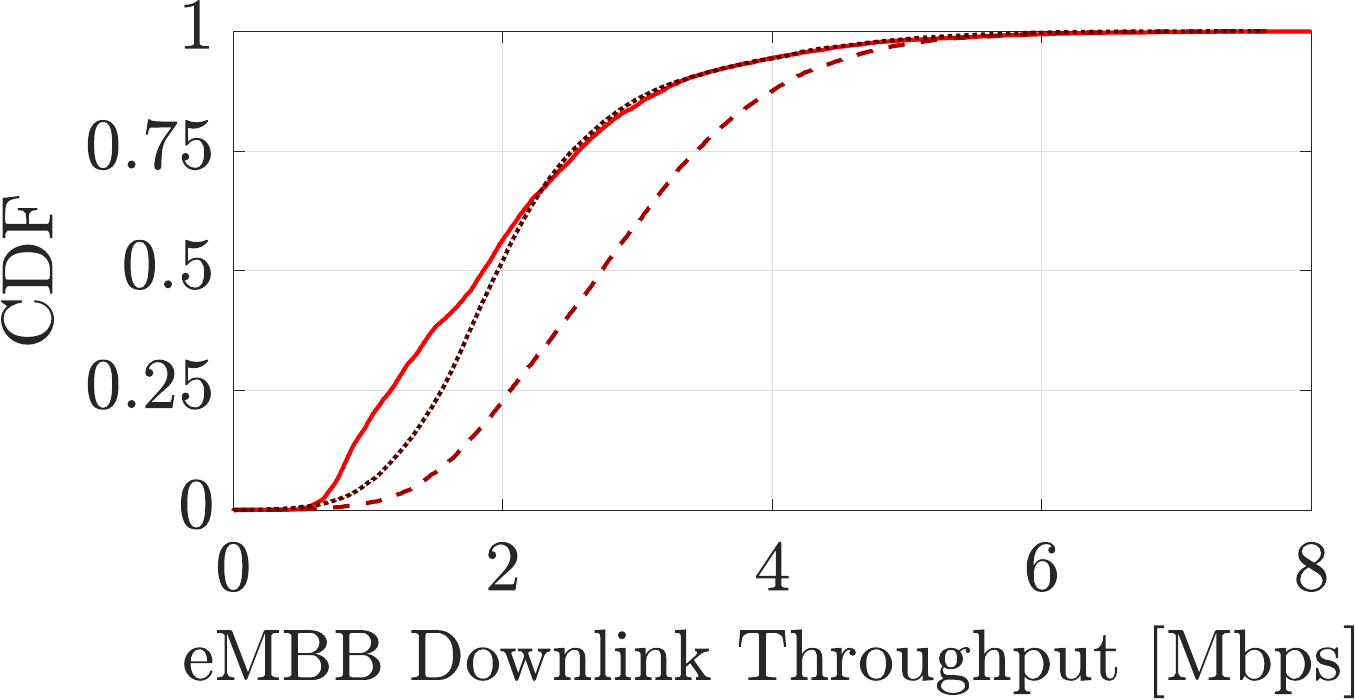}}
\label{fig:Figure9a}
\hfil
\subfigure[\gls{mmtc} Packets]{\includegraphics[height=2.85cm]{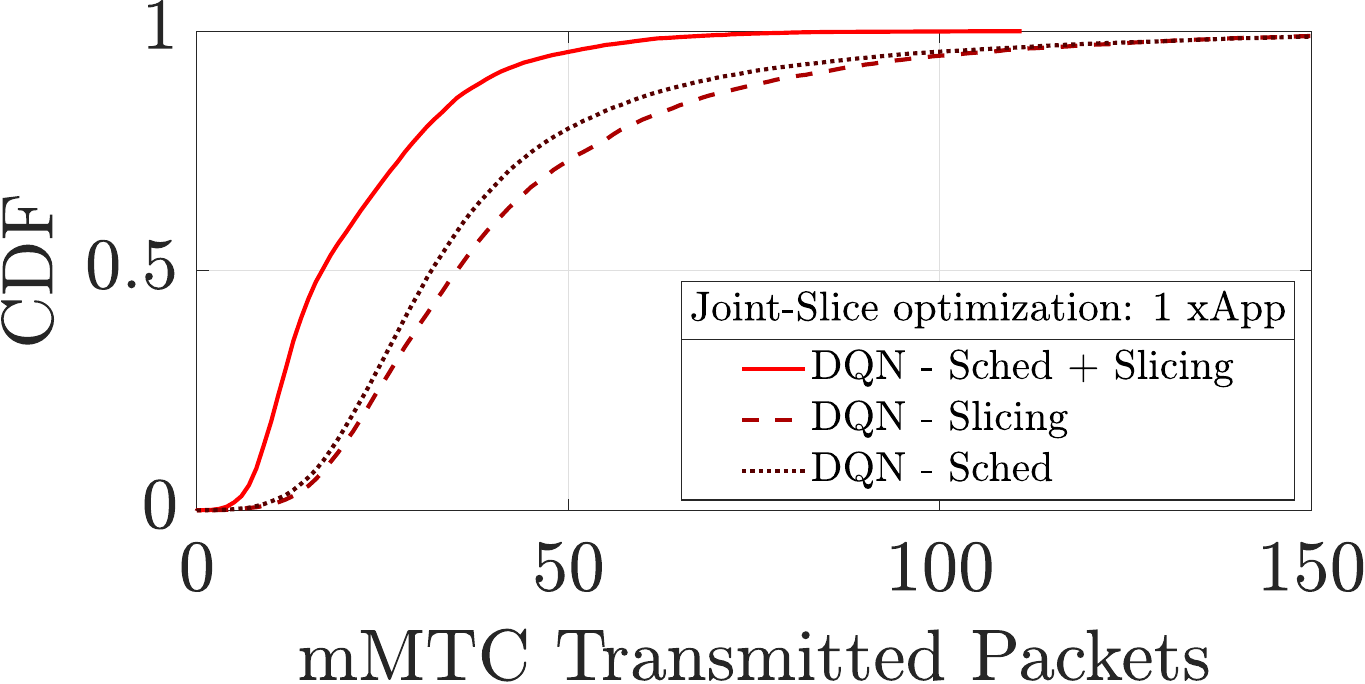}
\label{fig:Figure9b}}
\hfil
\subfigure[\gls{urllc} Buffer Occupancy]{\includegraphics[height=2.85cm]{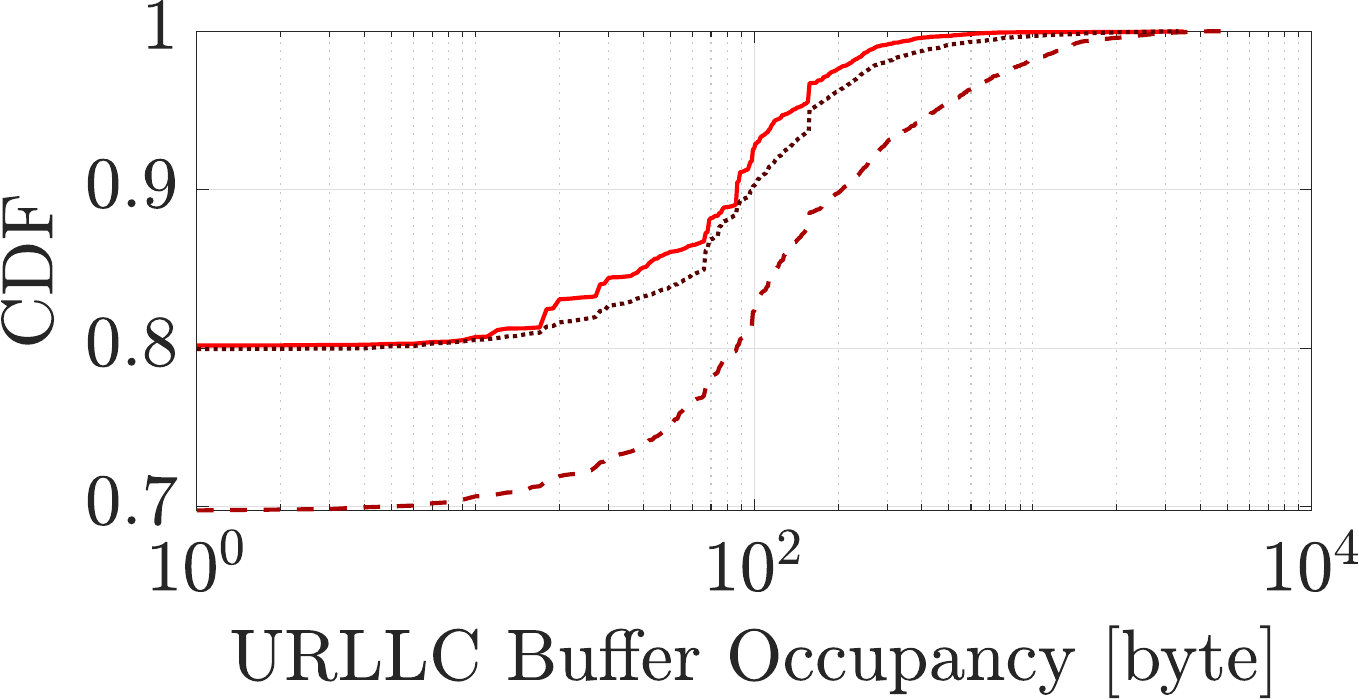}
\label{fig:Figure9c}}
\setlength\abovecaptionskip{-.02cm}
\caption{\textcolor{blue}{Performance evaluation under the \textit{Default} weight configuration for different actions spaces and discount factor $\gamma=0.95$ with the DQN DRL Architecture.}}
\label{Figure9-1a}
\vspace{-0.55cm}
\end{figure*}

\begin{figure}[t!]
\centering
\subfigure[\gls{embb} DL Throughput]{\includegraphics[width=1.6in]{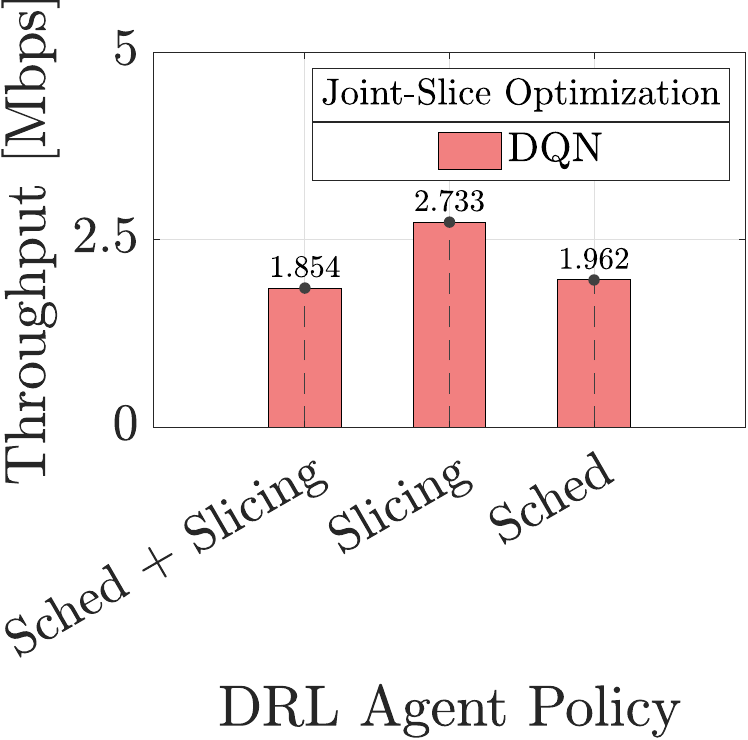}
\label{fig:Figure9d}}
\hfil
\subfigure[\gls{mmtc}  Packets]{\includegraphics[width=1.6in]{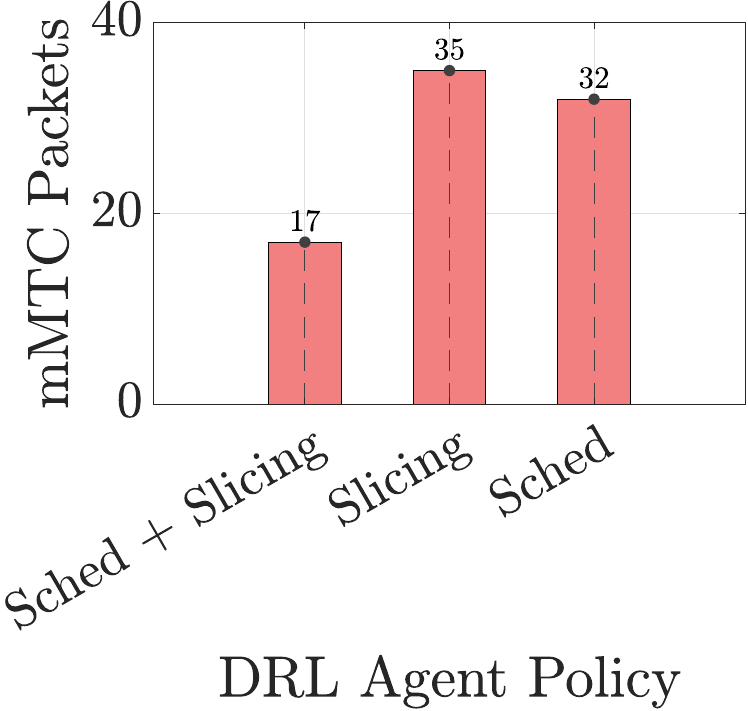}
\label{fig:barplot_mmtc_dqn}}
\setlength\abovecaptionskip{-.02cm}
\caption{\textcolor{blue}{Median values under the \textit{Default} weight configuration for different actions spaces and discount factor $\gamma=0.95$ with the DQN DRL Architecture.}}
\label{fig:Figure9e}
\vspace{-0.55cm}
\end{figure}

\begin{figure*}[t!]
\centering
\subfigure[\gls{embb} Throughput]{\includegraphics[height=2.85cm]{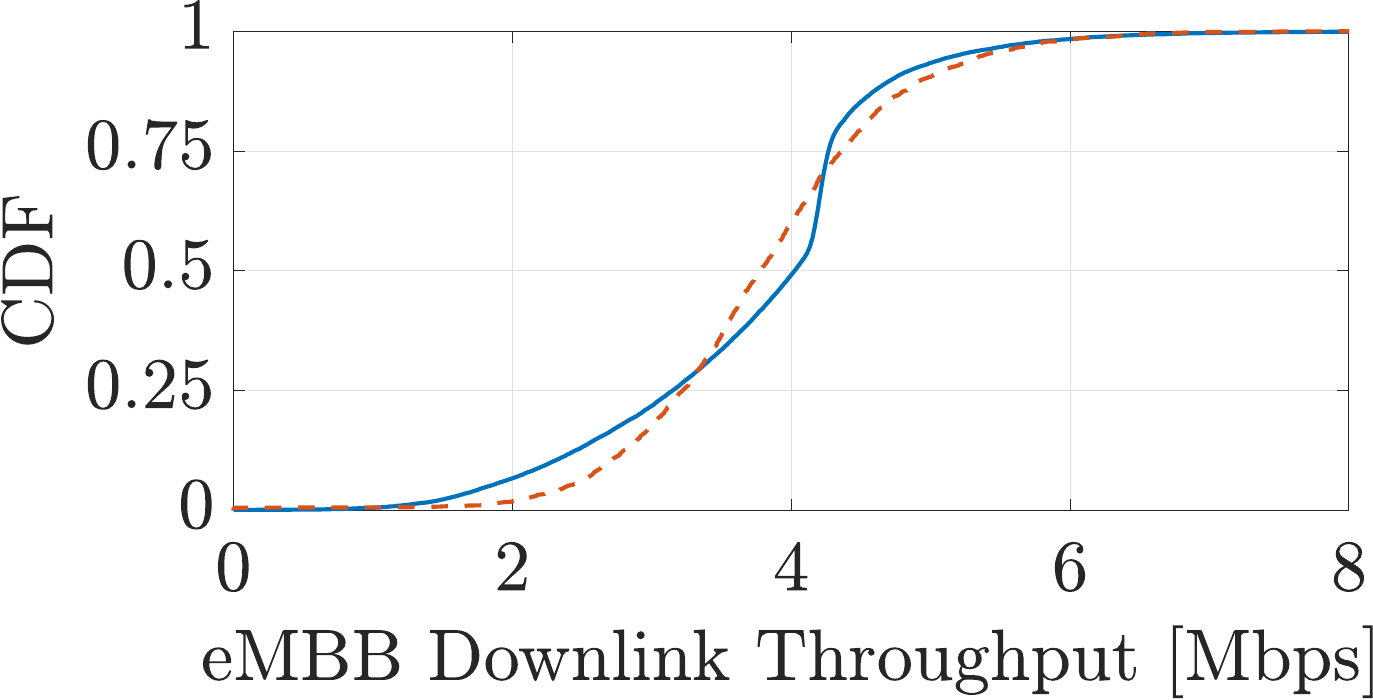}}
\label{fig:Figure8a}
\hfil
\subfigure[\gls{mmtc} Packets]{\includegraphics[height=2.85cm]{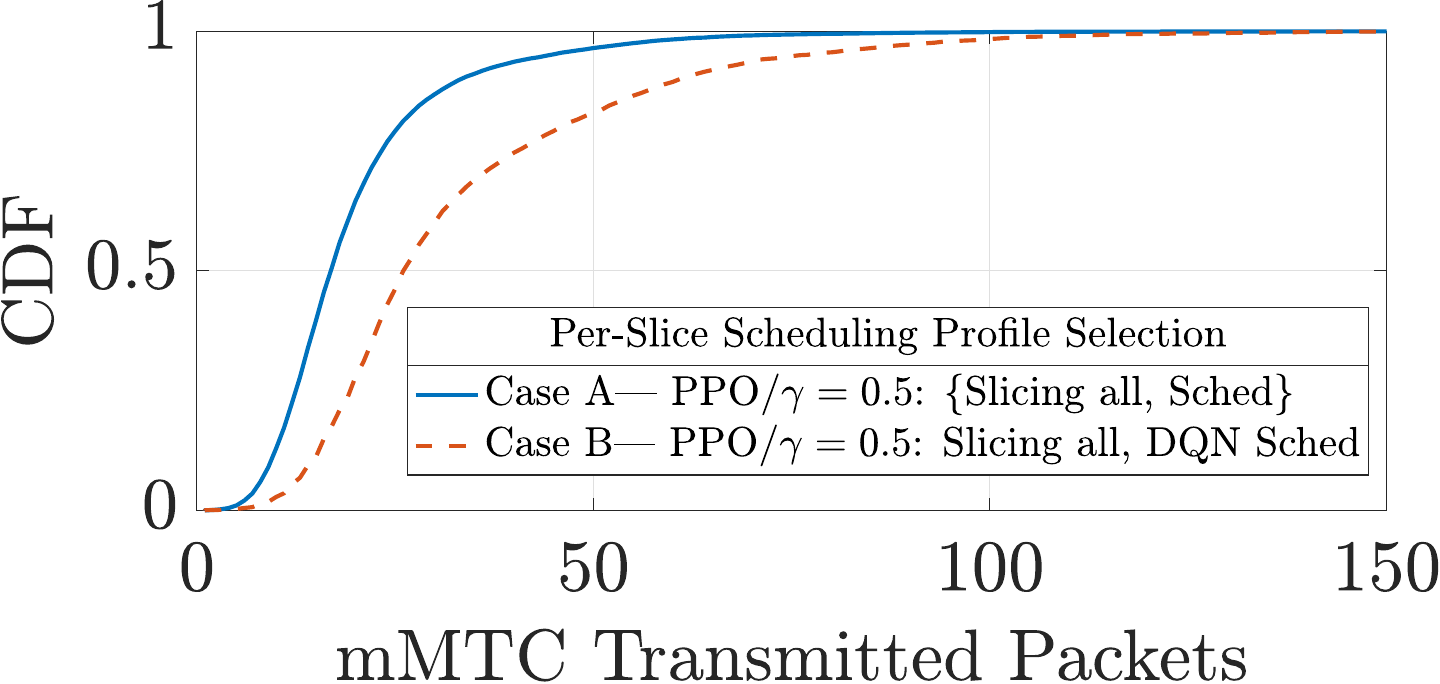}
\label{fig:Figure8b}}
\hfil
\subfigure[\gls{urllc} Buffer Occupancy]{\includegraphics[height=2.85cm]{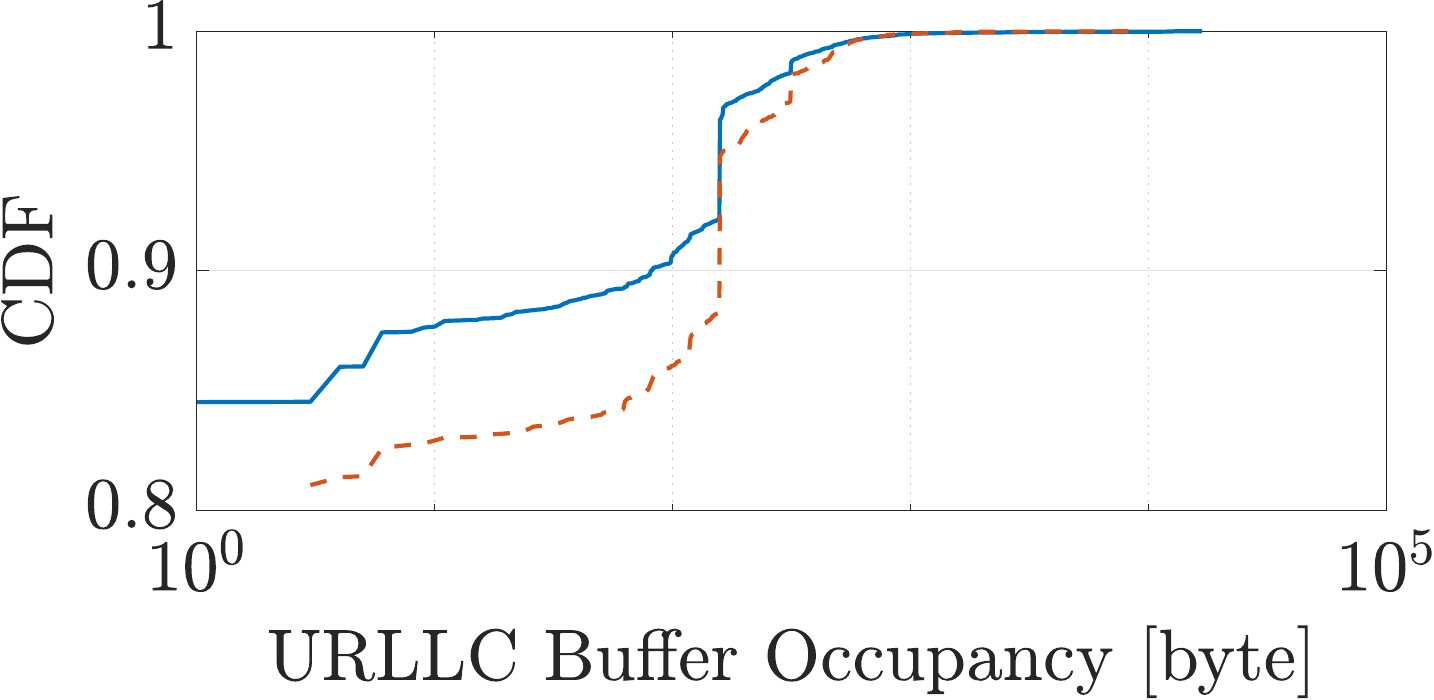}
\label{fig:Figure8c}}
\setlength\abovecaptionskip{-.02cm}
\caption{\textcolor{blue}{Performance evaluation with~$4$ xApps and per-slice scheduling profile selection under PPO and DQN Architectures.}}
\label{Figure8-1a}
\vspace{-0.55cm}
\end{figure*}

\begin{figure}[t!]
\centering
\subfigure[\gls{embb} Throughput]{\includegraphics[width=1.6in]{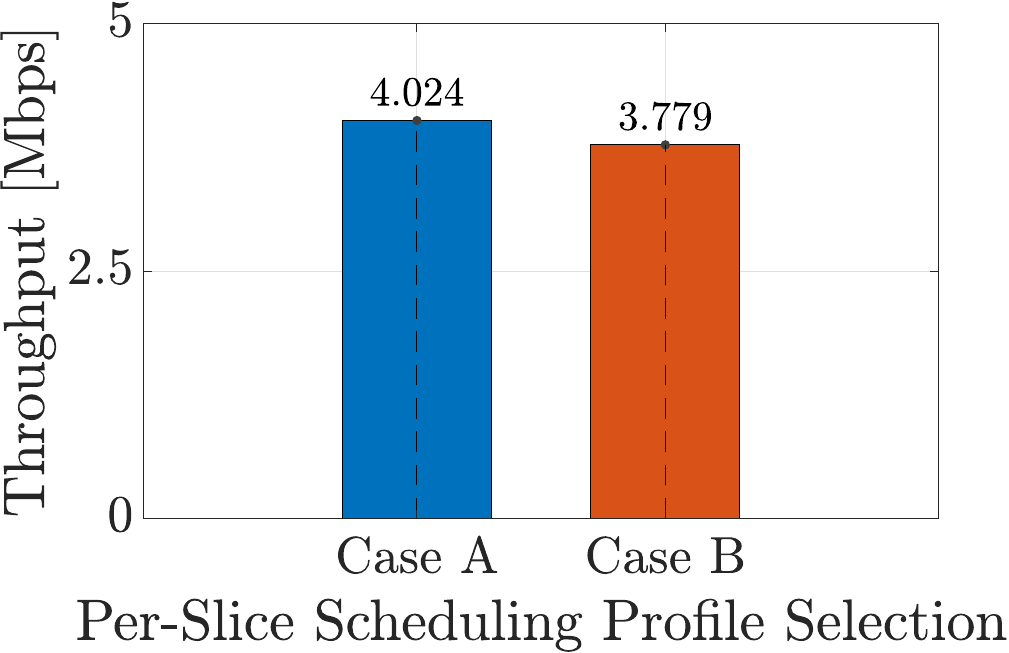}
\label{fig:NewEntry1}}
\hfil
\subfigure[\gls{mmtc} Packets]{\includegraphics[width=1.6in]{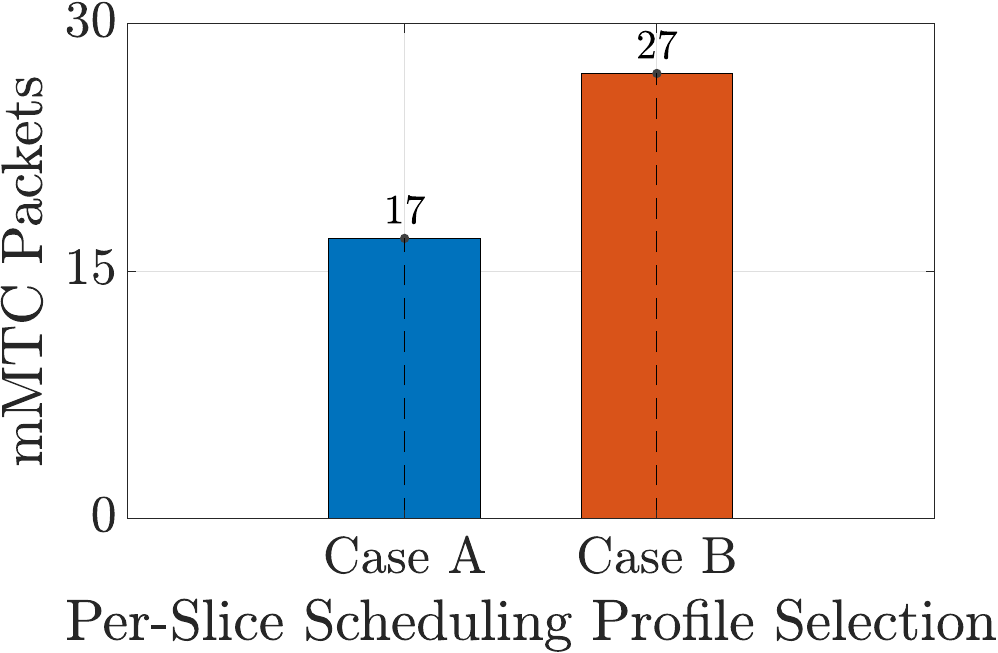}
\label{fig:barplot-4xapps-mmtc}}
\setlength\abovecaptionskip{-.02cm}
\caption{\textcolor{blue}{Median values obtained with a $4$-xApp setup and per-slice scheduling profile selection under PPO and DQN Architectures.}}
\label{Figure9-1b}
\vspace{-0.5cm}
\end{figure}

Our results show that \texttt{Sched \& Slicing} and \texttt{Slicing 0.5} favor \gls{embb} the most, with \texttt{Sched \& Slicing 0.5} being the best configuration among the ones considered.
Moreover, slicing is essential to ensure high throughput values (the four top-performing xApps for \gls{embb} include slicing as a control action). We also notice that prioritizing immediate rewards (i.e., $\gamma=0.5$) results in higher throughput values if compared to xApps embedding agents trained to maximize long-term rewards. This design option, when combined with a bigger action space (e.g., scheduling \& slicing) ultimately yields a higher throughput.

For the \gls{mmtc} slice, the \texttt{Slicing 0.99} xApp always yields the best performance. However, we notice that  
\texttt{Sched \& Slicing 0.5}, which is the best-performing xApp for \gls{embb}, yields the worst performance for \gls{mmtc}. 
Although a larger action space and a short-term reward design is ideal for \gls{embb} (e.g., \texttt{Sched \& Slicing 0.5}), we notice that this performance gain comes at the expense of the \gls{mmtc} slice.
Indeed, in Figs.~\ref{fig:Figure4d} and \ref{fig:Figure4e}, we observe that the higher the \gls{embb} performance, the lower the \gls{mmtc}'s. This is clearly illustrated when we compare the \blockquote{best} per-slice policies, respectively \texttt{Sched \& Slicing 0.5} (\gls{embb}) and \texttt{Slicing 0.99} (\gls{mmtc}). The former delivers the highest reported \gls{embb} throughput ($4.168$\:Mbps) but the lowest number of \gls{mmtc} packets ($16$\:packets), while the latter delivers the highest number of \gls{mmtc} packets ($37$\:packets) and one of the lowest measured \gls{embb} throughput values (i.e., $3.636$\:Mbps). 

\begin{table}[bt]
\centering
\small
\setlength\abovecaptionskip{-.1cm}
\caption{RAN Control Timers}
\begin{adjustbox}{width=1\linewidth}
\begin{tabularx}{\linewidth}{@{}>{\centering\arraybackslash}X@{\hspace{-0.2cm}}>{\centering\arraybackslash}X@{\hspace{-0.2cm}}>{\centering\arraybackslash}X@{\hspace{-0.2cm}}>{\centering\arraybackslash}X@{}}
\toprule
\multicolumn{1}{c}{\textbf{Control Time}} & \multicolumn{1}{c}{\textbf{Set 1}} & \multicolumn{1}{c}{\textbf{Set 2}}  & \multicolumn{1}{c}{\textbf{Set 3}} \\ \midrule 
DU Report & $1$\:s & $250$\:ms & $100$\:ms \\
\glspl{kpm} Log  & $250$\:ms & $250$\:ms & $100$\:ms \\
Action Update & $250$\:ms & $250$\:ms & $100$\:ms \\
\bottomrule
\end{tabularx}
\end{adjustbox}
\label{table:control-time}
\vspace{-.5cm}
\end{table}

Hence, \gls{embb}-\gls{mmtc} slices indicate a competitive behavior, since we cannot optimally satisfy both of them without loss in their respective rewards, as they compete for the amount of packets required for transmission. Our results show  that, in general, controlling scheduling only is not ideal as it strongly penalizes \gls{embb} performance with a modest improvement in terms of number of transmitted \gls{mmtc} packets.

\subsection{Impact of Hierarchical Decision-Making}\label{Section IV-B}

In this analysis, we
evaluate the effectiveness of making disjoint decisions to control scheduling and slicing policies.
We select the best performing single-action xApps from Table~\ref{table:default-comp-analysis}, i.e., \texttt{Slicing 0.5} and \texttt{Sched 0.99}, and we compare their execution at different timescales. The former, provides a good balance in terms of \gls{embb} throughput ($\sim 4$\:Mbps) and number of \gls{mmtc} packets, while the latter, provides the best performance for the \gls{mmtc} slice. With this design choice, we expect to maintain high performance for both \gls{embb} and \gls{mmtc}. 

\begin{table}
\centering
\setlength\abovecaptionskip{-.1cm}
\caption{Feasible \gls{prb} Allocation}
\label{tab:feasible_prb_allocation}
\begin{tabular}{@{}c@{\quad}c@{\quad}c@{}}
\hline
\textbf{\gls{embb}} & \textbf{\gls{mmtc}} & \textbf{\gls{urllc}} \\
\hline
 30 & 9 & 11 \\
 30 & 15 & 5 \\
 36 & 9 & 5 \\
 24 & 21 & 5 \\
 24 & 15 & 11 \\
 18 & 15 & 17 \\
 18 & 9 & 23 \\
 18 & 21 & 11 \\
 12 & 27 & 11 \\
 12 & 15 & 23 \\
 12 & 9 & 29 \\
 6 & 27 & 17 \\
 6 & 39 & 5 \\
 6 & 15 & 29 \\
 6 & 9 & 35 \\
 36 & 3 & 11 \\
\hline
\end{tabular}
\vspace{-.35cm}
\end{table}

We consider four setups, summarized in Table~\ref{table:hier-setups}. Each entry describes how frequently the \gls{bs} reports \glspl{kpm} to the \gls{ric}. For instance, in Setup~1, the xApp for slicing control receives data from the \gls{bs} every $1$\:s, while the scheduling agent receives the respective metrics every $10$\:s. Despite taking into account \gls{ran} telemetry reported every $1, 5$ or $10$\:s, the \gls{drl} decision-making process and the enforcement of a control policy on the \gls{bs} occur within a granularity of \textcolor{red}{tens of milliseconds}, and hence the intelligent control loops are still in compliance with the timescale requirements of the near-real-time \gls{ric}.

Results of this analysis are presented in Figs.~\ref{Figure4-2a} and~\ref{Figure4-2b}. From Fig.~\ref{Figure4a2}, \textit{Setup 3} delivers the best \gls{embb} performance, \textit{Setups} $1$ and $2$ perform almost equally, while \textit{Setup 4} performs the worst. For \gls{mmtc}, in Fig.~\ref{Figure4b2} we notice that all combinations perform similarly and deliver approximately $26$ packets, with \textit{Setup 1} and \textit{Setup 4} delivering an additional packet. From Fig.~\ref{Figure4c2}, we notice that all setups deliver the same performance for the \gls{urllc} slice and, despite not being reported in the figures, they all yield a median buffer occupancy of $0$\:byte, i.e., they maintain an empty buffer to ensure low latency values. In Fig.~\ref{Figure4d2}, we notice that \textit{Setups} $2$ and $3$ deliver the highest \gls{embb} throughput. In Fig.~\ref{Figure4e2}, instead, we notice that \textit{Setups} $1$ and $4$ deliver the highest number of transmitted \gls{mmtc} packets.

Our findings on hierarchical control verify \gls{embb}'s and \gls{mmtc}'s competitive behavior for individual reward maximization. 
Our results show that the rewards of \gls{embb} and \gls{mmtc} slices are competing with one another, as the best configuration for \gls{embb} corresponds to the worst configuration for \gls{mmtc}, and vice versa.  
Among all considered configurations, \textit{Setup 3} offers the best trade-off, as it delivers the highest throughput at the expense of a single \gls{mmtc} packet less being transmitted. 

\vspace{-0.25cm}

\subsection{Impact of Per-Slice Scheduling Profile Selection}\label{Section IV-C}

\textcolor{blue}{We extend our prior work~\cite{tsampazi2023comparative} by evaluating the coexistence of a multitude of xApps delivering services to an Open \gls{ran} multi-slice scenario. Specifically, we consider the case where a single xApp distributes the available \glspl{prb} to all three slices, while three xApps are each tasked with selecting a dedicated scheduling profile for a single slice. All four xApps operate at the same timescale as given in Set~$1$ of Table~\ref{table:control-time}, where \glspl{kpm} and actions are both logged and updated every~$250$ ms respectively, while fresh \gls{ran} data are reported by the \gls{du} within a granularity of $1$~s.}

\textcolor{blue}{We begin the evaluation by focusing on the \gls{dqn} algorithm and the case of the joint-slice optimization. All results were produced with agents trained under the \texttt{Default} weight configuration of Table~\ref{table:weight-confs-list} and with a discount factor of $\gamma=0.95$. The findings presented in Figs.~\ref{Figure9-1a} and~\ref{fig:Figure9e} indicate that in the \gls{embb} slice, \gls{dqn} underperforms compared to \gls{ppo} for both $\gamma\in\{0.5,0.99\}$,  with only \texttt{Slicing} delivering the highest reported throughput value at $2.733$\:Mbps, as reported in Fig.~\ref{fig:Figure9d}. With respect to \gls{mmtc}, Figs.~\ref{fig:Figure9b} and~\ref{fig:barplot_mmtc_dqn} depict a similar performance to the one shown in Figs.~\ref{fig:Figure4b} and~\ref{fig:Figure4e}, when examining the actions separately under their respective discount factor (i.e., $\gamma\in\{0.5,0.99\}$). In detail, we observe that in both cases, \texttt{Slicing} delivers the highest reported number of transmitted packets, followed by \texttt{Sched}, and \texttt{Sched \& Slicing}. It is also shown that the \texttt{Sched} xApp, in the case of \gls{dqn}, performs better in the \gls{mmtc} compared to the case with \gls{ppo}, delivering a median value of  $32$ transmitted packets, which is $\sim28$\% more compared to \texttt{Sched 0.99} with \gls{ppo}, as shown in Fig.~\ref{fig:Figure4e}. Finally, we observe that slicing is the action that delivers the best performance in both \gls{embb} and \gls{mmtc}. Indeed, the competitive behavior observed with \gls{ppo} is mitigated in the case of \gls{dqn}, as shown in Fig.~\ref{fig:Figure9e}, since the actions that deliver the highest \gls{embb} throughput also deliver the highest number of \gls{mmtc} transmitted packets. On the \gls{urllc}, all configurations once again delivered optimal performance, reporting a median buffer occupancy of $0$~byte.}

\textcolor{blue}{Both \gls{drl} architectures are model free, i.e., they do not rely on an explicit model of the environment. Instead, the \gls{drl} agent learns directly by interacting with the environment and makes decisions through trial and error.
\gls{ppo} employs a trust region method~\cite{schulman2017proximal, kim2022adaptive},
that helps in achieving more stable training, compared to \gls{dqn} which can suffer from divergence and slow convergence \cite{zhang2021minibatch}.
The trust region ensures that the old and current policy do not deviate significantly, which helps mitigate divergence issues. Actor and critic networks calculate the state-value, with the critic network reducing variance in action value estimates. This reduction in variance makes policy updates more reliable and exploration safer \cite{lillicrap2015continuous}. On the contrary, \gls{dqn} does not implement trust regions and uses the target network to achieve stability during training \cite{mnih2015human}. 
In environments affected by noisy observations and stochasticity, the aforementioned features, combined with the model's ability to explicitly model the policy's probability distribution over actions, 
make \gls{ppo} capable of adapting more efficiently to varying conditions. This ability to generalize better across tasks and environments can be achieved with \gls{dqn} but might require more fine-tuning and adjustments when transitioning to new tasks. Furthermore, \gls{dqn} relies on epsilon-greedy exploration, which can be less efficient, especially in complex environments \cite{hessel2018rainbow,fortunato2017noisy}. \gls{ppo}, being an on-policy algorithm, manages to optimize the current policy directly based on the most recent experiences. On the other hand, \gls{dqn}, an off-policy algorithm, estimates the value of actions independently of the current policy. On-policy methods often work better when the policy is changing during training, especially in stochastic environments. Therefore, in the latter case where rapid policy changes are observed, off-policy methods may struggle to deal successfully with them, and they may require more extensive replay buffers or further fine-tuning to handle noisy environments effectively\textcolor{red}{~\cite{schulman2015trust,fujimoto2019off,9068634}}.
}

\begin{figure*}[t!]
\centering
\subfigure[\gls{embb} Throughput]{\includegraphics[height=2.85cm]{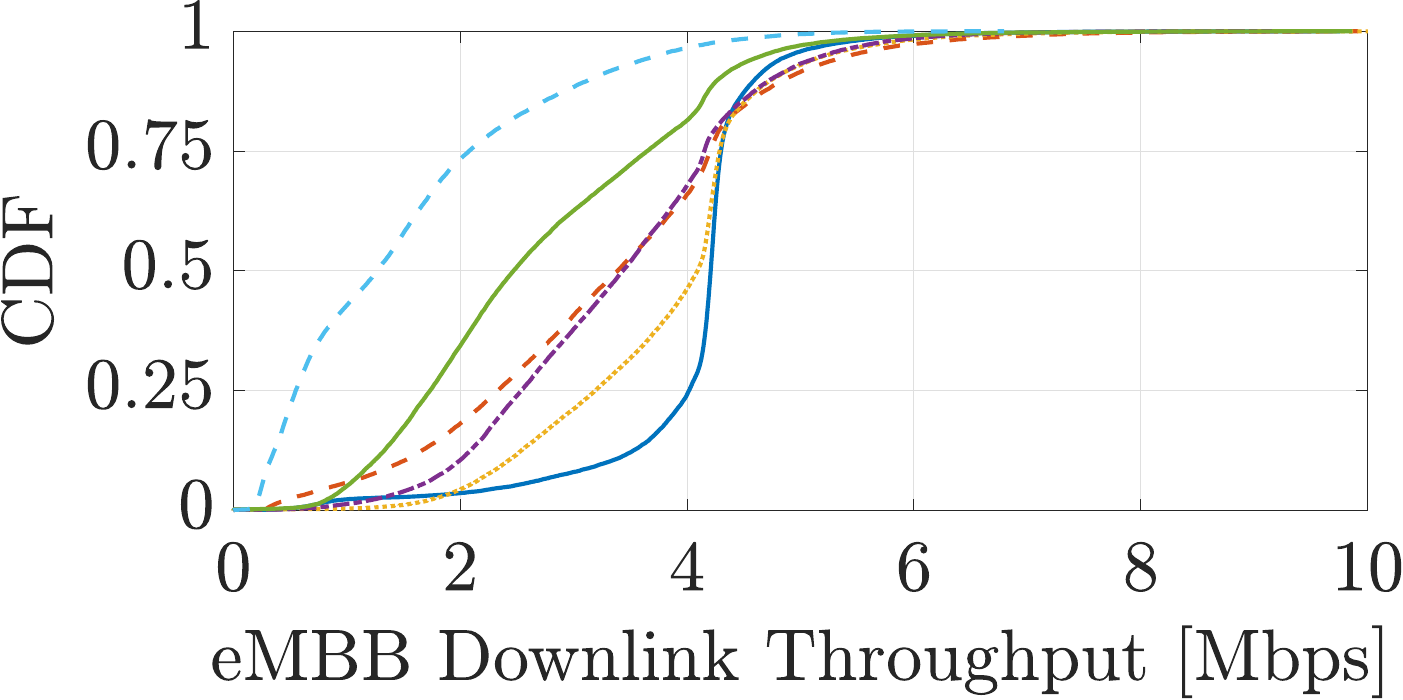}
\label{Figure4a3}}
\hfil
\subfigure[\gls{mmtc} Packets]{\includegraphics[height=2.85cm]{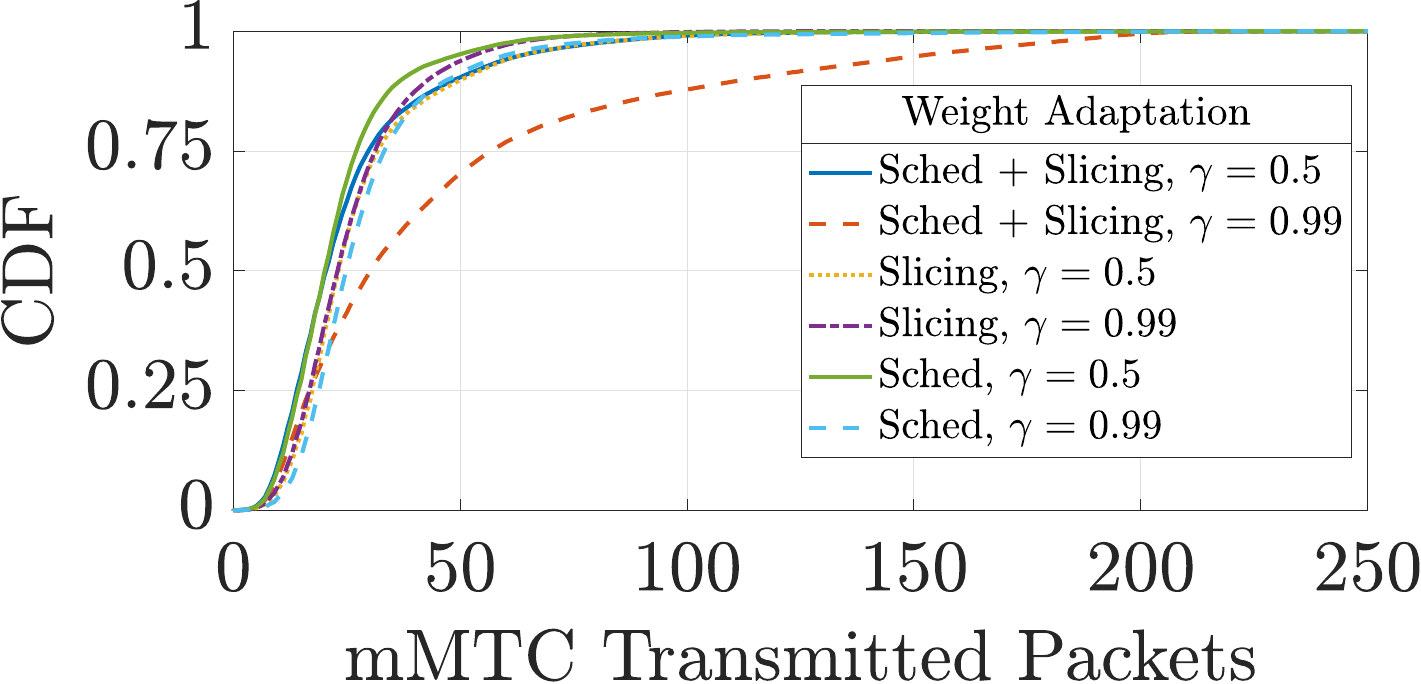}
\label{Figure4b3}}
\hfil
\subfigure[\gls{urllc} Buffer Occupancy]{\includegraphics[height=2.85cm]{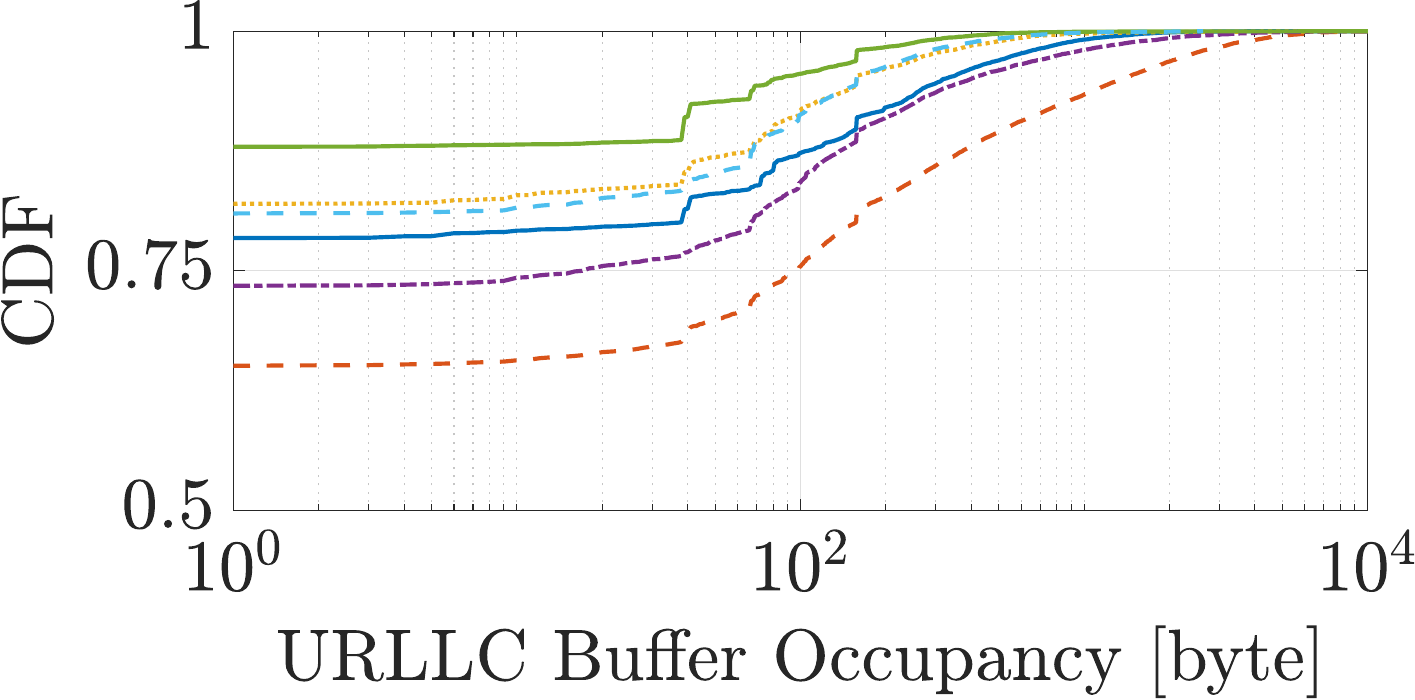}
\label{Figure4c3}}
\setlength\abovecaptionskip{-.02cm}
\caption{Performance evaluation under the \textit{Alternative} weight configuration for different actions spaces and discount factors \textcolor{blue}{with the PPO DRL Architecture.}}
\label{Figure4-3a}
\vspace{-0.55cm}
\end{figure*}

\begin{figure}[t!]
\centering
\subfigure[\gls{embb} DL Throughput]{\includegraphics[width=1.6in]{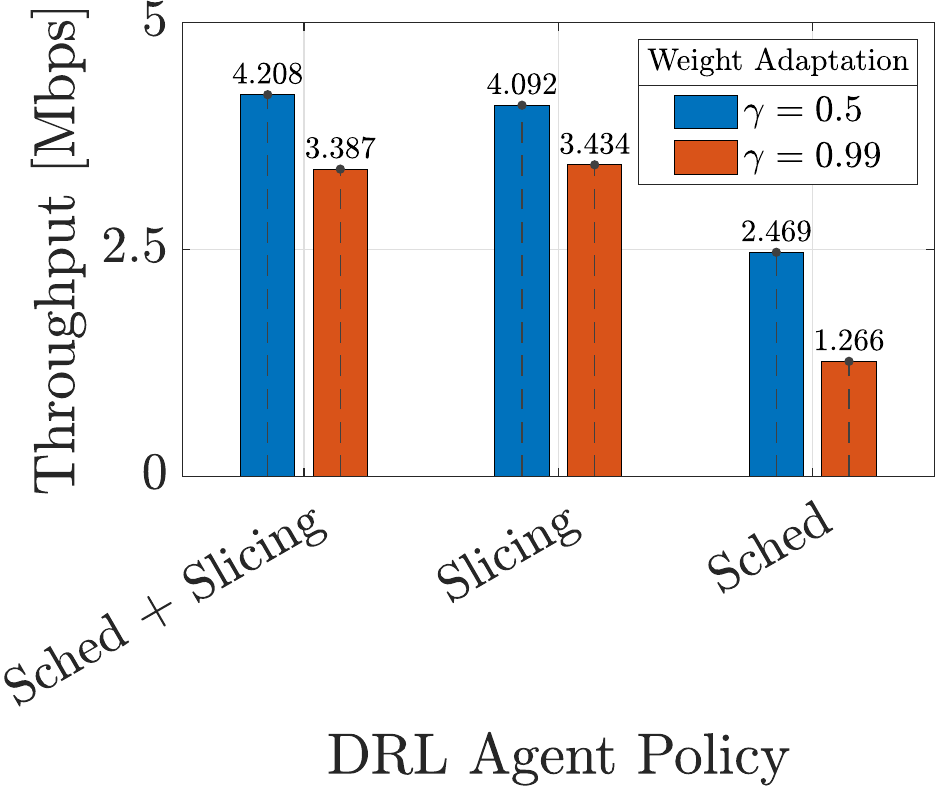}
\label{Figure4d3}}
\hfil
\subfigure[\gls{mmtc}  Packets]{\includegraphics[width=1.6in]{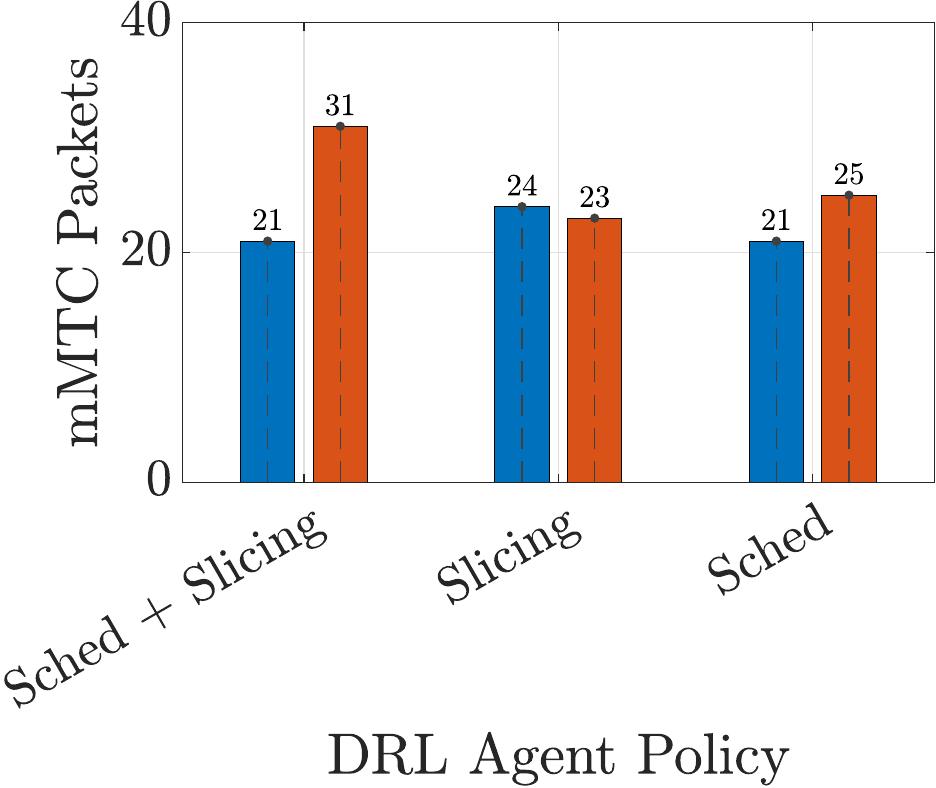}
\label{Figure4e3}}
\setlength\abovecaptionskip{-.02cm}
\caption{Median values under the \textit{Alternative} weight configuration for different actions spaces and discount factors \textcolor{blue}{with the PPO DRL Architecture.}}
\label{Figure4-3b}
\vspace{-0.55cm}
\end{figure}

\textcolor{blue}{Lastly, \gls{ppo} has been shown to work more efficiently in problems with high-dimensional state and action spaces, as demonstrated in~\cite{kozlica2023deep}, owing to its policy-based and trust region approaches. In our work, the state and action spaces are high-dimensional. In detail, a total of $43$~actions are considered when resource allocation is performed jointly for all slices. 
For scheduling, the size of the action space is $27$, while for slicing, it is $16$. The former size encompasses all possible combinations of scheduling policies for all three slices (i.e., \gls{rr}, \gls{wf}, and \gls{pf}), while the combination for \gls{prb} allocation is given in Table~\ref{tab:feasible_prb_allocation} for a total of $50$~\glspl{prb}.} Finally, the observation of the state, obtained through srsRAN, consists of $10$ independent measurements of $3$~\glspl{kpi}, i.e., \gls{dl} buffer occupancy, the \gls{dl} throughput and the number of transmitted packets on the \gls{dl}.

\textcolor{blue}{From the previous discussion, we have seen that \gls{dqn} performs poorly in the high-dimensional state and action spaces, such as the ones considered in this work. Therefore, we test its effectiveness in a reduced action space where the \gls{dqn} agent only controls scheduling decisions.
Specifically, we consider the case of one \texttt{Slicing} xApp that serves all slices simultaneously, while the remaining $3$~xApps are dedicated to each slice and control the scheduling profile for the corresponding slice only. For instance, one xApp selects a scheduling profile for the \gls{embb} slice, another one focuses on the \gls{mmtc} slice, and the last one controls the \gls{urllc} slice only.
This configuration allows for both testing and mixing different \gls{drl} architectures, such as \gls{ppo} and \gls{dqn}. It also enables experimentation with a setup in which the scheduling selection action follows the \gls{prb} allocation. Therefore, the preceding xApps will reconfigure a network where the action space is altered after the changes made  by the \texttt{Slicing} xApp.
}

\textcolor{blue}{In Figs.~\ref{Figure8-1a} and~\ref{Figure9-1b}, we present the results obtained from testing a total of~$4$ xApps that perform resource allocation. For the purposes of the experiments we consider two cases, namely \emph{Case A} and \emph{Case B}. In both cases, the \texttt{Slicing} xApp embeds a \gls{drl} agent trained with \gls{ppo} under the discount factor, $\gamma=0.5$. In \emph{Case A}, the remaining \texttt{Sched} xApps also embed \gls{ppo} agents trained with $\gamma=0.5$. In \emph{Case B}, the xApps embed \gls{dqn} agents with a discount factor of $\gamma=0.95$, as described in Section~\ref{Section IIA}. In Fig.~\ref{Figure8-1a}, we observe that \emph{Case A} performs better for the \gls{embb} slice. Indeed, the results reported in Fig.~\ref{Figure9-1b} indicate that \emph{Case A} achieves a median \gls{embb} throughput value of $4.024$\:Mbps, which is $\sim 6.5\%$ higher than that of~\emph{Case B}. Regarding the \gls{mmtc} slice, the median for the corresponding \gls{kpm} metric was measured at $27$ transmitted packets, representing an increase of $\sim60\%$ compared to \emph{Case A}. For \gls{urllc}, both cases delivered optimal performance with a median buffer occupancy of $0$~byte. We notice that \emph{Case A} performs similarly to the case of jointly performing slicing and scheduling control with $\gamma=0.5$, as shown in Figs.~\ref{Figure4-1a} and~\ref{Figure4-1b}, by delivering one additional \gls{mmtc} packet, as depicted in Fig.~\ref{Figure9-1b}. With regards to \emph{Case B}, the obtained results closely match those reported with Hierarchical Control under Setup~$1$ in Fig.~\ref{Figure4-2b}. Precisely, as depicted in Fig.~\ref{fig:barplot-4xapps-mmtc}, the two configurations achieve identical performance on the \gls{mmtc} slice, by achieving the same median value of transmitted packets (i.e., $27$~packets). For the \gls{embb} slice, the setup with $4$~xApps (i.e., \emph{Case B}) exhibits a minimal drop of~$\sim0.66\%$ in the median throughput value compared to the performance achieved with Setup~$1$. Therefore, \emph{Case B} provides a more flexible and heterogeneous setup where different \gls{drl} agents can coexist, without a significant performance degradation. Furthermore, the \gls{dqn} can still be leveraged in those cases where agents independently optimize the performance of each slice thanks to the reduced action space.
}

\begin{table}[tb]
\centering
\small
\setlength\abovecaptionskip{-.1cm}
\caption{Weight Design}
\begin{adjustbox}{width=0.65\linewidth}
\begin{tabular}{ccc}
\toprule
\textbf{$w_{eMBB}$} & \textbf{$w_{mMTC}$} & \textbf{$w_{URLLC}$} \\
\midrule
$\alpha_{eMBB}\cdot\dfrac{1}{A}$ & $\beta_{mMTC}\cdot\dfrac{1}{B}$ & $\gamma_{URLLC}\cdot \left( -\dfrac{1}{C} \right)$ \\
\bottomrule
\end{tabular}
\end{adjustbox}
\label{table:weight-design}
\vspace{-0.3cm}
\end{table}

\noindent
\subsection{Impact of Weight Configuration}\label{Section IV-D}

\begin{figure*}[t!]
\centering
\subfigure[\gls{embb} Throughput]{\includegraphics[height=2.85cm]{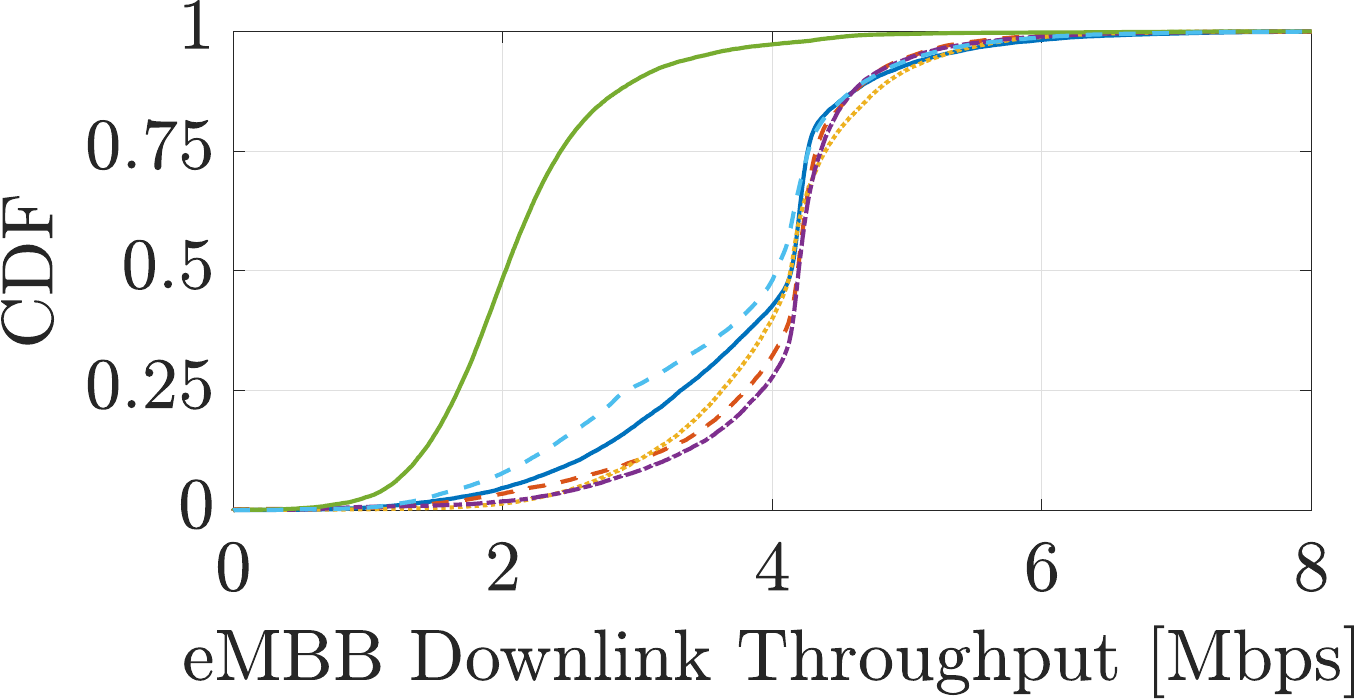}
\label{Figure5a3}}
\hfil
\subfigure[\gls{mmtc} Packets]{\includegraphics[height=2.85cm]{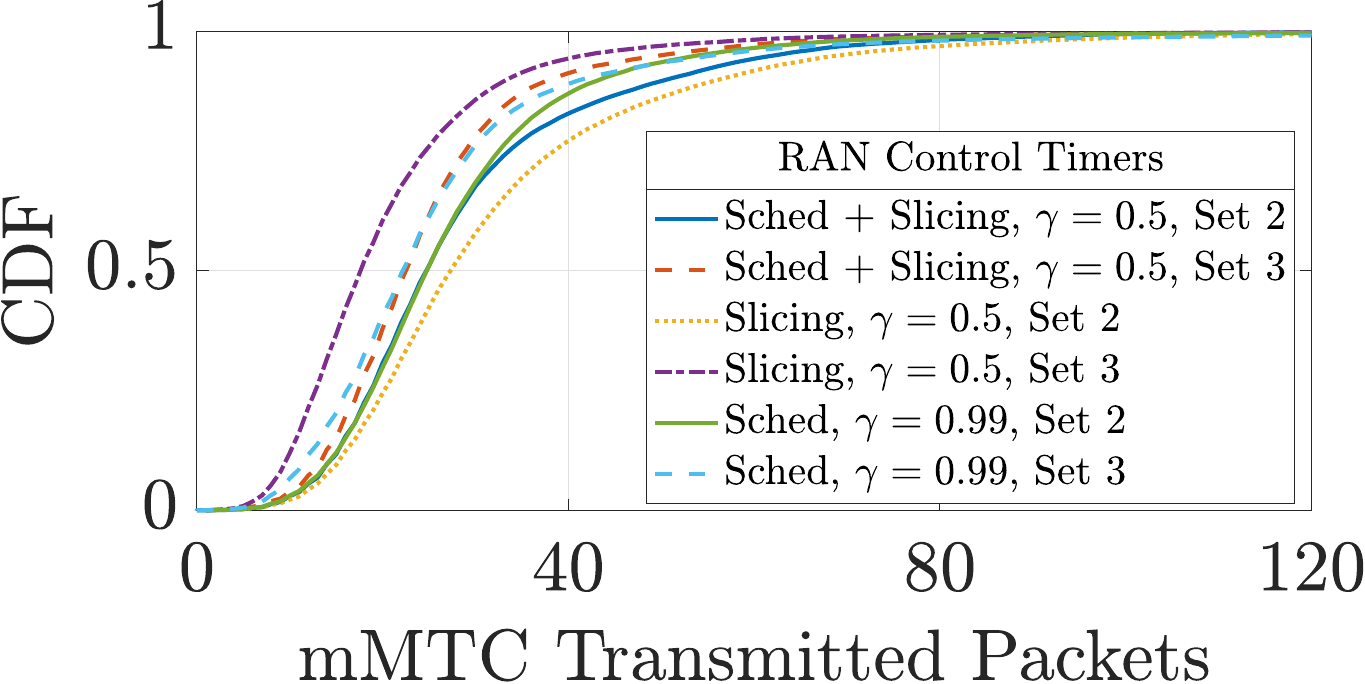}
\label{Figure5b3}}
\hfil
\subfigure[\gls{urllc} Buffer Occupancy]{\includegraphics[height=2.85cm]{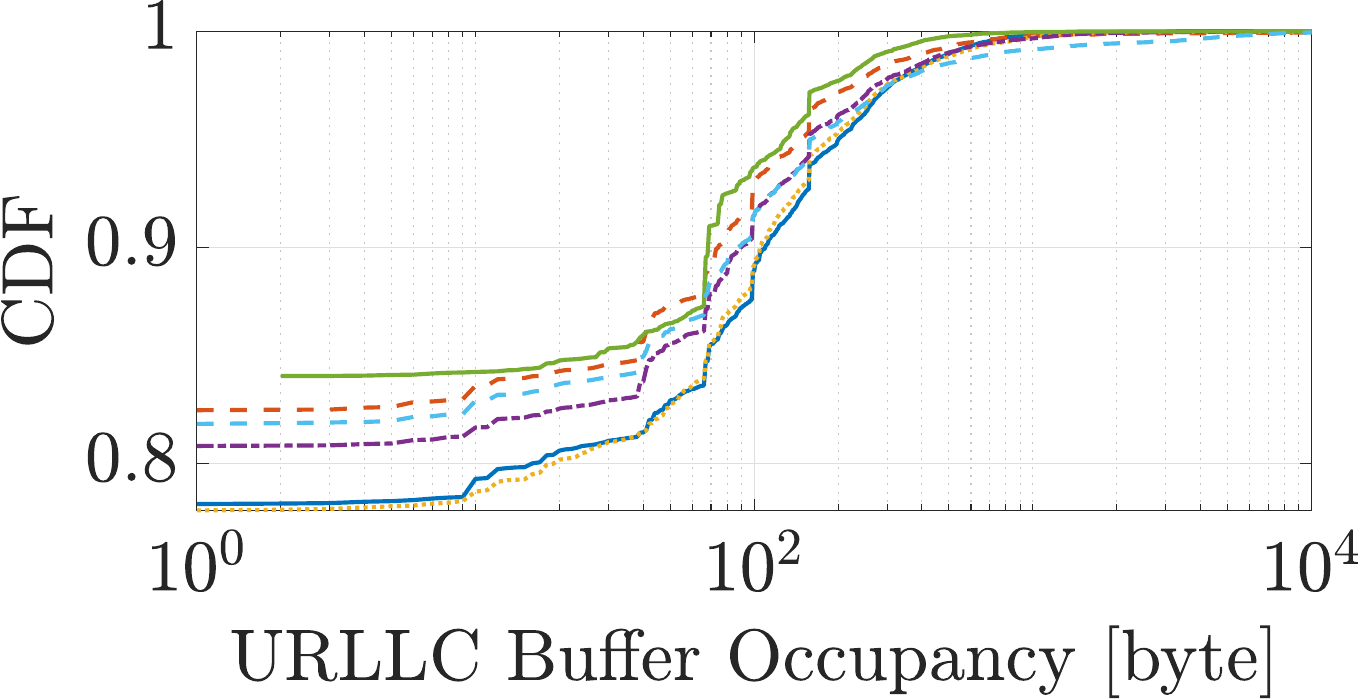}
\label{Figure5c3}}
\setlength\abovecaptionskip{-.02cm}
\caption{\textcolor{blue}{Performance evaluation under different action spaces, values of the $\gamma$ parameter and sets of \gls{ran} control timers with the PPO DRL Architecture.}}
\label{Figure5-3a}
\vspace{-0.55cm}
\end{figure*}

\begin{figure}[t!]
\centering
\subfigure[\gls{embb} DL Throughput]{\includegraphics[width=1.6in]{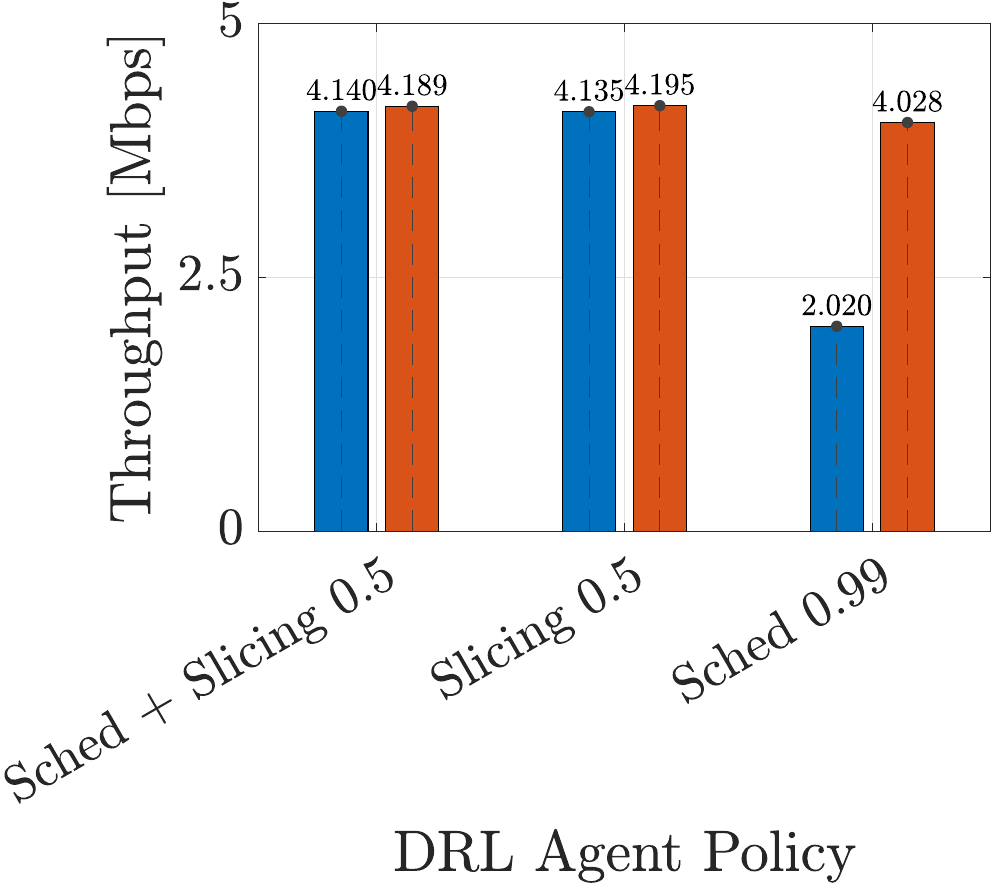}
\label{ran-timers-bar-throughput}}
\hfil
\subfigure[\gls{mmtc}  Packets]{\includegraphics[width=1.6in]{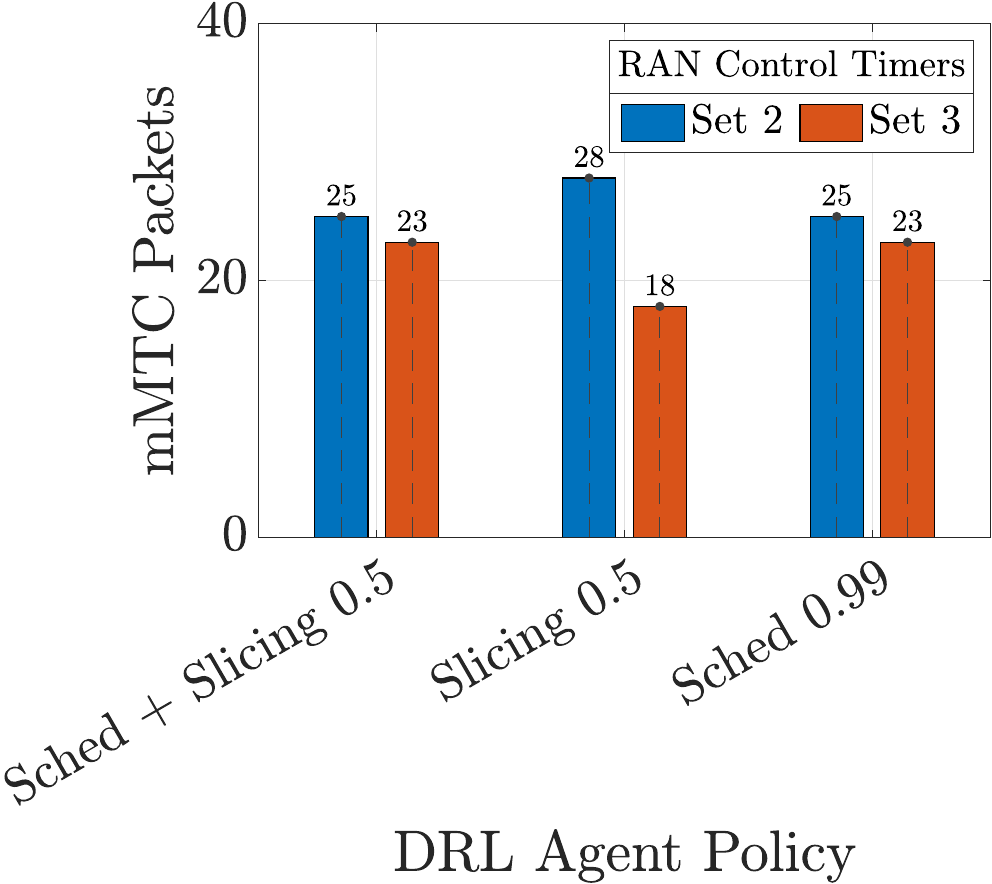}
\label{ran-timers-bar-mmtc}}
\setlength\abovecaptionskip{-.02cm}
\caption{\textcolor{blue}{Median values under different action spaces and \gls{ran} control timers with the PPO DRL Architecture.}}
\label{Figure10-1b}
\vspace{-0.55cm}
\end{figure}

In this study, we consider different weight configurations to compute the cumulative average reward function in Eq.~\eqref{eq:weighted_reward}. The considered configurations are reported in Table \ref{table:weight-confs-list}. The \textit{Alternative} weight configuration is computed by using the weights in Table
\ref{table:weight-design}, where $A, B, C, \alpha_{eMBB}, \beta_{mMTC}$, and $\gamma_{URLLC}$ are used to both scale and prioritize certain slices. Specifically, $A, B, C$ are used to scale the individual weights according to statistical information of corresponding \glspl{kpm}. For example, $A, B, C$ can represent either the average, minimum or maximum values reported \gls{kpm} per slice so as to scale the weight according to the dynamic range of the corresponding \gls{kpm}. Similarly, $\alpha_{eMBB}, \beta_{mMTC}$, and $\gamma_{URLLC}$ can be used to give priority to one slice or the other.

We set $\alpha_{eMBB}=1000$, $\beta_{mMTC}=456$ and $\gamma_{URLLC}=1$. As a reference for $A$, $B$ and $C$, we choose the historically maximum reported \gls{kpm} values for each slice, i.e., $A=13.88$ Mbps, $B=304$, and $C=20186$ byte.

Based on these steps, we derive their respective weights $w_{eMBB}$, $w_{mMTC}$, $w_{URLLC}$. For example, the weight of \gls{mmtc} can be computed as $w_{mMTC} = \beta_{mMTC}\cdot\dfrac{1}{B} = 456/304 = 1.5$, as reported in the \textit{Alternative} configuration in Table \ref{table:weight-confs-list}.
The goal of comparing the two \textit{Default} and \textit{Alternative} weight configurations is to explore and understand the dynamics between \gls{mmtc} and \gls{embb} and the overall impact on the network performance. Specifically, since previous results have shown that the \gls{mmtc} can be penalized by the \gls{embb} slice, with the \textit{Alternative configuration} we aim at giving the former a weight that is $6\times$ larger than the \textit{Default} configuration.

Results for the \textit{Alternative} configuration are reported in Figs. \ref{Figure4-3a} and \ref{Figure4-3b}.
In Fig. \ref{Figure4a3}, \texttt{Sched \& Slicing 0.5} delivers the best \gls{embb} performance. Similarly to the results presented in Section \ref{Section IV-A}, \texttt{Scheduling \& Slicing 0.5} and \texttt{Slicing 0.5} are the best choices, with short-term reward design being ideal for \gls{embb}. In Fig. \ref{Figure4b3}, the \textit{Alternative} weight configuration results in \texttt{Scheduling \& Slicing 0.99} being the best \gls{mmtc} choice and long-term rewards are better for \gls{mmtc} users. For \gls{urllc}, all policies perform well, with \texttt{Scheduling \& Slicing 0.5} performing slightly better compared to \texttt{Scheduling \& Slicing 0.99}.

Figs.~\ref{Figure4d3} and~\ref{Figure4e3}, confirm that controlling scheduling alone does not improve performance in general. Similarly to our previous analysis, a high \gls{embb} performance (i.e., \texttt{Sched \& Slicing 0.5}) results in a degraded \gls{mmtc} performance.
However, 
if compared with the \texttt{Default}, the \texttt{Alternative} weight configuration achieves a $31.25\%$ increase for \gls{mmtc}, with the same equally good \gls{urllc} performance and a $1\%$ throughput increase for \gls{embb} users.

\begin{table}[htb]
\centering
\small
\setlength\abovecaptionskip{-.1cm}
\caption{Design Options Catalog}
\begin{adjustbox}{width=0.85\linewidth}
\begin{tabular}{@{}l@{\hspace{4.2 mm}}l@{\hspace{4.2 mm}}l@{\hspace{4.2 mm}}l@{}}
\toprule
$\textbf{Option 1}$ & $\texttt{Sched \& Slicing 0.5 - Alternative}$ \\
$\textbf{Option 2}$ & $\texttt{Slicing 0.5 - Default}$ \\
$\textbf{Option 3}$ & $\texttt{Hierarchical Control - Setup 1}$ \\
$\textbf{Option 4}$ & $\texttt{Slicing 0.99 - Default}$ \\
\bottomrule
\end{tabular}
\end{adjustbox}
\label{table:design-catalogue}
\vspace{-0.15cm}
\end{table}

In Table~\ref{table:design-catalogue} we summarize the design options that \textcolor{blue}{have delivered a} good overall performance \textcolor{blue}{so far}. Table~\ref{table:final-designs} indicates \gls{embb} and \gls{mmtc}'s dynamic and competitive relation. Option $2$ brings balance, in terms of throughput and transmitted packets, Option $1$ favors \gls{embb}, and Option $4$ boosts \gls{mmtc} but with a significant decrease in the \gls{qos} of the \gls{embb} slice.

\begin{table}[htb]
\centering
\small
\setlength\abovecaptionskip{-.15cm}
\caption{Design Options}
\begin{adjustbox}{width=0.85\linewidth}
\begin{tabular}
{@{}l@{\hspace{0.1mm}}c@{\hspace{0.1mm}}c@{\hspace{0.1mm}}c@{}}
\toprule
\multicolumn{1}{c}{\textbf{}} & \multicolumn{1}{c}{\textbf{eMBB} [Mbps]} & \multicolumn{1}{c}{\textbf{mMTC} [packet]} & \multicolumn{1}{c}{\textbf{URLLC} [byte]} \\
\midrule
$\textbf{Option 1}$ & $4.208$& $21$ & $0$ \\
$\textbf{Option 2}$ & $4.114$ & $26$ & $0$ \\
$\textbf{Option 3}$ & $3.804$ & $27$ & $0$ \\
$\textbf{Option 4}$ & $3.636$ & $37$ & $0$ \\
\bottomrule
\end{tabular}
\end{adjustbox}
\label{table:final-designs}
\vspace{-0.5cm} 
\end{table}

\subsection{Impact of RAN Control Timers}\label{Section IV-E}

\begin{figure*}[t!]
\centering
\subfigure[\gls{embb} Throughput]{\includegraphics[height=2.85cm]{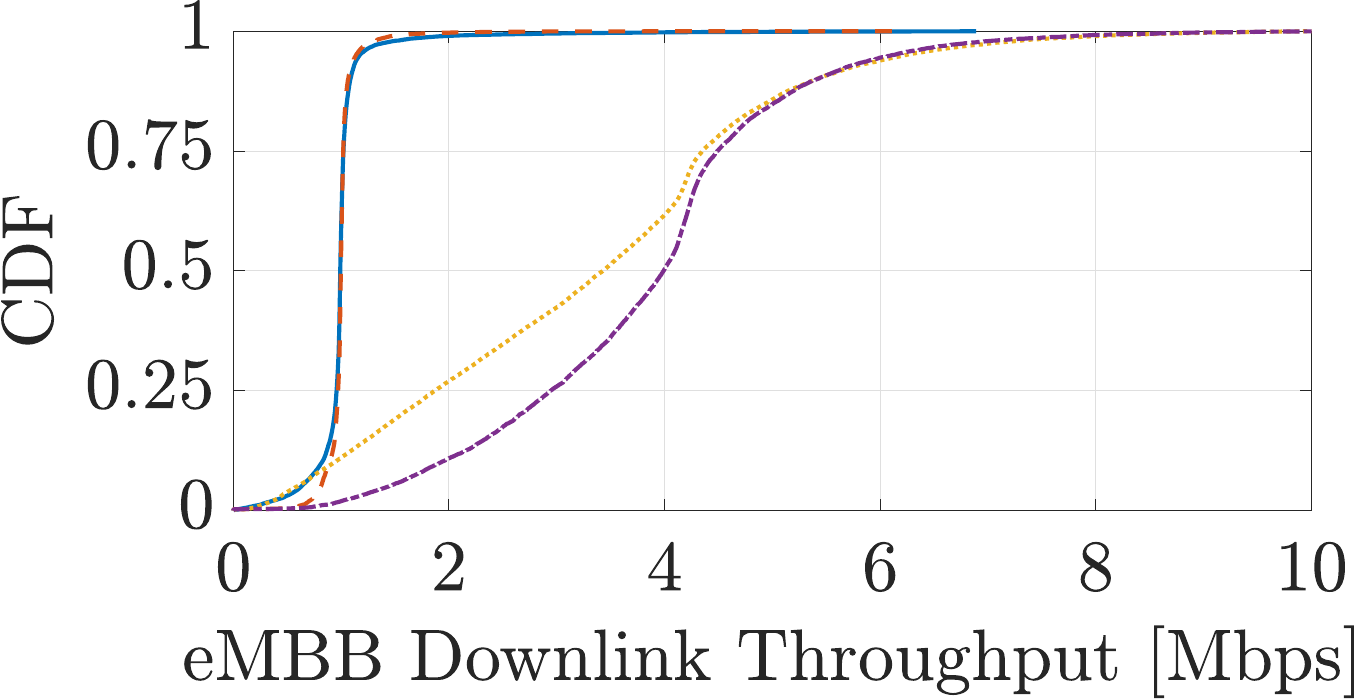}
\label{Figure6a2}}
\hfil
\subfigure[\gls{mmtc} Packets]{\includegraphics[height=2.85cm]{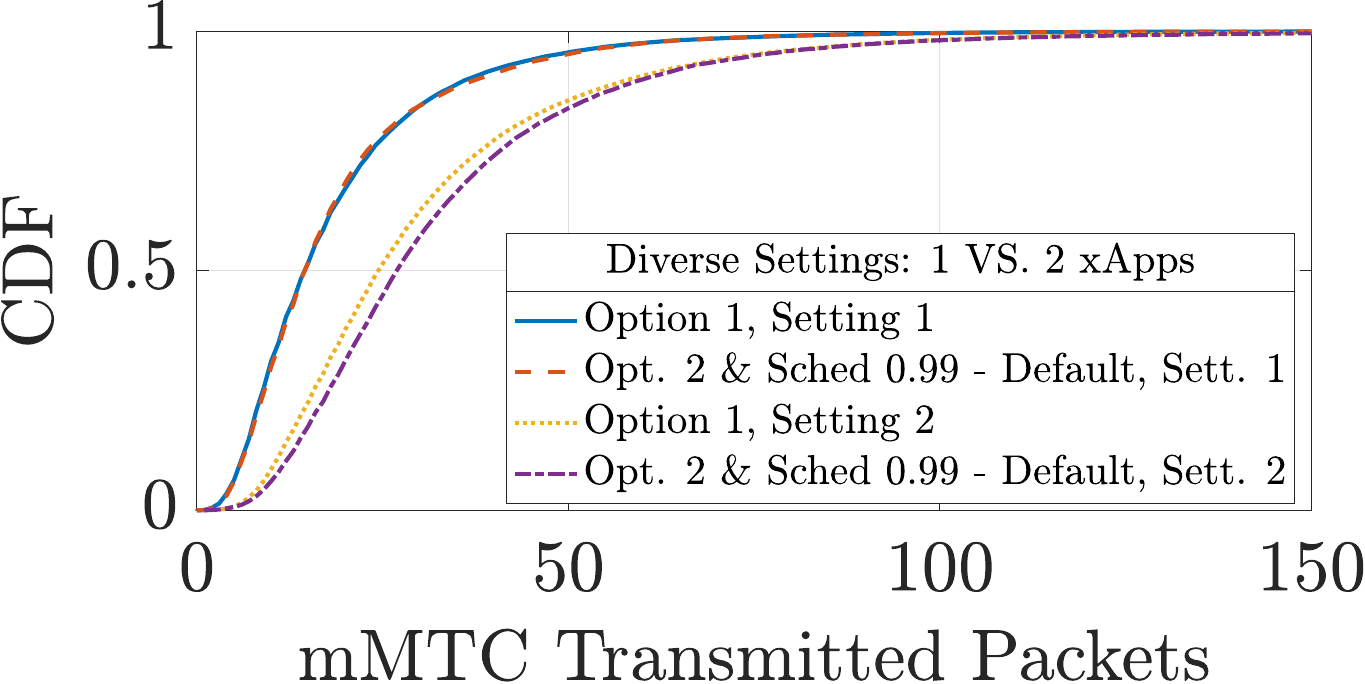}
\label{Figure6b2}}
\hfil
\subfigure[\gls{urllc} Buffer Occupancy]{\includegraphics[height=2.85cm]{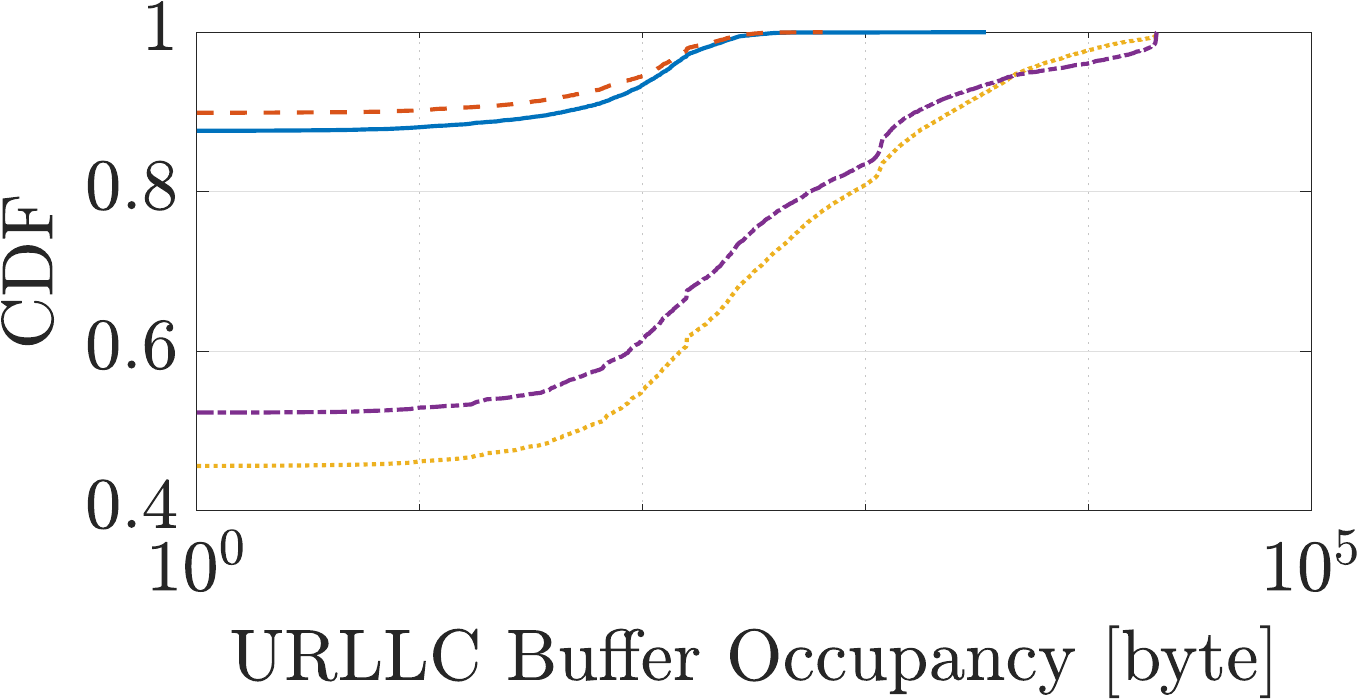}
\label{Figure6c2}}
\setlength\abovecaptionskip{-.02cm}
\caption{\textcolor{blue}{Performance evaluation under mobility with Settings~$1$ and~$2$ of Table~\ref{table:net-settings} and the PPO DRL Architecture.}}
\label{Figure6-2a}
\vspace{-0.25cm}
\end{figure*}

\begin{figure}[t!]
\centering
\subfigure[\gls{embb} Throughput]{\includegraphics[width=1.692in]{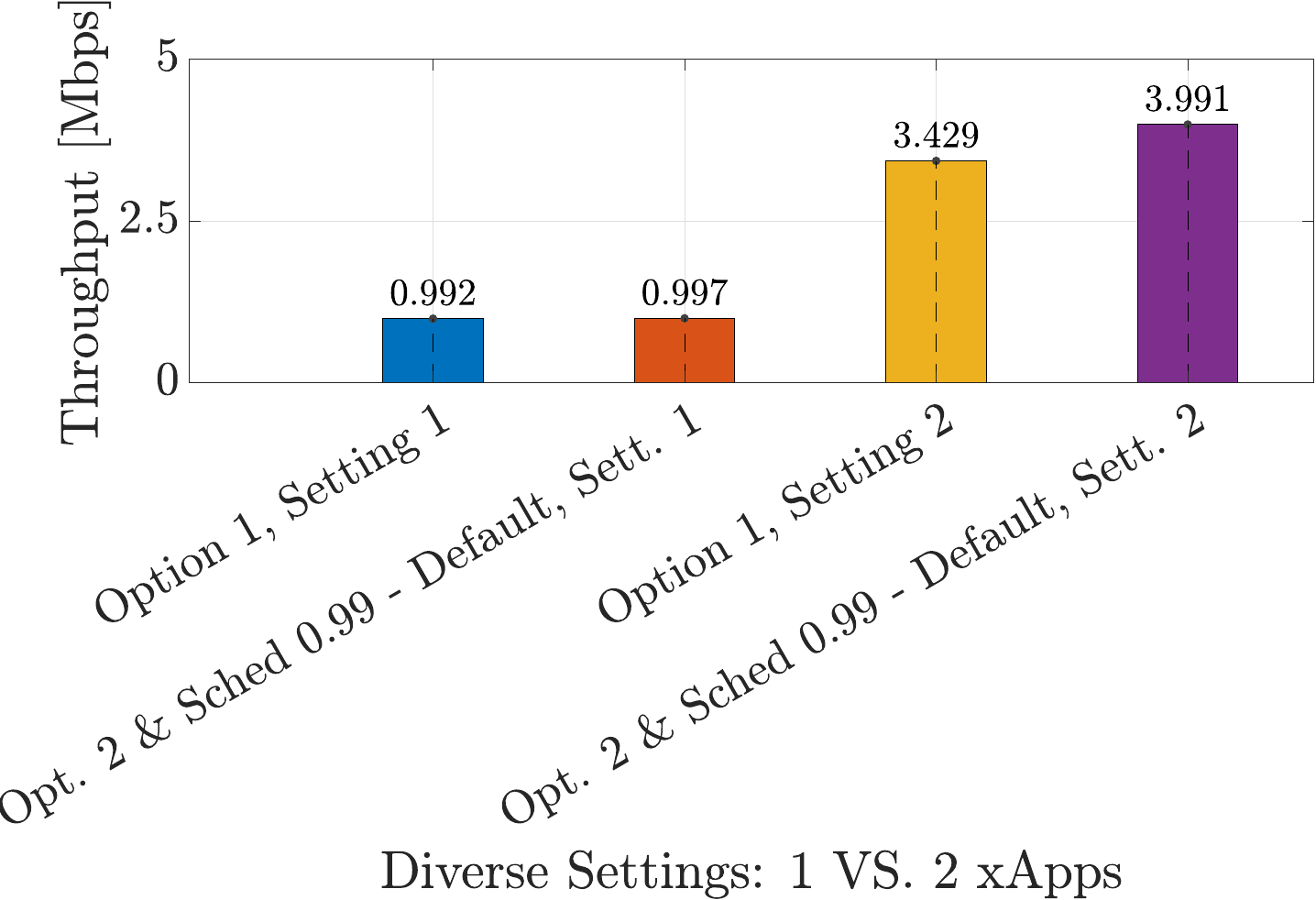}
\label{mob_barplot_troughput}}
\hfil
\subfigure[\gls{mmtc} Packets]{\includegraphics[width=1.692in]{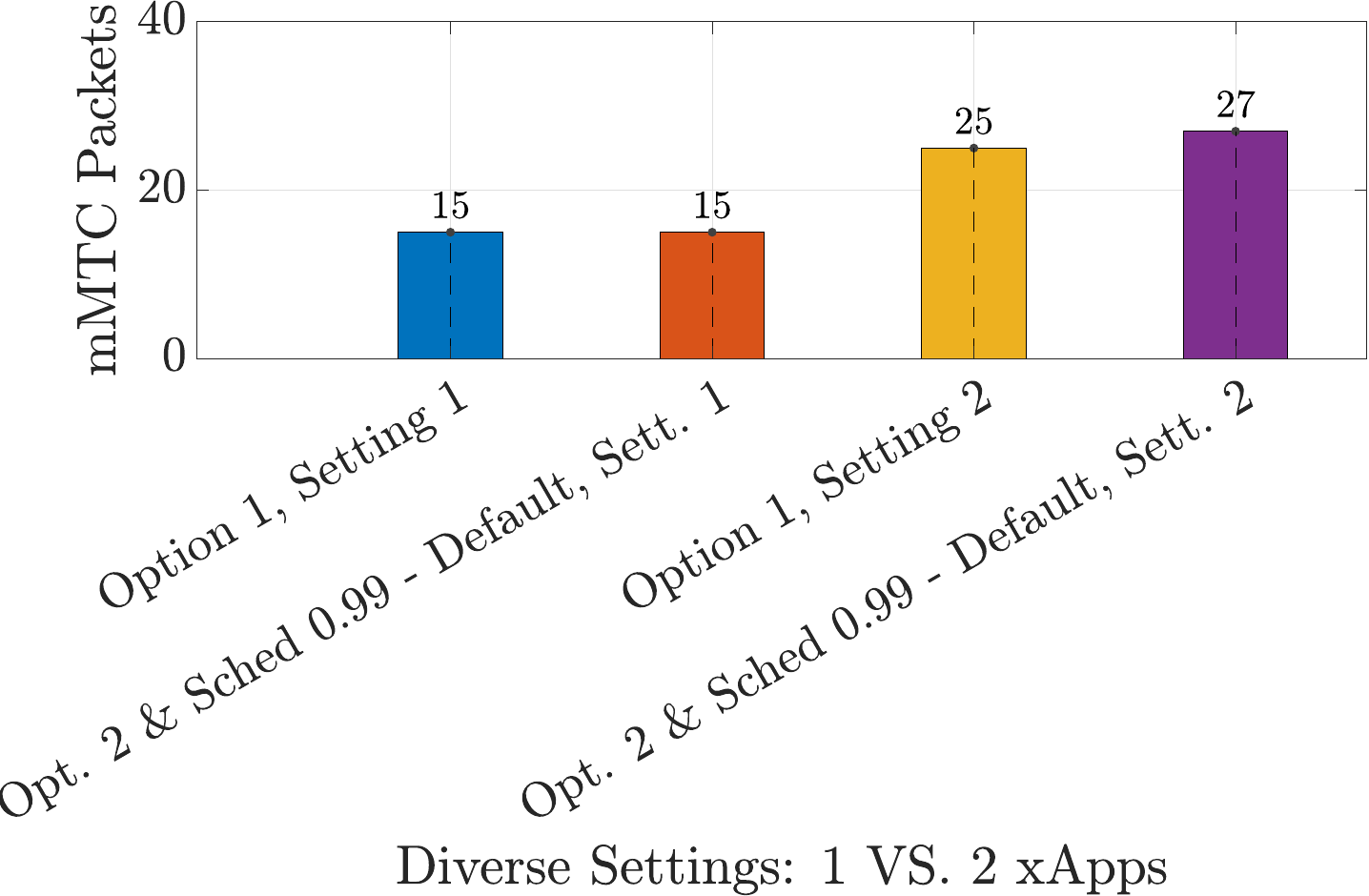}
\label{mob_barplot_mmtc_packets}}
\setlength\abovecaptionskip{-.02cm}
\caption{\textcolor{blue}{Median values obtained under mobility with Settings~$1$ and~$2$ of Table~\ref{table:net-settings} and the PPO DRL Architecture.}}
\label{Figure12-1b}
\vspace{-0.25cm}
\end{figure}

\begin{figure*}[t!]
\centering
\subfigure[\gls{embb} Throughput]{\includegraphics[height=2.85cm]{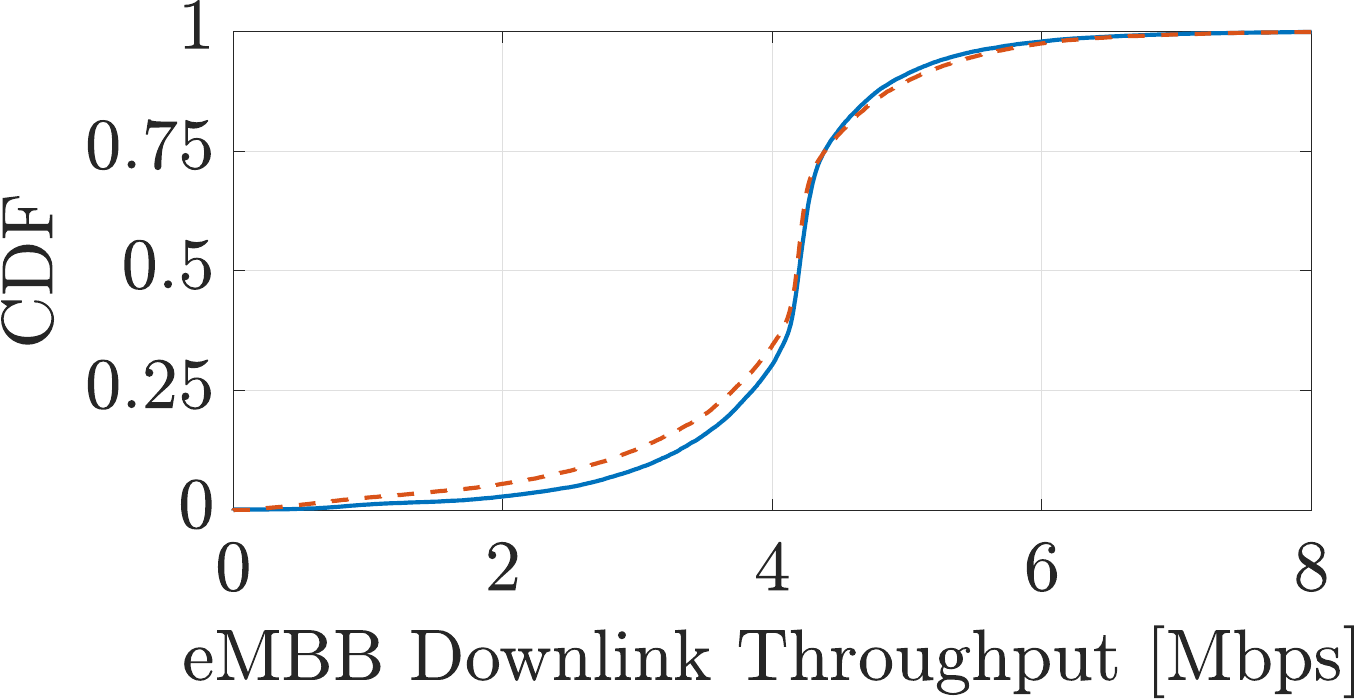}
\label{comp2-embb-throughput}}
\hfil
\subfigure[\gls{mmtc} Packets]{\includegraphics[height=2.85cm]{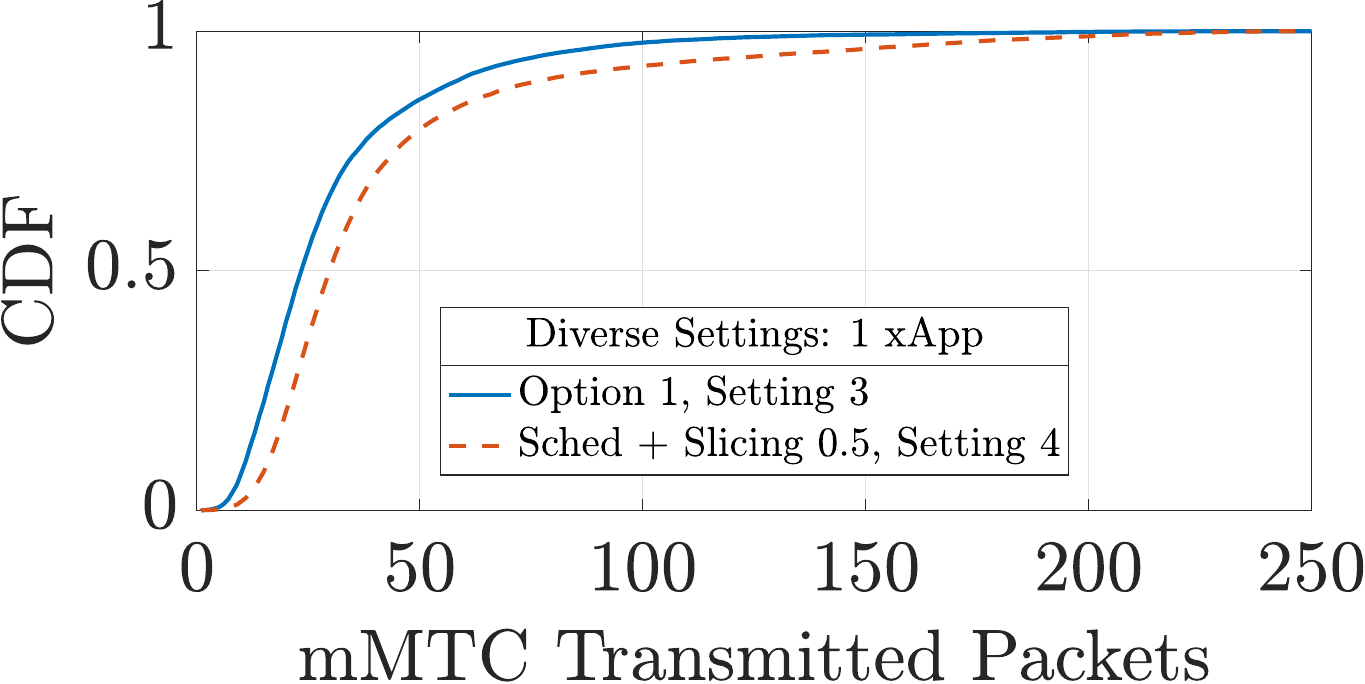}
\label{comp2-tx-pkts-mmtc}}
\hfil
\subfigure[\gls{urllc} Buffer Occupancy]{\includegraphics[height=2.85cm]{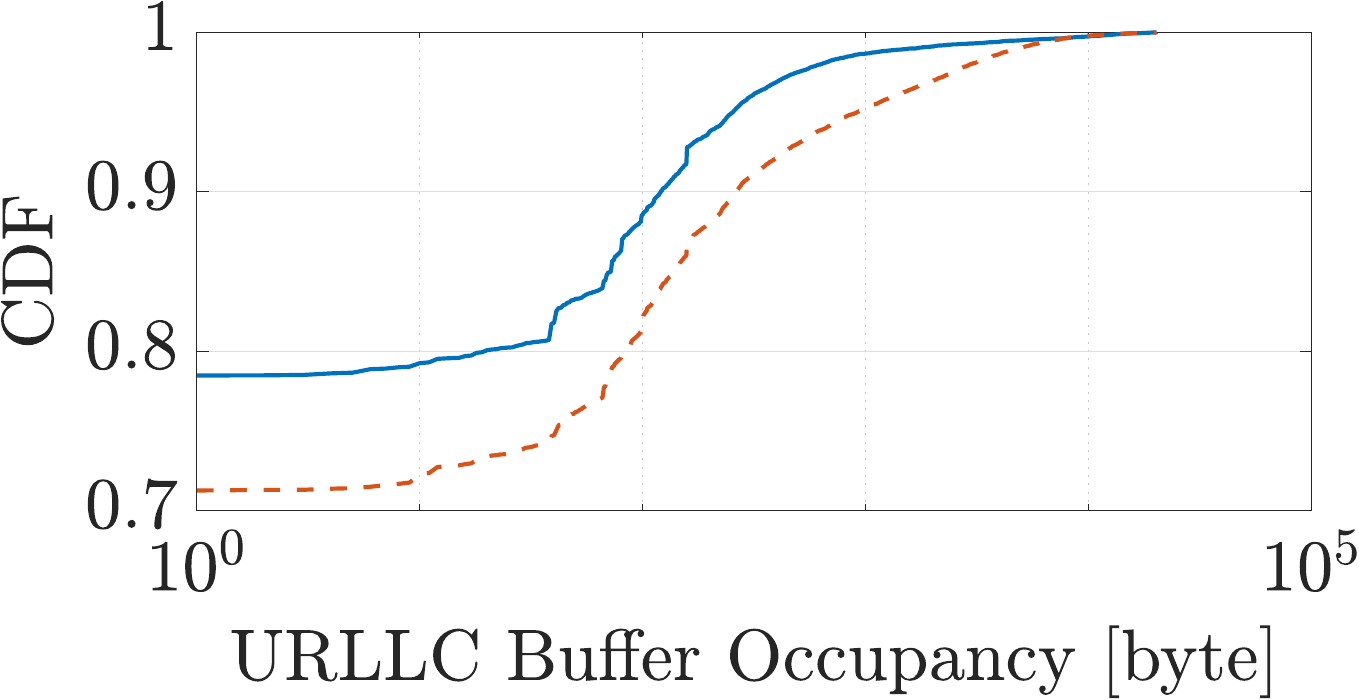}
\label{Figure7c2}}
\setlength\abovecaptionskip{-.02cm}
\caption{Performance evaluation focusing on the case of joint-slice optimization with a single xApp and the PPO DRL Architecture under Settings~$3$ and~$4$ of Table~\ref{table:net-settings}.} 
\label{Figure7-2a}
\vspace{-0.55cm}
\end{figure*}

\begin{figure}[t!]
\centering
\subfigure[\gls{embb} Throughput]{\includegraphics[width=1.6in]{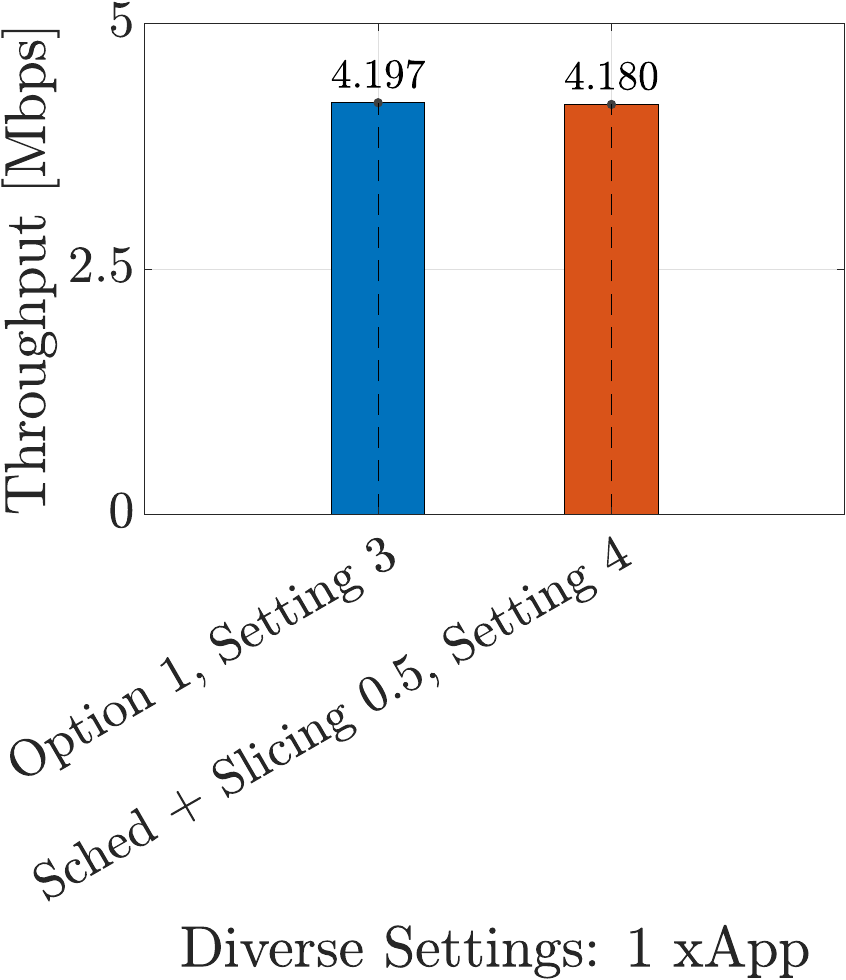}
\label{comp2-embb-throughput_barplot}}
\hfil
\subfigure[\gls{mmtc}  Packets]{\includegraphics[width=1.6in]{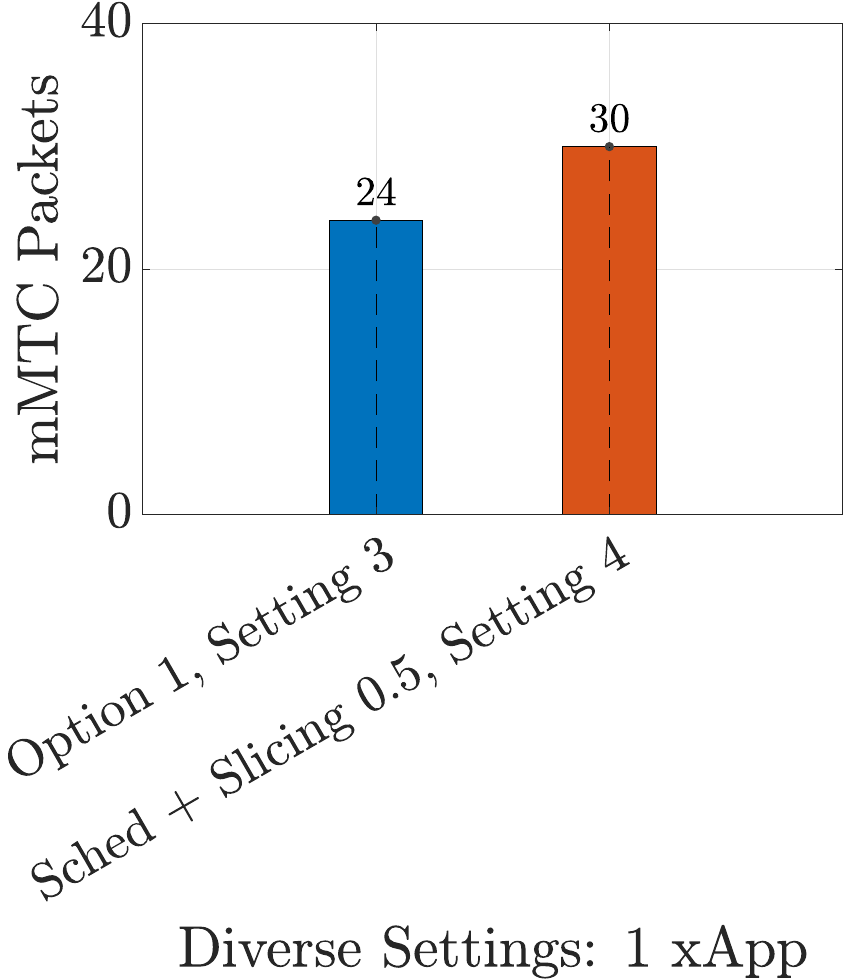}
\label{comp2-urllc-bufsize_barplot}}
\setlength\abovecaptionskip{-.02cm}
\caption{\textcolor{blue}{Median values obtained when focusing on the case of joint-slice optimization with a single xApp and the PPO DRL Architecture under Settings~$3$ and~$4$ of Table~\ref{table:net-settings}.}}
\label{Figure11-1b}
\vspace{-1.01cm}
\end{figure}

\textcolor{blue}{We look into the case where we control three different \gls{ran} control timers, as reported in Table~\ref{table:control-time}, with the focus primarily steered onto Sets~$2$ and~$3$. In the latter cases, the \gls{kpm} collection granularity (i.e., \glspl{kpm} Log time) matches both the Action Update and the \gls{kpm} report granularities on the \gls{bs}'s \gls{du}. In detail, \glspl{kpm} are directly streamed into the \gls{ric}, upon collected, with an action being sent back through the E2 termination. This is in contrast with the case of Set $1$, where the \gls{du} report time is equal to $1$ s, and hence some of the freshly captured data may not be reported. The findings of this evaluation are presented in Figs.~\ref{Figure5-3a} and~\ref{Figure10-1b}, and all configurations concern \gls{drl} agents trained under the \texttt{Default} weight configuration presented in Table~\ref{table:weight-confs-list}.}

\textcolor{blue}{In Fig.~\ref{Figure5a3}, it is observed that all configurations result in a comparable high performance, delivering a median throughput value of $\sim 4$\:Mbps (Fig.~\ref{ran-timers-bar-throughput}) with the only exception of \texttt{Sched 0.99} under Set.~$2$ which delivers poor throughput performance.
The aforementioned configuration yields a performance similar to the one reported in Fig.~\ref{fig:Figure4d}, where \texttt{Sched 0.99} is evaluated under the time granularities of Set.~$1$. Indeed, this combination achieves the same number of transmitted packets (i.e., $25$\:packets) with a negligible drop on the \gls{embb} throughput, which is now reported at $\sim 2.02$ Mbps, a value approximately $5\%$ less than the $2.125$\:Mpbs reported with Set~$1$. 
Additionally, the \texttt{Sched 0.99} xApp, when tested under Set~$3$, attains the maximum reported throughput value for controlling scheduling using a single xApp, \textcolor{red}{(reaching $4.028$\:Mbps, a $\sim99\%$ increase compared to $2.02$ Mbps achieved with Set~$2$).} The latter is achieved while transmitting  23 packets, which is two packets fewer than the maximum number of transmitted packets ever recorded when controlling scheduling alone, and with \gls{ppo}-implemented agents. Based on these findings, scheduling should be controlled using small time granularities (i.e., $\sim 100$\:ms) to ensure high \gls{embb} throughput, while guaranteeing no performance degradation on the \gls{mmtc} slice. Furthermore, the results illustrated in Figs.~\ref{Figure5b3} and~\ref{ran-timers-bar-mmtc} indicate that both scheduling and slicing, as well as scheduling under Sets $2$ and $3$, can achieve the same performance on the \gls{mmtc}. However, \texttt{Sched \& Slicing 0.5} delivers higher \gls{embb} throughput, and hence, is preferred compared to \texttt{Sched 0.99}. In consistency with the reported findings so far, and when focusing on the three action profiles individually, we observe \gls{mmtc}'s and \gls{embb}'s competitive behavior (e.g., they heavily compete to gain radio resources). This trend is clearly illustrated in Figs.~\ref{Figure5a3},~\ref{Figure5b3},~\ref{ran-timers-bar-throughput}, and~\ref{ran-timers-bar-mmtc}, where \texttt{Slicing 0.5} under Set~$3$ achieves the highest throughput value at $4.195$\:Mbps, but the lowest number of transmitted packets (i.e., $18$ packets). The reported results in Figs.~\ref{ran-timers-bar-throughput} and~\ref{ran-timers-bar-mmtc} also indicate that Set~$3$ primarily boosts \gls{embb}'s performance, compared to Set~$2$ which is a better fit for \gls{mmtc} \glspl{ue}. Finally, in Fig.~\ref{Figure5c3}, we note that all configurations for the investigated set of granularities deliver identical performance on the \gls{urllc} slice, and although not shown in the figures, they consistently maintain a median buffer occupancy of 0 byte to ensure low latency. It is also observed that both \texttt{Sched \& Slicing 0.5} and \texttt{Slicing 0.5} under Set~$2$, perform slightly better on the aforementioned slice, compared to \texttt{Sched 0.99} under Set~$2$, which performs slightly worse.}


\begin{table}[htb]
\centering
\small
\setlength\abovecaptionskip{-.1cm}
\caption{\gls{ml} Agent \& Network Condition Settings}
\begin{adjustbox}{width=1\linewidth}
\begin{tabular}{@{}cccccl@{}}
\toprule
\multicolumn{1}{c}{\shortstack{\textbf{Setting} \\ \textbf{ID}}} & \multicolumn{1}{c}{\shortstack{\textbf{\gls{ue}} \\ \textbf{Speed [m/s]}}}
& \multicolumn{1}{c}{\shortstack{\textbf{Traffic} \\ \textbf{Load}}}
& \multicolumn{1}{c}{\shortstack{\textbf{\gls{ran} Control} \\ \textbf{Timers}}}
& \multicolumn{1}{c}{\shortstack{\textbf{Weight} \\ \textbf{Configuration}}} \\
\midrule
\centering \textbf{1} & 3 & \text{Profile 1} & \text{Set 1} & \texttt{Alternative/Default} \\
\centering \textbf{2} & 3 & \text{Profile 2} & \text{Set 1} & \texttt{Alternative/Default} \\
\centering \textbf{3} & 0 & \text{Profile 2} & \text{Set 1} & \texttt{Alternative} \\
\centering \textbf{4} & 0 & \text{Profile 2} & \text{Set 2} & \texttt{Default} \\
\bottomrule
\end{tabular}
\end{adjustbox}
\label{table:net-settings}
\vspace{-.4cm}
\end{table}

\section{Out-of-Sample Experimental Evaluation}\label{sec:experimental-evaluation2}

\textcolor{blue}{In the Out-of-Sample experimental performance evaluation, we focus on the case of \textbf{Location 2}, and we test the framework's performance in static and mobile scenarios under two different traffic loads. Recall that our agents have been trained using data from \textbf{Location 1} only. Information regarding the traffic profiles and the \gls{ue} mobility can be found in Tables~\ref{table:traffic-profiles} and~\ref{table:net-settings}.}

\textcolor{blue}{In Figs.~\ref{Figure6-2a} and~\ref{Figure12-1b}, we present the results collected when testing the framework under mobile \gls{ue} conditions and diverse traffic profiles for Settings~$1$ and $2$ as defined in Table~\ref{table:net-settings}. We juxtapose the configuration that delivered the highest \gls{embb} throughput (i.e., $4.208$\:Mbps), among those tested in this work, namely Option 1: \texttt{Sched \& Slicing 0.5 - Alternative} with the case of two xApps which jointly optimize the performance of all slices, namely Option 2: \texttt{Slicing 0.5}, and \texttt{Sched 0.99}. Finally, the set of granularities which are considered for this evaluation are given by Set~$1$ of Table~\ref{table:control-time}.}

\textcolor{blue}{
In Figs.~\ref{Figure6a2},~\ref{Figure6b2}, and~\ref{Figure6c2}, we observe that under the traffic conditions of Setting~$1$, both configurations deliver almost identical performance. This is clearly depicted in Figs.~\ref{mob_barplot_troughput} and~\ref{mob_barplot_mmtc_packets}, where the mentioned xApps achieve the same median values for \gls{embb} throughput and number of \gls{mmtc} transmitted packets. However, when the two configurations are tested under Setting~$2$ and specifically under Traffic Profile $2$, a slight performance degradation is observed in the case of joint-slice optimization with $1$~xApp. In detail, Figs.~\ref{Figure6a2} and~\ref{mob_barplot_troughput} indicate that the performance of \gls{embb} is optimized when using $2$~xApps, resulting in $\sim16$\% more throughput compared to the case with a single xApp. In the \gls{mmtc} slice, both configurations deliver similar performance, as shown in Fig.~\ref{Figure6b2}. However, when focusing on the former setup, it results in delivering two more packets as shown in Fig.~\ref{mob_barplot_mmtc_packets}. On the \gls{urllc}, it is noted that most of the configurations under both Settings resulted in a median buffer occupancy of $0$~byte. However, in the case of joint optimization with $1$~xApp, this value was reported at $52$~byte. Hence, in case of \gls{ue} mobility, \texttt{Sched \& Slicing 0.5 - Alternative} underperforms compared to joint optimization with \texttt{Slicing 0.5}, and \texttt{Sched 0.99}. Based on the reported findings, the simultaneous operation of two xApps can more successfully tackle  possible \gls{ue} disconnections caused by mobile conditions, and therefore enhance network performance.}

\textcolor{blue}{In Figs.~\ref{Figure7-2a} and~\ref{Figure11-1b} we include the results collected when testing the optimization framework under Settings~$3$ and $4$ in \textbf{Location 2}. In this evaluation, we once again compare Option 1: \texttt{Sched \& Slicing 0.5 - Alternative} with \texttt{Sched \& Slicing 0.5 - Default}. It is noted that the main difference between these two setups pertains to the fact that the former is evaluated under granularity Set~$1$, while the latter under Set~$2$. This comparison allows us to assess how collecting and reporting \glspl{kpm} as well as updating actions at smaller timescales (i.e., $250$\:ms) will impact the effectiveness of the optimization framework, under the traffic load defined in Profile~$2$. Additionally, the latter xApp is chosen due to the fact that, among all configurations evaluated in Section~\ref{Section IV-E} in terms of \gls{ran} control timers, it has managed to provide the most significant performance boost in the network. In detail, when tested under Set~$2$, 
it has improved the performance of \gls{mmtc} by $\sim 56$\:\%, achieving a median value of $25$~packets, as shown in Fig.~\ref{ran-timers-bar-mmtc}. This is in contrast to when it was evaluated under Set~$1$, where it delivered a median value of $16$~packets (Fig.~\ref{fig:Figure4e}). With regards to the \gls{embb} slice, it still manages to maintain good overall performance, delivering a median throughput value of $4.140$\:Mbps (Fig.~\ref{ran-timers-bar-throughput}).}

\textcolor{blue}{The results reported in Figs.~\ref{comp2-embb-throughput} and~\ref{comp2-embb-throughput_barplot} indicate that both configurations deliver similar performance in the \gls{embb} slice, with Option 1: \texttt{Sched \& Slicing 0.5 - Alternative} performing slightly better. However, on the \gls{mmtc} slice, \texttt{Sched \& Slicing 0.5 - Default} performs better, as depicted in Fig.~\ref{comp2-tx-pkts-mmtc}, by achieving a median value of $30$ transmitted packets (Fig.~\ref{comp2-urllc-bufsize_barplot}). \gls{urllc} \glspl{ue}, once again, achieve zero latency (i.e., buffers are emptied), and, as a result, the respective figures are omitted.}

\section{Broader Evaluation of \pandora}\label{sec:generalization-pandora}

\begin{figure*}[t!]
\centering
\subfigure[\gls{embb} \gls{prb} Ratio]{\includegraphics[height=2.85cm]{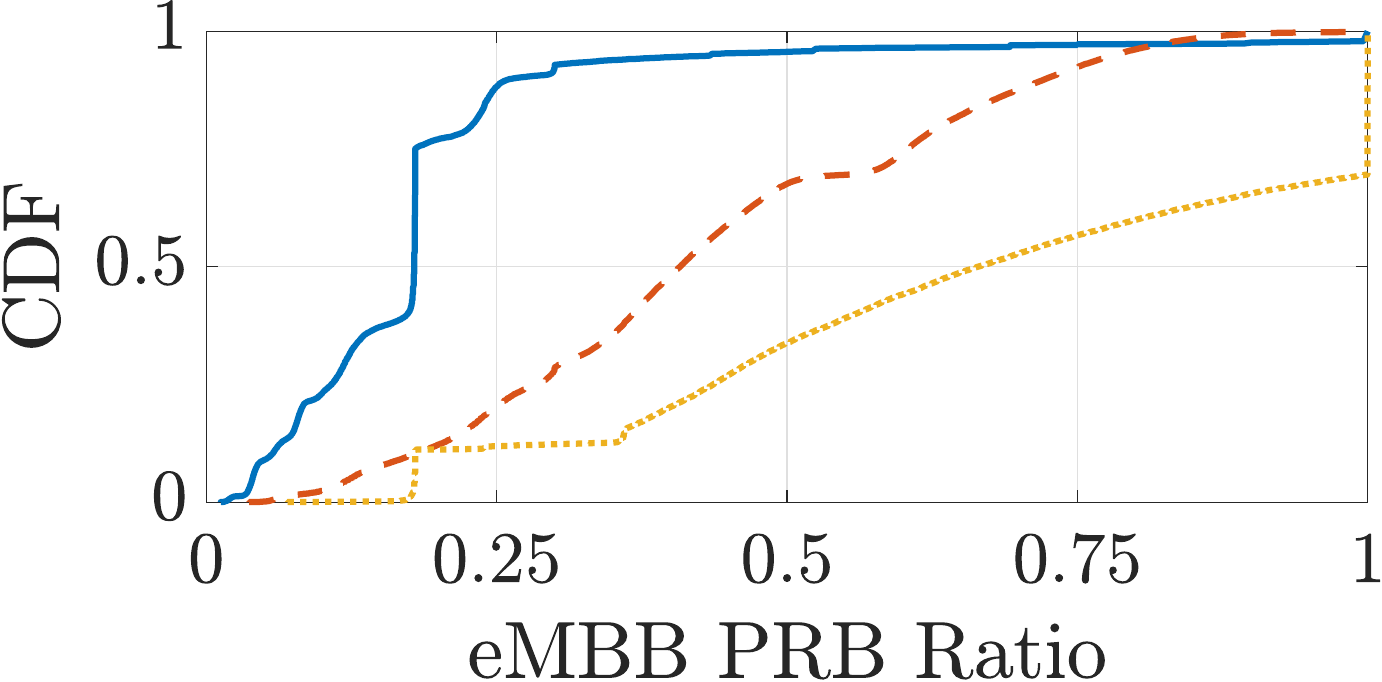}
\label{fig:Figure-prbratio-1}}
\hfil
\subfigure[\gls{mmtc} \gls{prb} Ratio]{\includegraphics[height=2.85cm]{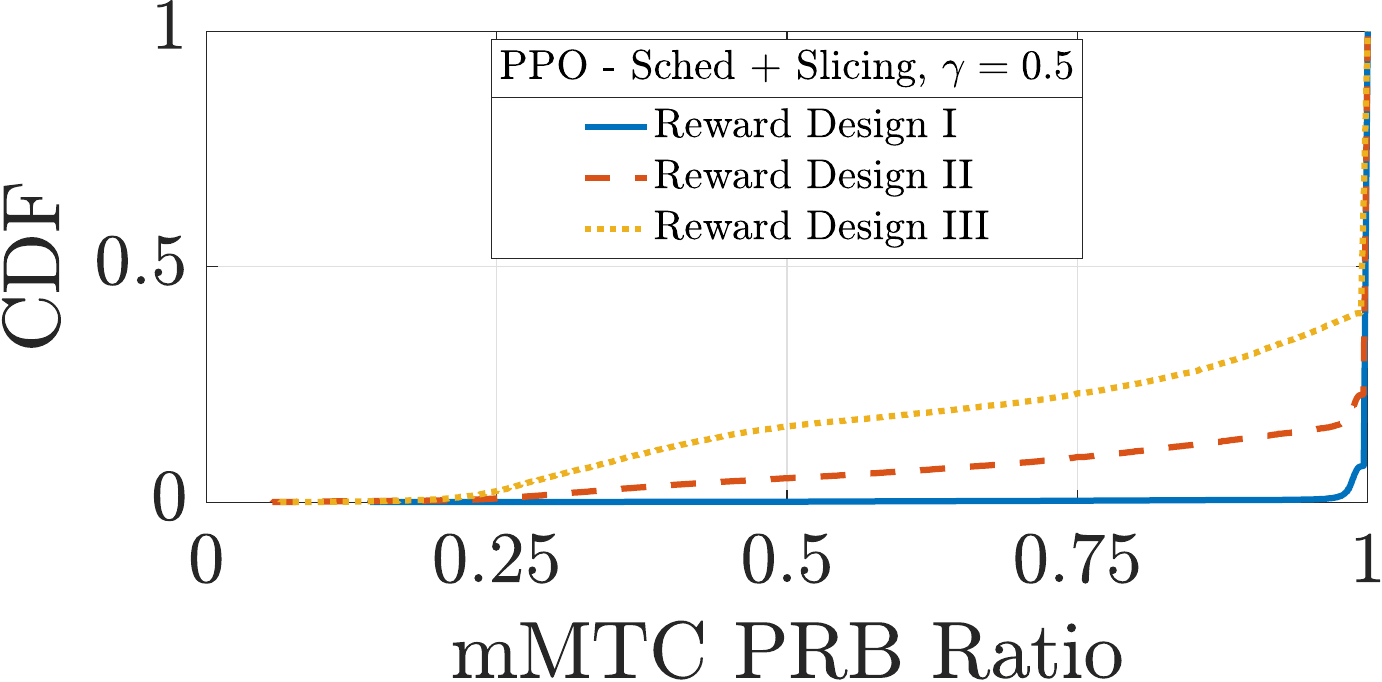}
\label{fig:Figure-prbratio-2}}
\hfil
\subfigure[\gls{urllc} \gls{prb} Ratio]{\includegraphics[height=2.85cm]{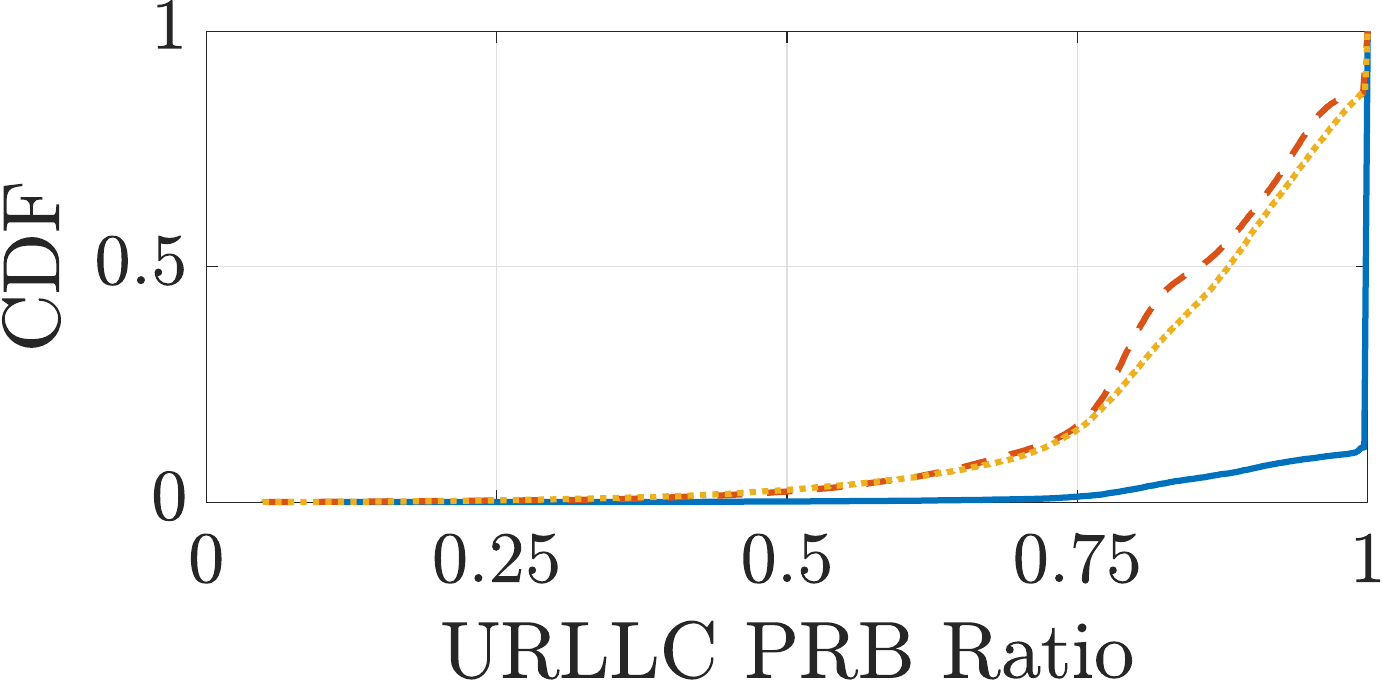}
\label{fig:Figure-prbratio-3}}
\setlength\abovecaptionskip{-.02cm}
\caption{\textcolor{red}{\gls{ue} Satisfaction expressed in the form of PRB ratio for the Reward Designs of Table~\ref{table:reward-design}.}}
\label{Figure-prbratio-reward}
\vspace{-0.55cm}
\end{figure*}

\begin{figure*}[t!]
\centering
\subfigure[\gls{embb} Throughput]{\includegraphics[height=2.85cm]{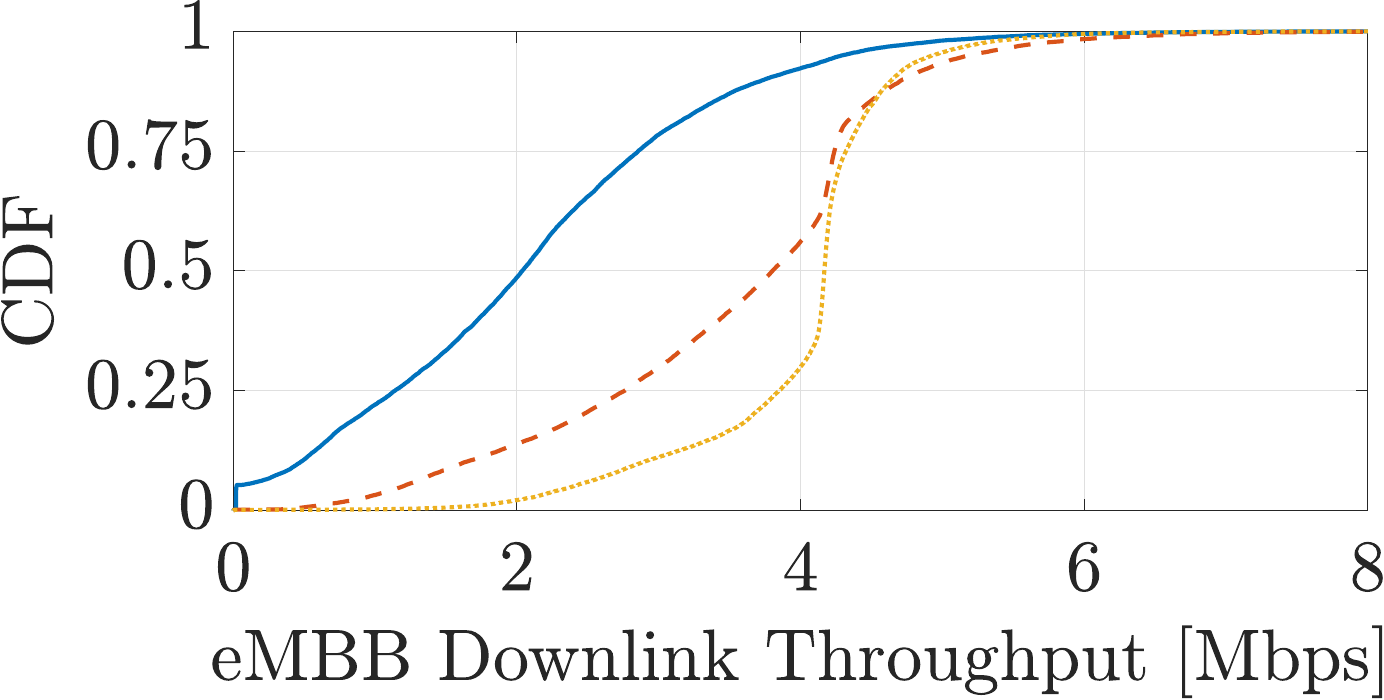}
\label{fig:Figure-prbratio-1-embb}}
\hfil
\subfigure[\gls{mmtc} Packets]{\includegraphics[height=2.85cm]{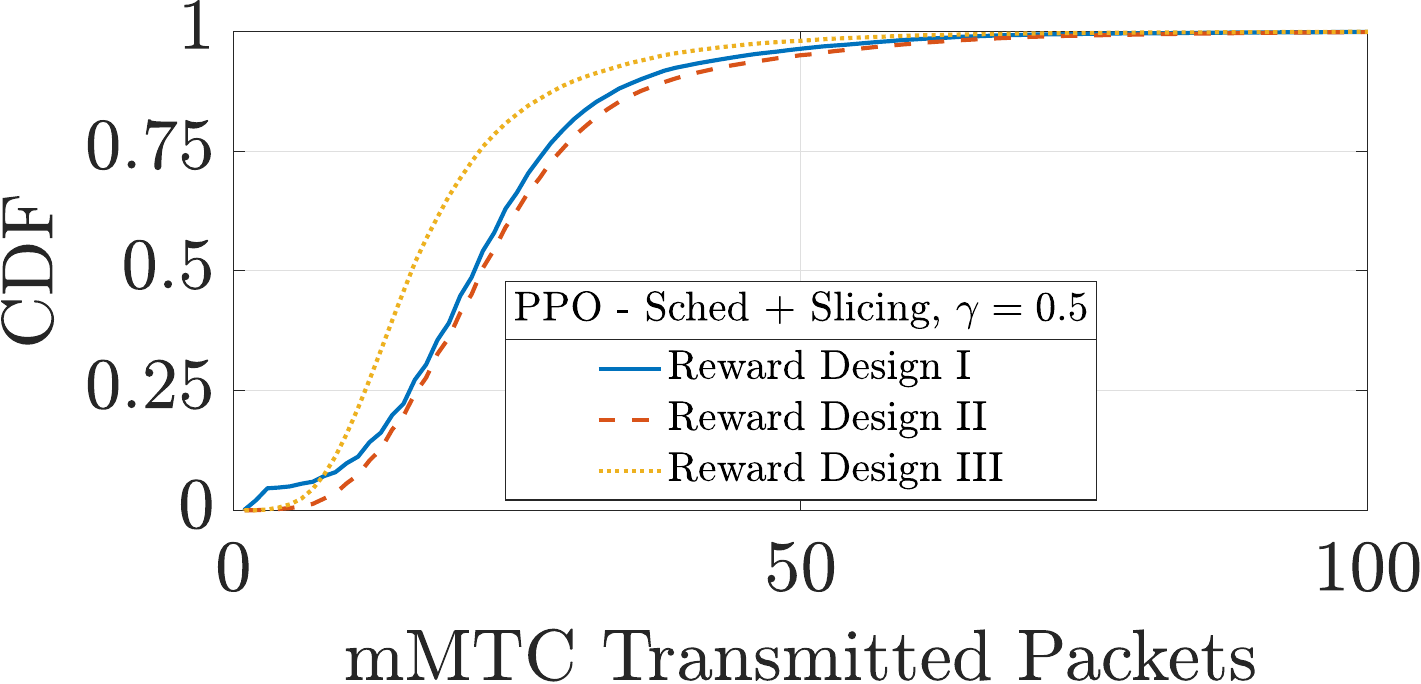}
\label{fig:Figure-prbratio-2-mmtc}}
\hfil
\subfigure[\gls{urllc} Buffer Occupancy]{\includegraphics[height=2.85cm]{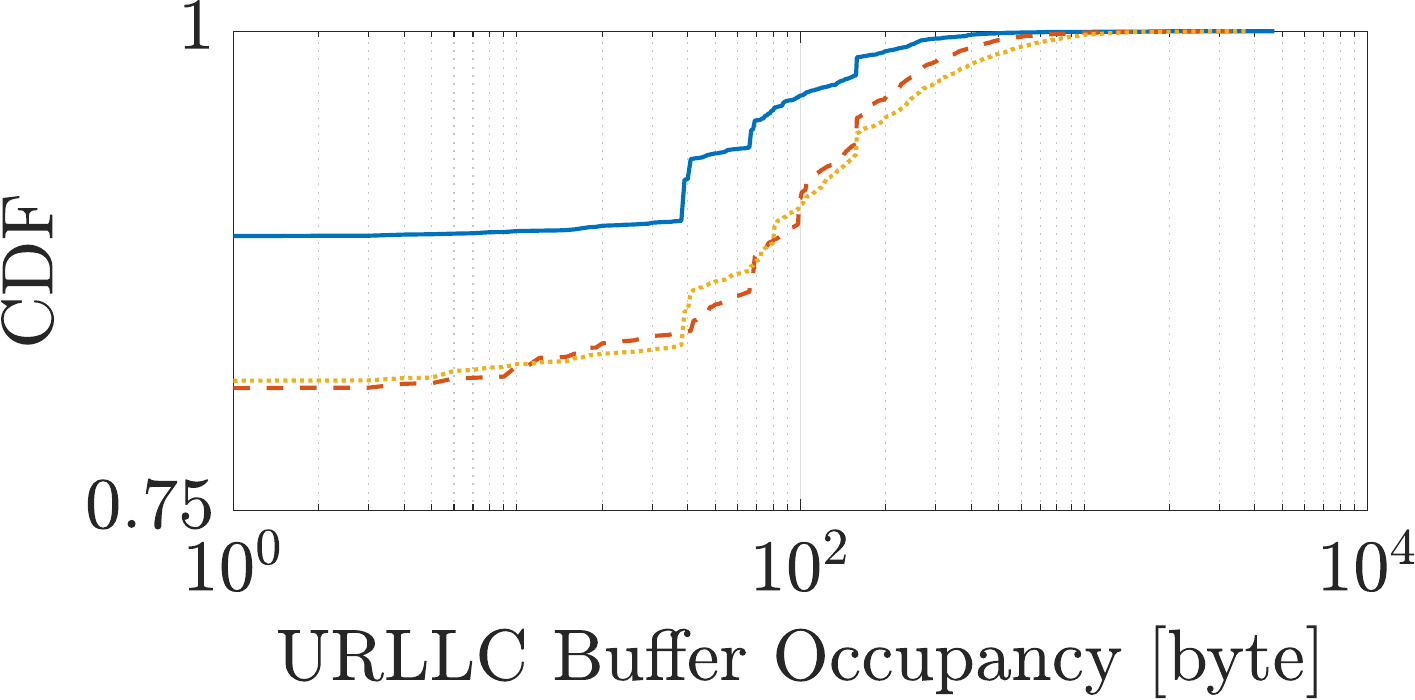}
\label{fig:Figure-prbratio-3-urllc}}
\setlength\abovecaptionskip{-.02cm}
\caption{\textcolor{red}{Impact of the Reward Designs from Table~\ref{table:reward-design} on \gls{dl} \gls{embb} Throughput, \gls{dl} \gls{mmtc} and \gls{tx} Packets}.}
\label{Figure-prbratio-reward-cdfs}
\vspace{-0.7cm}
\end{figure*}

\begin{figure}[t!]
\centering
\subfigure[\gls{embb} DL Throughput]{\includegraphics[width=1.6in]{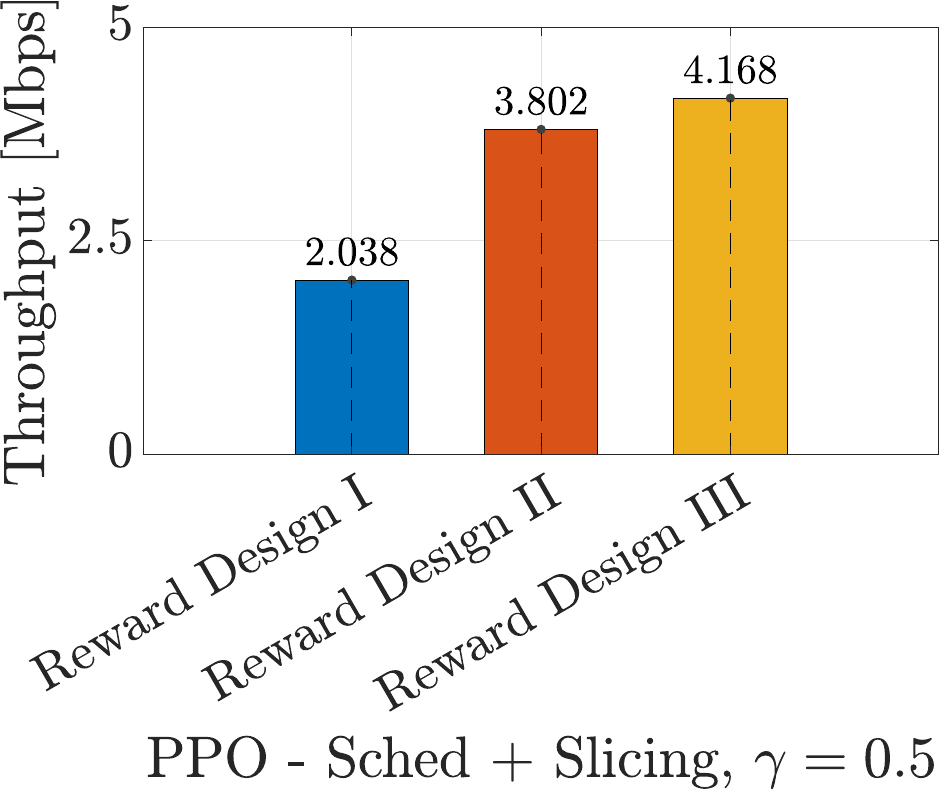}
\label{fig:ebarplot_prbratiReward_embb}}
\hfil
\subfigure[\gls{mmtc}  Packets]{\includegraphics[width=1.6in]{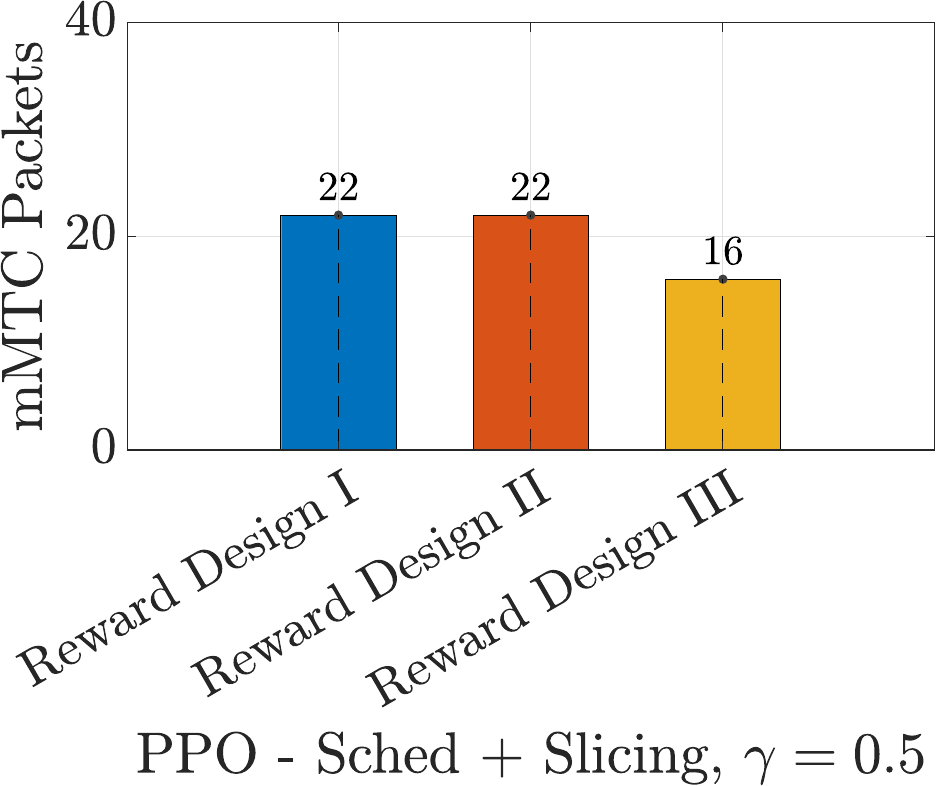}
\label{fig:barplot_prbratiReward_mmtc}}
\setlength\abovecaptionskip{-.02cm}
\caption{\textcolor{red}{Impact of the Reward Designs from Table~\ref{table:reward-design} on the median values of \gls{dl} \gls{embb} throughput, \gls{dl} \gls{mmtc} \gls{tx} Packets, and \gls{dl} \gls{urllc} buffer occupancy}.}
\label{barplot_prbratiReward}
\vspace{-0.65cm}
\end{figure}

\textcolor{red}{The goal of this section is to provide additional insights into \pandora by: (i) considering a broader set of evaluation metrics aimed at shedding light on resource utilization and 
\gls{ue} satisfaction; and (ii) testing across scenarios with a larger number of \glspl{ue}. Specifically, in the latter case, we examine the performance of our xApps in a scenario where we add additional \glspl{ue} to the the \gls{mmtc} slice. In line with the previous sections, we will evaluate \pandora in setups previously seen during the training process (i.e., in-sample experimental evaluation in Location~$1$), as well as unseen conditions (i.e., out-of-sample experimental evaluation in Location~$2$). All subsequent results concern joint optimization with \texttt{Sched \& Slicing 0.5} and are tested using Sets $1$ and $2$ of \gls{ran} control timers from Table~\ref{table:control-time} and the Traffic Profiles listed in Table~\ref{table:traffic-profiles}, both in static and mobile scenarios (e.g., $3$~m/s).}

\textcolor{red}{It is noted that so far 
we have considered three main reward functions for the respective slices (i.e., maximization of \gls{dl} \gls{embb} throughput and \gls{mmtc} \gls{tx} Packets, and minimization of \gls{dl} buffer occupancy in the \gls{urllc}).
In order to demonstrate \pandora's capabilities as well as to compare the reward functions evaluated so far, we additionally craft two new reward designs, which are defined in Table~\ref{table:reward-design}. In \emph{Reward Design I}, all slices are given the same priority (i.e., they are all assigned a weight of~$1$). The reward for all slices is the maximization of the \gls{prb} Ratio, defined as $\text{\small \gls{prb} Ratio}=\frac{\text{Sum of Granted \glspl{prb}}}{\text{Sum of Requested \glspl{prb}}}, \text{where \gls{prb} Ratio} \in [0, 1]$, that represents the amount of the allocated \glspl{prb} per slice. For \emph{Reward Design II}, we consider the maximization of the \gls{embb} throughput and the minimization of the \gls{urllc} buffer occupancy,
with their weights defined in Table~\ref{table:weight-confs-list} under the \texttt{Default} weight configuration. For the \gls{mmtc} slice, we consider the maximization of the \gls{prb} Ratio and
the slice's weight is set to $0.5$. Lastly, \emph{Reward Design III} pertains to the joint scheduling and slicing optimization discussed and evaluated in Section~\ref{Section IV-A} and shown in Table~\ref{table:reward-design}.
The results of this experimental evaluation are illustrated in Figs.~\ref{Figure-prbratio-reward},~\ref{Figure-prbratio-reward-cdfs},~\ref{barplot_prbratiReward} and were collected under Traffic Profile~$2$ (i.e., $4$~Mbps \gls{embb} throughput, $44.6$~kbps \gls{mmtc} throughput and $89.3$~kbps \gls{urllc} throughput), in Location~$1$ (i.e, $50$~m radius from the \gls{bs}), and Set~$1$ of \gls{ran} control timers (see Table~\ref{table:control-time}), for a total of $6$ \glspl{ue}, uniformly distributed across the slices.}

\textcolor{red}{In Fig.~\ref{Figure-prbratio-reward}, we show the \gls{prb} Ratio for the three Reward Designs defined in Table~\ref{table:reward-design}. We observe that with \emph{Reward Design I}, \gls{embb} \glspl{ue} experience lower levels of \emph{satisfaction} (Fig.~\ref{fig:Figure-prbratio-1}), in terms of the number of allocated \glspl{prb}. On the contrary, both on the \gls{mmtc} and \gls{urllc} slices (Figs.~\ref{fig:Figure-prbratio-2} and~\ref{fig:Figure-prbratio-3}), \emph{Reward Design I} results in significant \gls{ue} satisfaction. The latter is due to the fact that both \gls{mmtc} and \gls{urllc} have lower traffic requirements to satisfy (i.e., $44.6$~kbps and $89.3$~kbps throughput, respectively), compared to \gls{embb} (i.e., $4$~Mbps throughput). 
With \emph{Reward Design II}, we observe an improvement in \gls{ue} satisfaction levels on the \gls{embb}, meaning that \glspl{ue} are granted the resources they request, along with equally good performance in \gls{mmtc} and \gls{urllc}.
The results with \emph{Reward Design III} indicate that this combination of weight configuration (i.e., tailored to prioritize one slice over the other) and slice-specific rewards effectively result in good overall performance without penalizing one slice over the other. Notably, \emph{Reward Design III} is the configuration under which \gls{embb} \glspl{ue} experience more satisfaction. Finally, we can observe the competitive behavior among \gls{embb} and \gls{mmtc}. Indeed, the reward designs that delivers a higher \gls{ue} satisfaction for \gls{embb} (Fig.~\ref{fig:Figure-prbratio-1}) also delivers lower satisfaction for \gls{mmtc} (Fig.~\ref{fig:Figure-prbratio-2}).
}

\textcolor{red}{In Figs.~\ref{Figure-prbratio-reward-cdfs} and~\ref{barplot_prbratiReward}, we observe the impact of the Reward Designs explored in Fig.~\ref{Figure-prbratio-reward} on the three \glspl{kpi} of interest, namely \gls{dl} \gls{embb} throughput, \gls{mmtc} \gls{tx} packets, and \gls{urllc} buffer occupancy. In detail, Figs.~\ref{fig:Figure-prbratio-1},~\ref{fig:Figure-prbratio-1-embb} and~\ref{fig:ebarplot_prbratiReward_embb} demonstrate a similar trend, since a higher value of \gls{prb} Ratio yields in higher throughput. A similar trend is observed in Figs.~\ref{fig:Figure-prbratio-2},~\ref{fig:Figure-prbratio-2-mmtc} and~\ref{fig:barplot_prbratiReward_mmtc} for the \gls{mmtc} slice, with \emph{Reward Designs I} \& \emph{II} demonstrating similar performance (i.e., a median value of~$22$ \gls{tx} packets as illustrated in Fig~\ref{fig:barplot_prbratiReward_mmtc}). These results indicate that the maximization of the \gls{prb} Ratio poses as ideal reward for the aforementioned slice but with a penalty on the \gls{embb}. Specifically, even though \emph{Reward Designs I} \& \emph{II} result in $\sim37.5\%$ improvement on the performance of the \gls{mmtc}, they result in lower \gls{embb} throughput values as depicted in Fig.~\ref{fig:ebarplot_prbratiReward_embb}. On the \gls{urllc}, all \emph{Reward Designs} result in empty buffers, with \emph{Reward Design I} slightly resulting in higher performance, as seen in Fig.~\ref{fig:Figure-prbratio-3-urllc}.}

\textcolor{red}{In Fig.~\ref{Figure-cdfs_prbs}, we include the distribution of actions in terms of selected \glspl{prb} for the xApp Catalog presented in Table~\ref{table:xapp-catalog-comp}. The performance evaluation results indicate that agents can make decisions resulting in diverse action distributions (summarized in Table~\ref{tab:feasible_prb_allocation}), which is due to varying rewards and design choices, impacting the performance achieved by the system.}

\textcolor{red}{We now consider the case where we increase the number of \glspl{ue} allocated to the \gls{mmtc} slice as illustrated by \emph{Use Cases II} \& \emph{III}. Specifically, \emph{Use Case II} involves a total of $7$ \glspl{ue}, while in \emph{Use Case III}, the setup comprises a total of $8$ \glspl{ue}, with $3$ and $4$ \glspl{ue} allocated to the \gls{mmtc} slice, respectively. \emph{Use Case I}, corresponds to a setup of $6$ \glspl{ue} in total, and all deployments have been tested under Traffic Profile~$1$ of Table~\ref{table:traffic-profiles} and Set~$1$ of \gls{ran} control timers of Table~\ref{table:control-time} in Location~$2$ (i.e., $20$~m radius from the \gls{bs}). Finally, the xApp under evaluation corresponds to \emph{Reward Design III} of Table~\ref{table:reward-design}.}

\begin{table}[htb]
\centering
\Huge
\textcolor{red}{
\setlength\abovecaptionskip{-.1cm}
\caption{\gls{drl} Reward Design Catalog for \texttt{Sched \& Slicing 0.5} with \gls{ppo} in Location~$1$ and Set~$1$ of \gls{ran} control timers.}
\begin{adjustbox}{width=1.01\linewidth}
\begin{tabular}{@{}c|ccccl@{}} 
\toprule
\thead{\textbf{\Huge Reward} \\ \textbf{\Huge Design ID}} & & \thead{\textbf{\Huge eMBB}} & \thead{\textbf{\Huge mMTC}} & \thead{\textbf{\Huge URLLC}} \\
\midrule
\textbf{I} & \shortstack{\emph{\textbf{Weight Config.}} \\ \emph{\textbf{Slice Reward}}} & \shortstack{$1$ \\ Max. \gls{prb} Ratio}  &  \shortstack{$1$ \\ Max. \gls{prb} Ratio}  & \shortstack{$1$ \\ Max. \gls{prb} Ratio}  &  \\
\midrule
\textbf{II} & \shortstack{\emph{\textbf{Weight Config.}} \\ \emph{\textbf{Slice Reward}}} & \shortstack{$72.0440333$ \\ Max. \gls{dl} Throughput}  & \shortstack{$0.5$ \\ Max. \gls{prb} Ratio}  & \shortstack{$0.00005$ \\ Min. \gls{dl} Buffer Occupancy} \\
\midrule
\textbf{III} & \shortstack{\emph{\textbf{Weight Config.}} \\ \emph{\textbf{Slice Reward}}} & \shortstack{$72.0440333$ \\ Max. \gls{dl} Throughput} & \shortstack{$0.229357798$ \\ Max. \gls{dl} \gls{tx} Packets}  &  \shortstack{$0.00005$ \\ Min. \gls{dl} Buffer Occupancy} \\
\bottomrule
\end{tabular}
\end{adjustbox}
\label{table:reward-design}}
\vspace{-.35cm}
\end{table}

\begin{figure*}[t!]
\centering
\subfigure[\gls{embb} \glspl{prb}]{\includegraphics[height=2.85cm,]{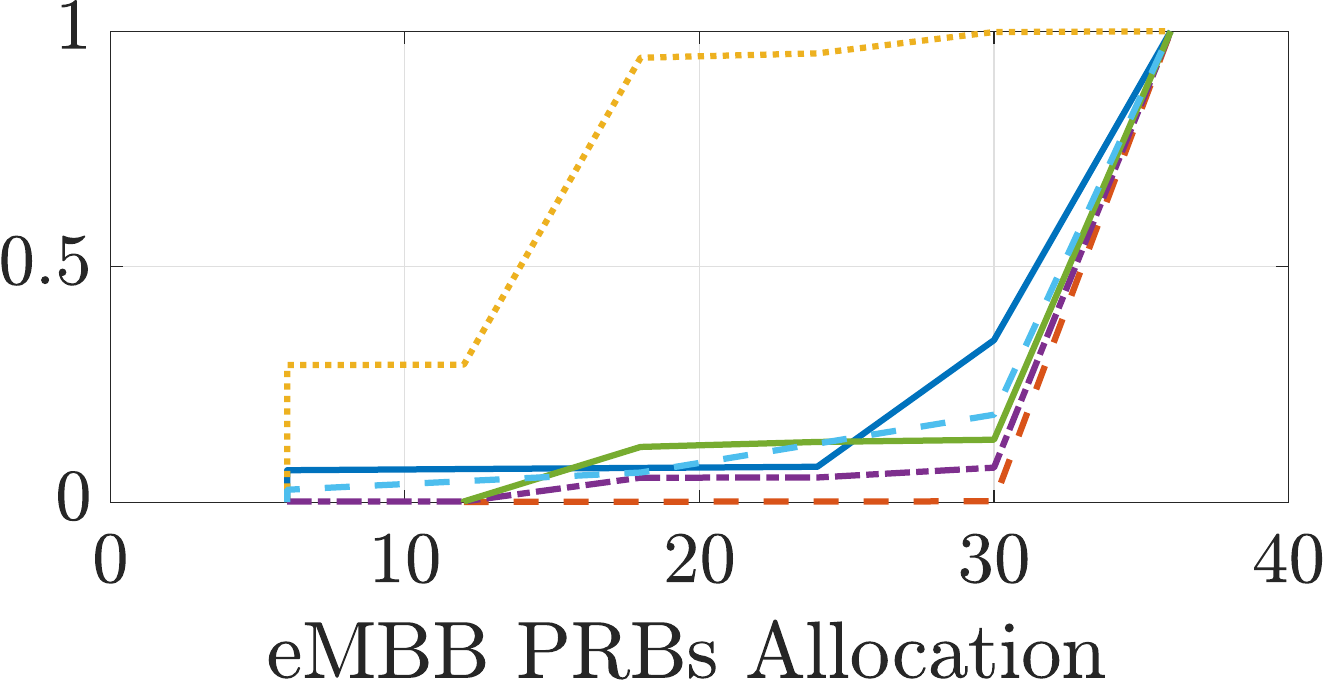}
\label{fig:Figure-embb_prbs}}
\hfil
\subfigure[\gls{mmtc} \glspl{prb}]{\includegraphics[height=2.85cm]{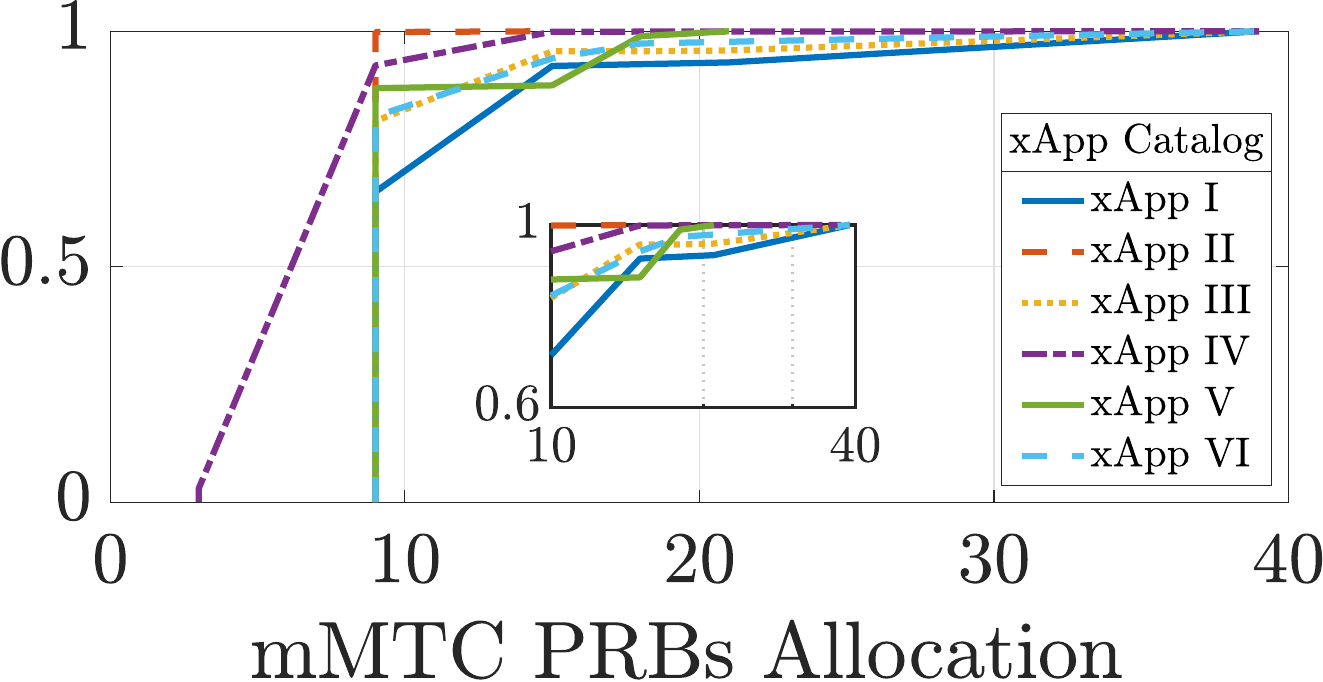}
\label{fig:Figure-mmtc_prbs}}
\hfil
\subfigure[\gls{urllc} \glspl{prb}]{\includegraphics[height=2.85cm]{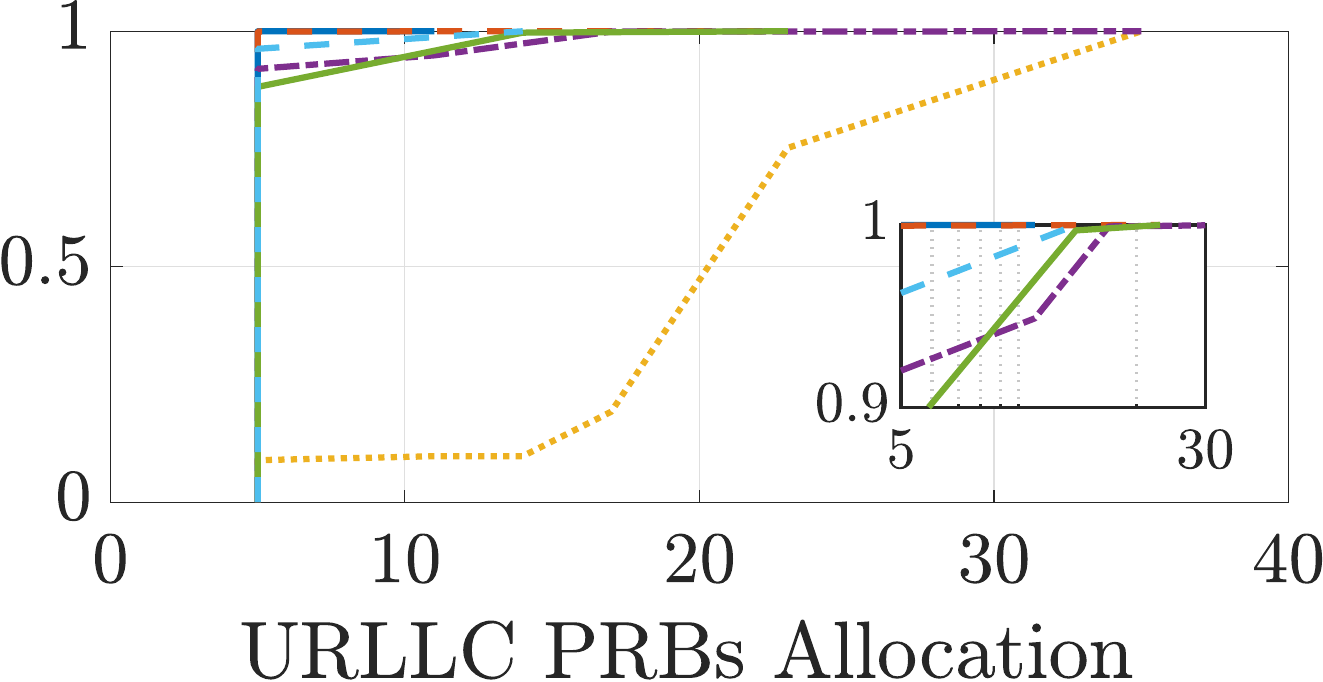}
\label{fig:Figure-urllc_prbs}}
\setlength\abovecaptionskip{-.02cm}
\caption{\textcolor{red}{Selection of \gls{prb} Actions for the three slices from the xApp Catalog of Table~\ref{table:xapp-catalog-comp}.}}
\label{Figure-cdfs_prbs}
\vspace{-0.55cm}
\end{figure*}

\begin{figure*}[t!]
\centering
\subfigure[\gls{embb} \gls{prb} Ratio]{\includegraphics[height=2.85cm]{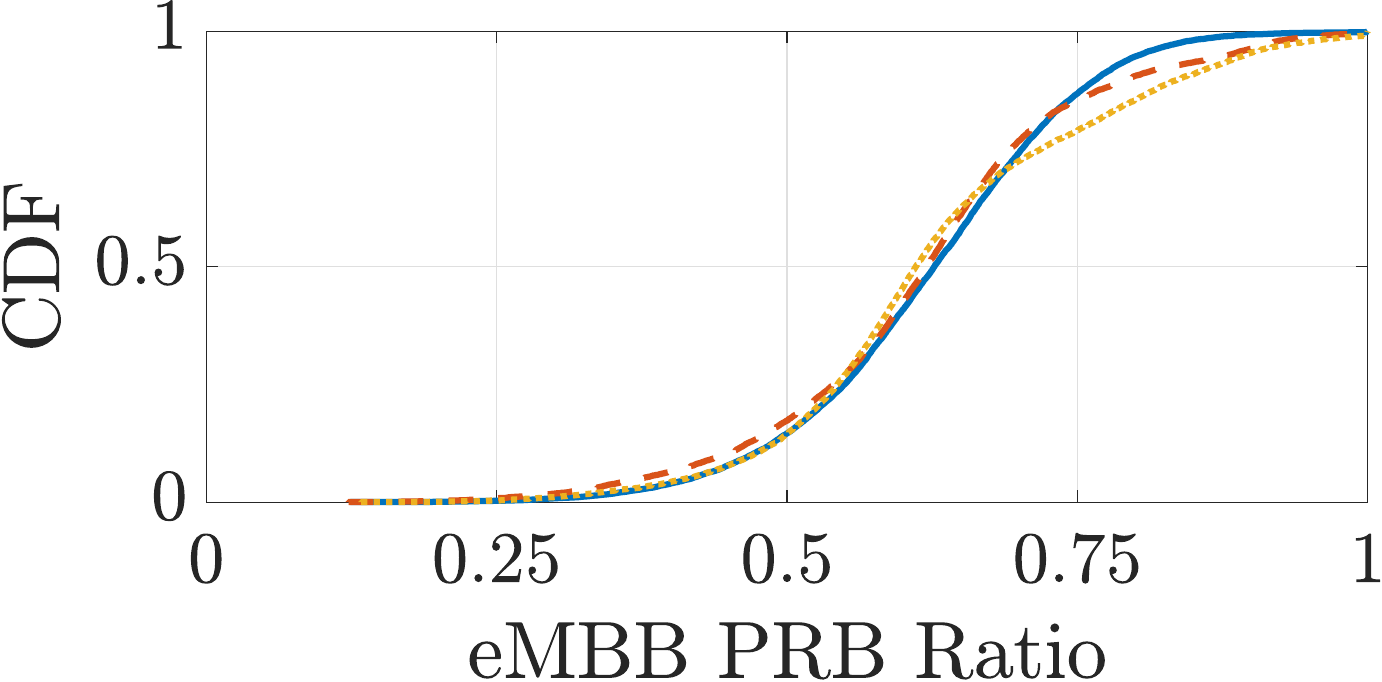}
\label{fig:Figure-prbratio-1-embb_moreues}}
\hfil
\subfigure[\gls{mmtc} \gls{prb} Ratio]{\includegraphics[height=2.85cm]{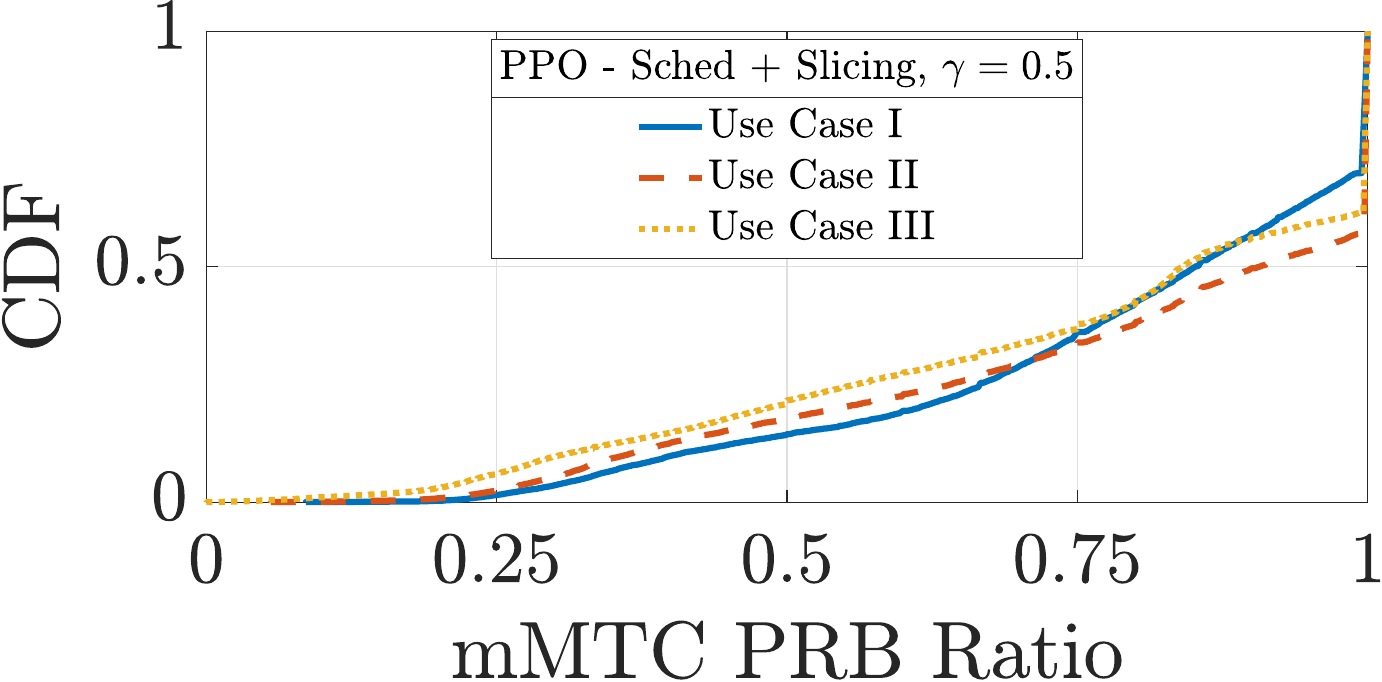}
\label{fig:Figure-prbratio-2-mmtc_moreues}}
\hfil
\subfigure[\gls{urllc} \gls{prb} Ratio]{\includegraphics[height=2.85cm]{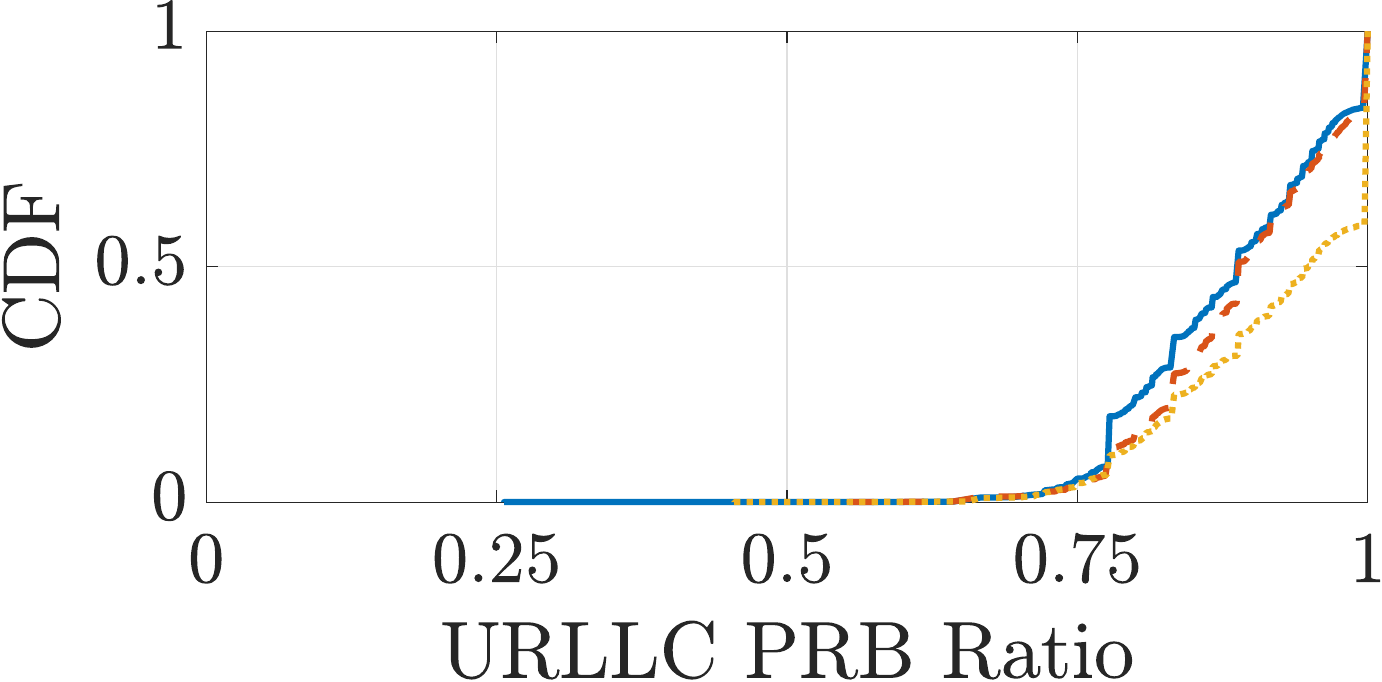}
\label{fig:Figure-prbratio-3-urllc_moreues}}
\setlength\abovecaptionskip{-.02cm}
\caption{\textcolor{red}{Resource Utilization and \gls{ue} Satisfaction for a network deployment described by three different Use Cases.}}
\label{Figure-prbratio-reward-cdfs_moreues}
\vspace{-0.65cm}
\end{figure*}

\begin{figure*}[t!]
\centering
\subfigure[\gls{embb} Throughput]{\includegraphics[height=2.85cm]{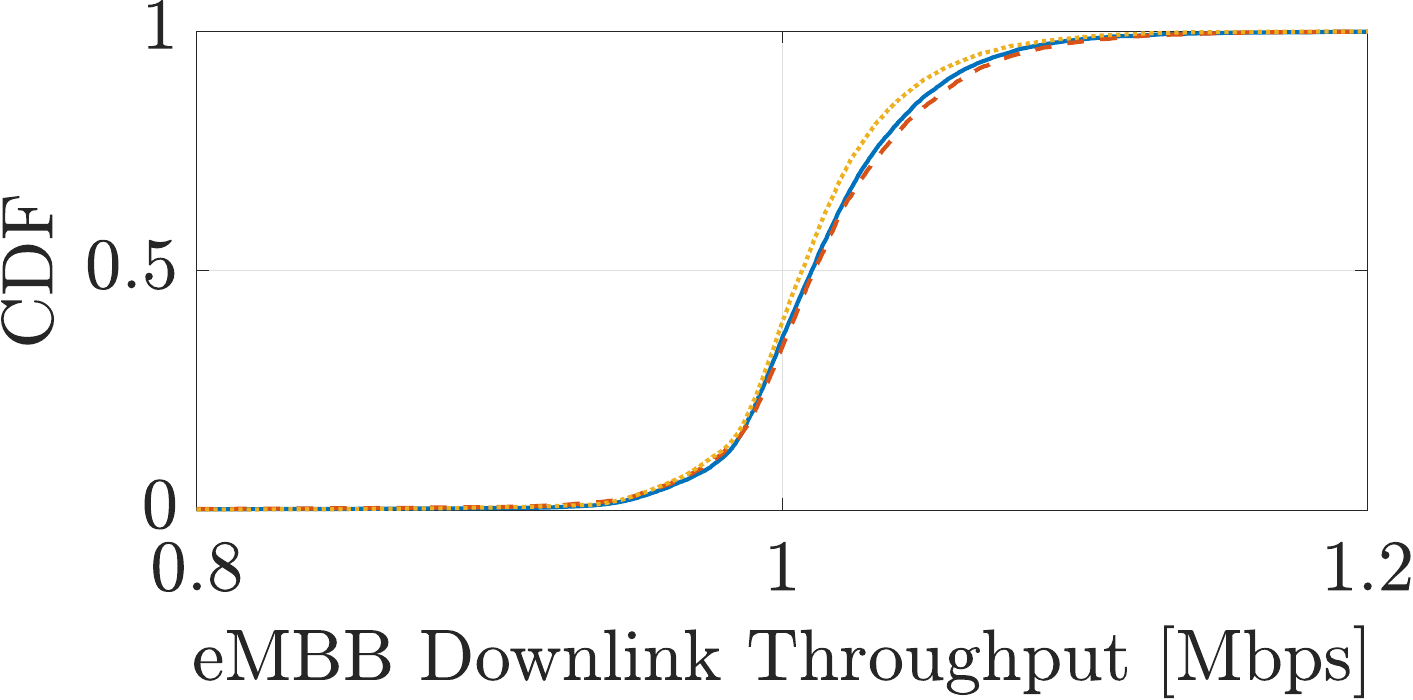}
\label{fig:Figure-embb_moreues}}
\hfil
\subfigure[\gls{mmtc} Packets]{\includegraphics[height=2.85cm]{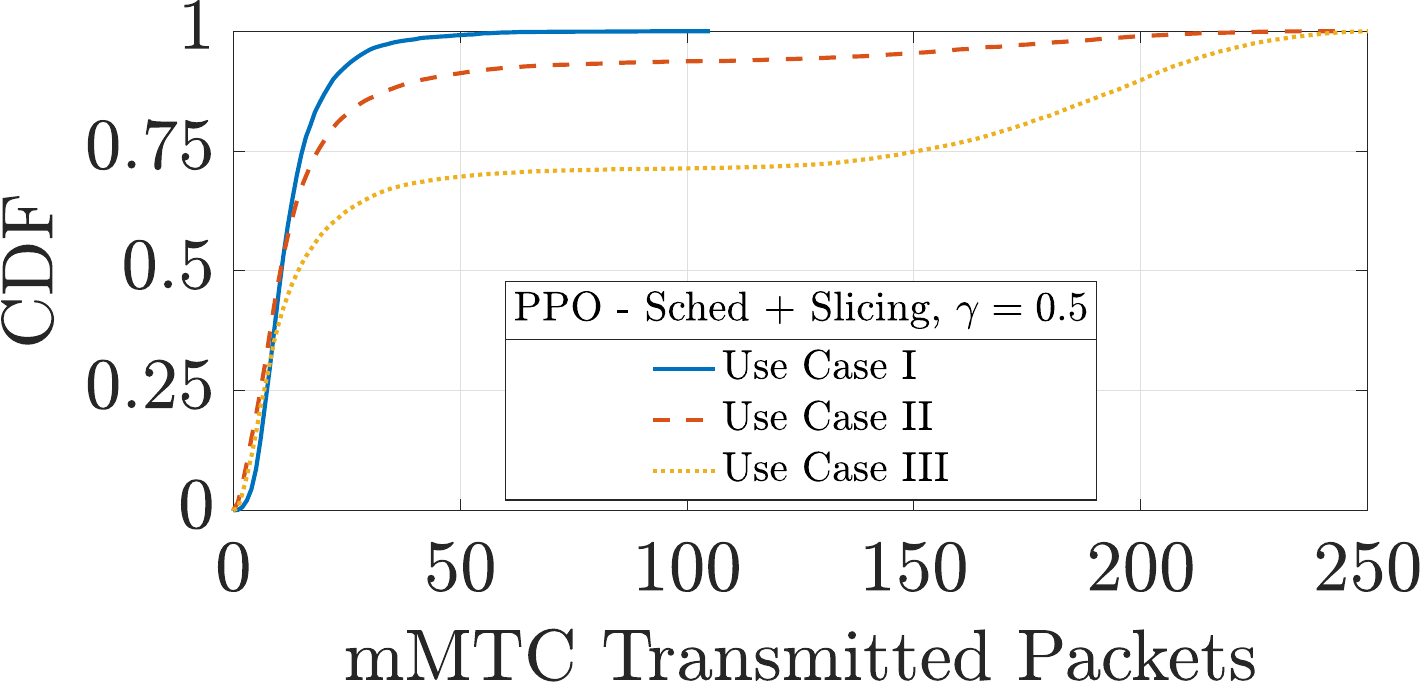}
\label{fig:Figure-mmtc_moreues}}
\hfil
\subfigure[\gls{urllc} Buffer Occupancy]{\includegraphics[height=2.85cm]{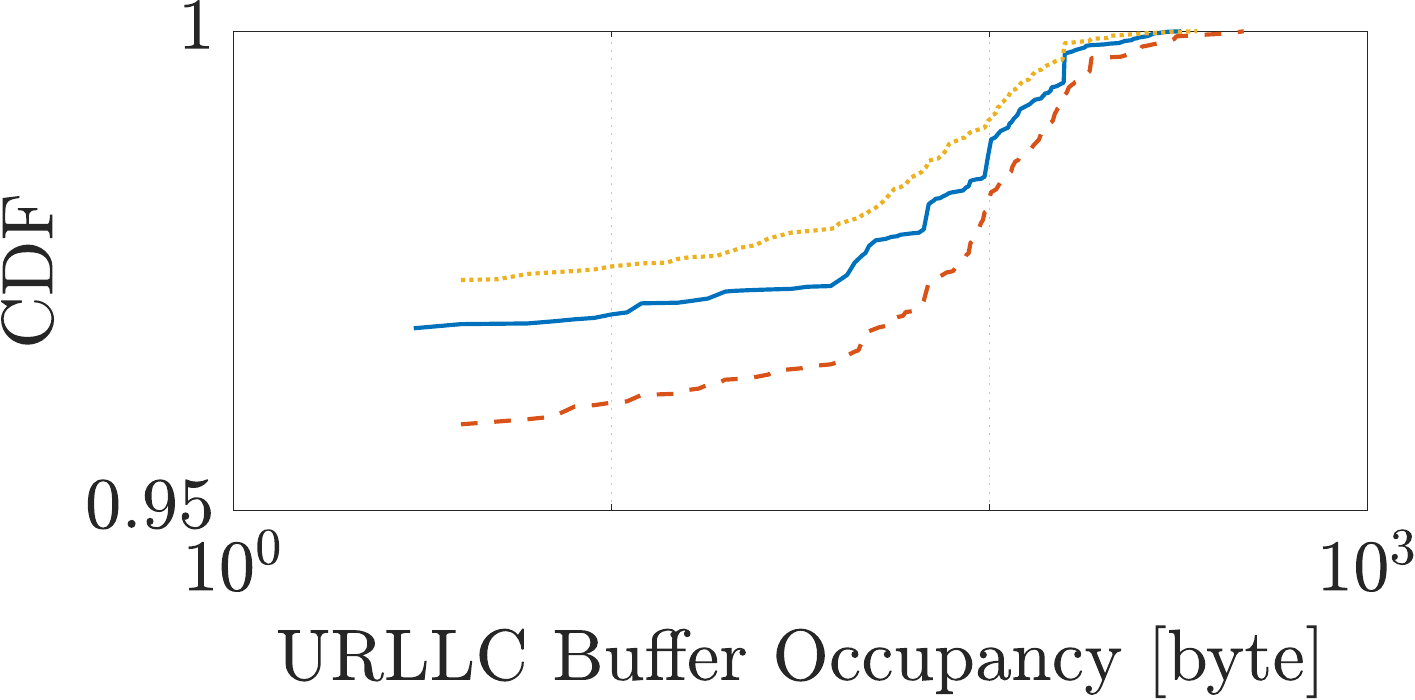}
\label{fig:Figure-urllc_moreues}}
\setlength\abovecaptionskip{-.02cm}
\caption{\textcolor{red}{Impact of the network deployment described by three different Use Cases on \gls{dl} \gls{embb} Throughput, \gls{dl} \gls{mmtc} \gls{tx} Packets, and \gls{dl} \gls{urllc} Buffer Occupancy}.}
\label{Figure-cdfs_moreues}
\vspace{-0.6cm}
\end{figure*}

\begin{figure}[t!]
\centering
\subfigure[\gls{embb} DL Throughput]{\includegraphics[width=1.6in]{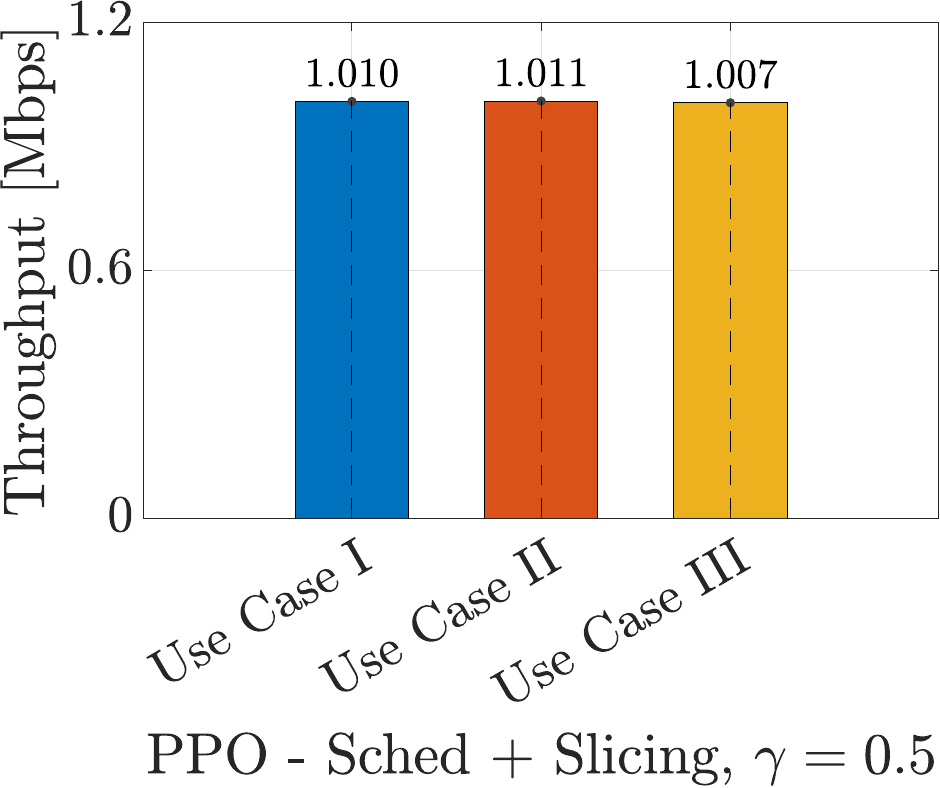}
\label{fig:barplot_iot_embb}}
\hfil
\subfigure[\gls{mmtc}  Packets]{\includegraphics[width=1.6in]{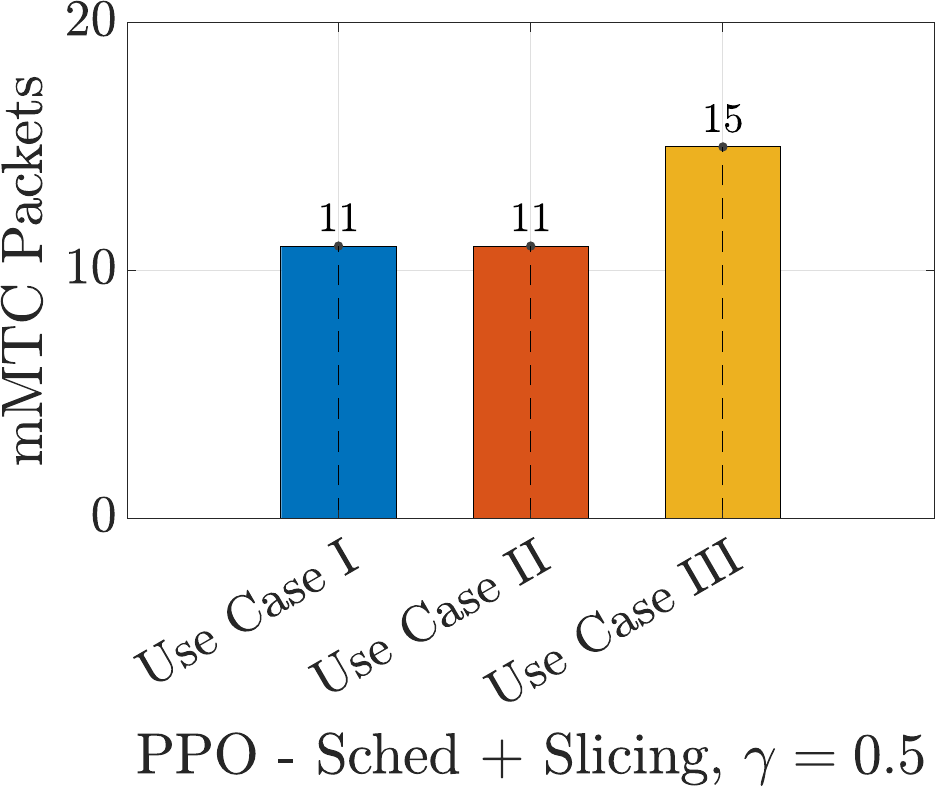}
\label{fig:barplot_iot_mmtc}}
\setlength\abovecaptionskip{-.02cm}
\caption{\textcolor{red}{Impact of the network deployment described by three Use Cases on the median values of \gls{dl} \gls{embb} Throughput and \gls{dl} \gls{mmtc} \gls{tx} Packets}.}
\label{barplot_iot}
\vspace{-0.75cm}
\end{figure}

\textcolor{red}{Figs.~\ref{Figure-prbratio-reward-cdfs_moreues},~\ref{Figure-cdfs_moreues} and~\ref{barplot_iot} demonstrate that even when the number of \glspl{ue} increase, xApps trained using \pandora still deliver good performance, with all \emph{Use Cases} resulting in good resource utilization (Fig.~\ref{Figure-prbratio-reward-cdfs_moreues}) for the three slices. Similarly, the improved \gls{ue} satisfaction is reflected on the \gls{embb} \gls{dl} throughput (Fig.~\ref{fig:Figure-embb_moreues}), with all \emph{Use Cases} enjoying a median throughput value of $\sim1$~Mbps (Fig.~\ref{fig:barplot_iot_embb}). For the \gls{mmtc} slice, and specifically in Fig.~\ref{fig:Figure-mmtc_moreues}, we observe that as the number of \glspl{ue} increase, the number of generated \gls{tx} packets also increases, with \emph{Use Case III} reporting the highest median value of $15$ \gls{dl} \gls{tx} packets (Fig.~\ref{fig:barplot_iot_mmtc}). In line with the results presented in the previous evaluation sections, the performance on the \gls{urllc} remains high (i.e., empty buffers), with \emph{Use Case III} performing slightly better as seen in Figs.~\ref{fig:Figure-prbratio-3-urllc_moreues} and~\ref{fig:Figure-urllc_moreues}.}
\vspace{-.15cm}

\vspace{-0.25cm}

\section{Conclusions} \label{conclusion}

\newcommand{\MyHuge}{\fontsize{42}{48}\selectfont}
\begin{table}[htb]
\centering
\MyHuge
\textcolor{red}{
\setlength\abovecaptionskip{-.1cm}
\caption{xApp Catalog for \texttt{Sched \& Slicing 0.5} with the \gls{ppo} \gls{drl} Architecture.}
\begin{adjustbox}{width=1.04\linewidth}
\begin{tabular}{@{}c@{\hspace{1.5pt}}|cccc|l@{}}
\toprule
\thead{\MyHuge \textbf{xApp} \\ \MyHuge \textbf{ID}} & & \thead{\MyHuge \textbf{eMBB}} & \thead{\MyHuge \textbf{mMTC}} & \thead{\MyHuge \textbf{URLLC}} &  \thead{\MyHuge \textbf{Testing} \\ \MyHuge \textbf{Conditions}}   \\
\midrule
\textbf{I} & \shortstack{\emph{\textbf{Weight Config.}} \\ \emph{\textbf{Slice Reward}} \\ \emph{\textbf{Traffic}}} & \shortstack{\(\text{\MyHuge{72.0440333}}\) \\ Max. \gls{dl} Throughput \\ \(\text{\MyHuge{1}}\)~Mbps}  &  \shortstack{\(\text{\MyHuge{1.5}}\) \\ Min. \gls{dl} \gls{tx} Packets \\\ \(\text{\MyHuge{30}}\)~kbps}  & \shortstack{\(\text{\MyHuge{0.00005}}\) \\ Min. \gls{dl} Buffer Occupancy \\ \(\text{\MyHuge{10}}\)~kbps}  & \shortstack{\emph{mobility}, \emph{\gls{ran} Control} \\ \emph{Timer Set~\(\text{\MyHuge{1}}\)}, \emph{Location 2}} \\
\midrule
\textbf{II} & \shortstack{\emph{\textbf{Weight Config.}} \\ \emph{\textbf{Slice Reward}} \\ \emph{\textbf{Traffic}}} & \shortstack{\(\text{\MyHuge{72.0440333}}\) \\ Max.  \gls{dl} Throughput \\ \(\text{\MyHuge{4}}\)~Mbps}  & \shortstack{\(\text{\MyHuge{0.229357798}}\) \\ Max. \gls{dl} \gls{tx} Packets \\ \(\text{\MyHuge{44.6}}\)~kbps}  & \shortstack{\(\text{\MyHuge{0.00005}}\) \\ Min.  \gls{dl} Buffer Occupancy \\ \(\text{\MyHuge{89.3}}\)~kbps} & \shortstack{\emph{\gls{ran} Control} \\ \emph{Timer Set~\(\text{\MyHuge{2}}\)}, \emph{Location 2}}  \\
\midrule
\textbf{III} & \shortstack{\emph{\textbf{Weight Config.}} \\ \emph{\textbf{Slice Reward}} \\ \emph{\textbf{Traffic}}} & \shortstack{\(\text{\MyHuge{1}}\) \\ Max.  \gls{prb} Ratio \\ \(\text{\MyHuge{4}}\)~Mbps} & \shortstack{\(\text{\MyHuge{1}}\) \\ Max. \gls{prb} Ratio  \\ \(\text{\MyHuge{44.6}}\)~kbps}  &  \shortstack{\(\text{\MyHuge{1}}\) \\ Max. \gls{prb} Ratio  \\ \(\text{\MyHuge{89.3}}\)~kbps} & \shortstack{\emph{\gls{ran} Control} \\ \emph{Timer Set~\(\text{\MyHuge{1}}\)}, \emph{Location 1}} \\
\midrule
\textbf{IV} & \shortstack{\emph{\textbf{Weight Config.}} \\ \emph{\textbf{Slice Reward}} \\ \emph{\textbf{Traffic}}} & \shortstack{\(\text{\MyHuge{72.0440333}}\) \\ Max.  \gls{dl} Throughput \\ \(\text{\MyHuge{4}}\)~Mbps} & \shortstack{\(\text{\MyHuge{0.5}}\) \\ Max. \gls{prb} Ratio  \\ \(\text{\MyHuge{44.6}}\)~kbps}  &  \shortstack{\(\text{\MyHuge{0.00005}}\) \\ Min. \gls{dl} Buffer Occupancy \\ \(\text{\MyHuge{89.3}}\)~kbps} & \shortstack{ \emph{\gls{ran} Control} \\ \emph{Timer Set~\(\text{\MyHuge{1}}\)}, \emph{Location 1}} \\
\midrule
\textbf{V} & \shortstack{\emph{\textbf{Weight Config.}} \\ \emph{\textbf{Slice Reward}} \\ \emph{\textbf{Traffic}}} & \shortstack{\(\text{\MyHuge{72.0440333}}\) \\ Max.  \gls{dl} Throughput \\ \(\text{\MyHuge{4}}\)~Mbps} & \shortstack{\(\text{\MyHuge{0.229357798}}\) \\ Max. \gls{dl} \gls{tx} Packets \\ \(\text{\MyHuge{44.6}}\)~kbps}  &  \shortstack{\(\text{\MyHuge{0.00005}}\) \\ Min. \gls{dl} Buffer Occupancy \\ \(\text{\MyHuge{89.3}}\)~kbps} & \shortstack{\emph{\gls{ran} Control} \\ \emph{Timer Set~\(\text{\MyHuge{1}}\)}, \emph{Location 1}}  \\
\midrule
\textbf{VI} & \shortstack{\emph{\textbf{Weight Config.}} \\ \emph{\textbf{Slice Reward}} \\ \emph{\textbf{Traffic}}} & \shortstack{\(\text{\MyHuge{72.0440333}}\) \\ Max.  \gls{dl} Throughput \\ \(\text{\MyHuge{4}}\)~Mbps} & \shortstack{\(\text{\MyHuge{1.5}}\) \\ Max. \gls{dl} \gls{tx} Packets \\ \(\text{\MyHuge{44.6}}\)~kbps}  &  \shortstack{\(\text{\MyHuge{0.00005}}\) \\ Min. \gls{dl} Buffer Occupancy \\  \(\text{\MyHuge{89.3}}\)~kbps} & \shortstack{\ \emph{\gls{ran} Control} \\ \emph{Timer Set~\(\text{\MyHuge{1}}\)}, \emph{Location 1}}  \\
\bottomrule
\end{tabular}
\end{adjustbox}
\label{table:xapp-catalog-comp}}
\vspace{-.85cm}
\end{table}

\textcolor{blue}{In this paper, we presented \pandora, a \textcolor{red}{comprehensive} evaluation with insights on how to design \gls{drl} agents for \gls{ran} control. \pandora leverages a framework to streamline and automate \gls{drl} agent training and xApp on-boarding for \textcolor{red}{extensive} evaluation and testing of \gls{drl} agents with Open RAN in Colosseum.} We investigated the impact of \gls{drl} design 
on the performance of an Open \gls{ran} system controlled by xApps embedding \gls{drl} agents that make decisions in near-real-time to compute efficient slicing and scheduling control policies.
We benchmarked \textcolor{red}{$23$}~xApps trained using \gls{drl} agents with different actions spaces, \textcolor{blue}{architectures}, reward design and decision-making timescales. \textcolor{blue}{Additionally, we tested the \gls{drl} agents under various traffic and mobility conditions, considering different hierarchical setups and time granularities in locations both observed and unseen during the training phase.}
Our experimental results show that network slices with similar objectives (e.g., maximizing throughput and number of transmitted packets) might result in a competitive behavior that can be mitigated using proper weight and reward configurations, and possibly different architectures. Additionally, we have explored how the fine-tuning of \textcolor{red}{\gls{ran} control timers can boost the performance of various xApps tasked with altering different action spaces.} Based on the reported findings, the \pandora evaluation shows that 
\gls{drl} agents adapt well to network conditions encountered during the \gls{ai}/\gls{ml} training phase as well as unforeseen conditions, \textcolor{red}{while ensuring high system resource utilization}.


\bibliographystyle{IEEEtran}
\bibliography{IEEEabrv,ref}
\vspace{-1.85cm}
\begin{IEEEbiography}[{\includegraphics[width=1in,height=1.1in,clip,keepaspectratio]{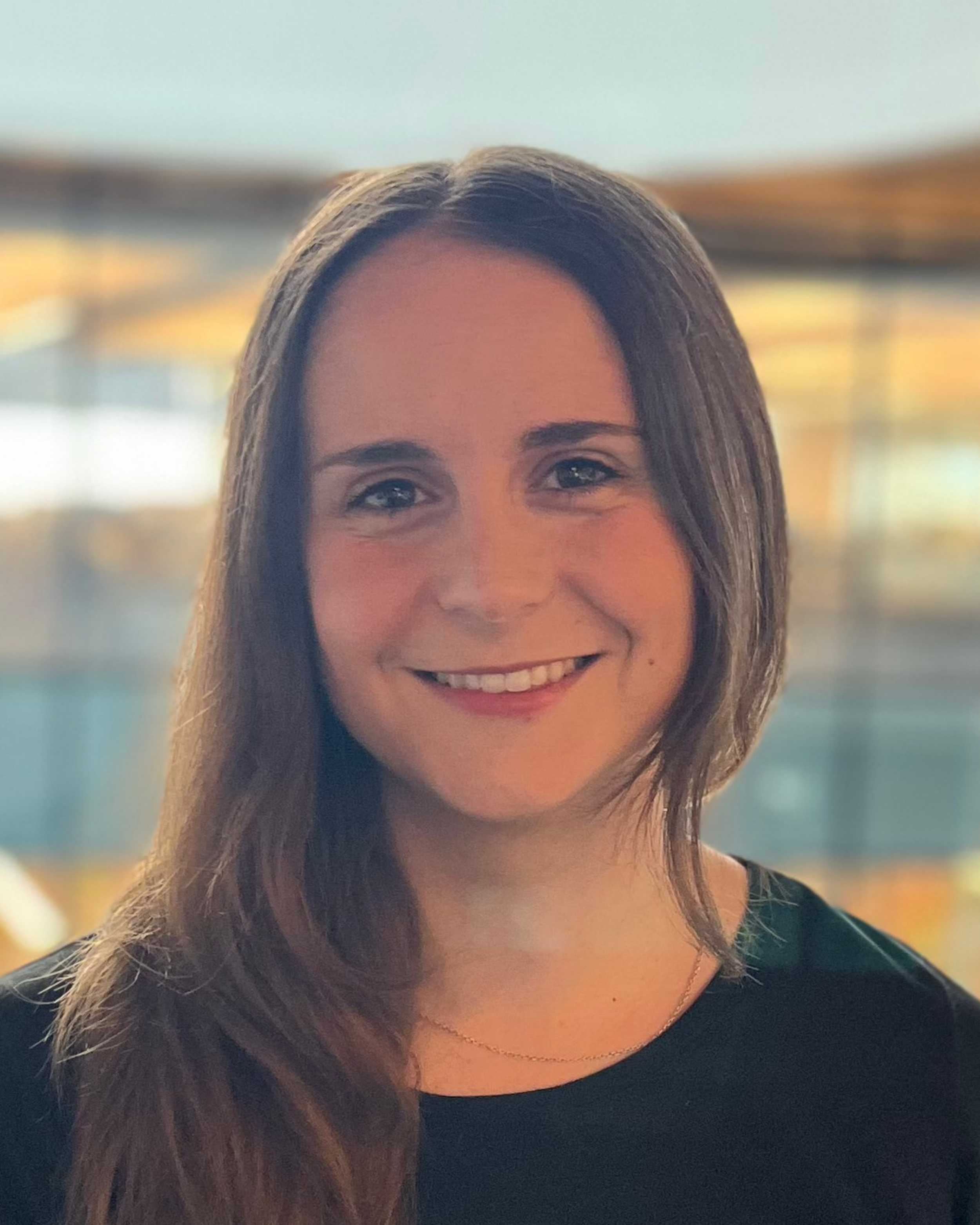}}]{Maria Tsampazi}
holds a MEng ('21) in ECE from National Technical University of Athens, Greece. She is a Ph.D. Candidate in Electrical Engineering at the Institute for the Wireless Internet of Things. Her research lies on NextG networks and intelligent resource allocation in Open RAN. She has received academic awards sponsored by the U.S. National Science Foundation, the IEEE Communications Society, and Northeastern University, and is a 2024 recipient of the National Spectrum Consortium Women in Spectrum Scholarship. She has previously collaborated with both government and industry, including organizations such as the U.S. Department of Transportation and Dell Technologies.
\end{IEEEbiography}

\vspace{-1.75cm}

\begin{IEEEbiography}[{\includegraphics[width=1in,height=1.1in,clip,keepaspectratio]{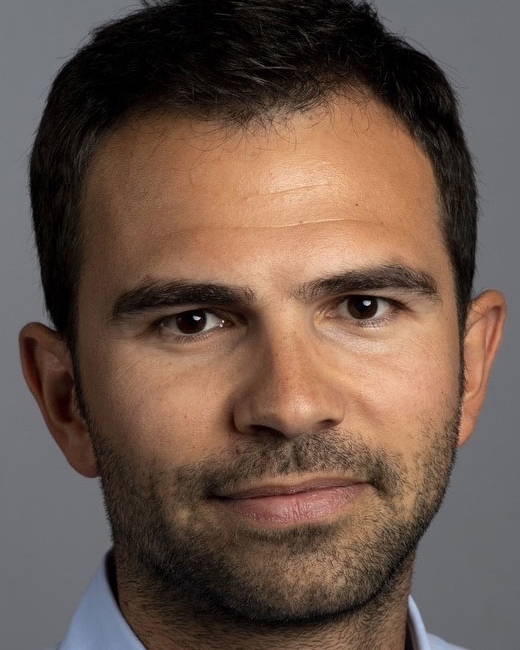}}]{Salvatore D'Oro} is a Research Assistant Professor at Northeastern University. He received
his Ph.D. from the University of Catania
and is an area editor of Elsevier
Computer Communications journal. He serves
on the TPC of
IEEE INFOCOM, IEEE CCNC \& ICC and
IFIP Networking. He is one of the contributors to
OpenRAN Gym for AI/ML applications in the Open RAN.
His research interests include optimization, AI \& network slicing for NextG Open RANs.
\end{IEEEbiography}

\vspace{-1.75cm}

\begin{IEEEbiography}[{\includegraphics[width=1in,height=1.1in,clip,keepaspectratio]{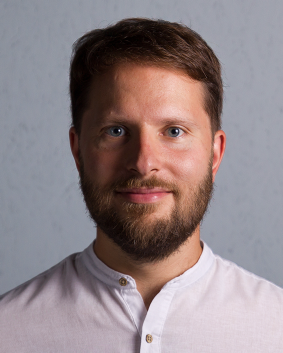}}]{Michele Polese} is a Research Assistant Professor at Northeastern University. He received his Ph.D. from the University of Padova. His research is funded by the US NSF, OUSD, NTIA, and the O-RAN ALLIANCE. He holds several best paper awards and the '22 Mario Gerla Award for Research in Computer Science. He is an Associate Technical Editor for the IEEE Communications Magazine, and a Guest Editor in JSAC's Special Issue on Open RAN.
\end{IEEEbiography}

\vspace{-1.75cm}

\begin{IEEEbiography}[{\includegraphics[width=1in,height=1.1in,clip,keepaspectratio]{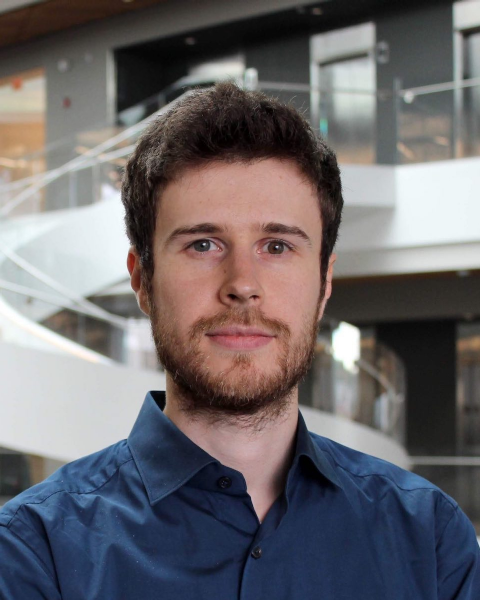}}]{Leonardo Bonati}
is an Associate Research Scientist at the Institute for the Wireless Internet of Things, Northeastern University. He received a Ph.D.\ degree in Computer Engineering from Northeastern University in 2022. His research focuses on softwarized approaches for the Open RAN of next generation of cellular networks, on O-RAN-managed networks, and on network automation and orchestration.
\end{IEEEbiography}

\vspace{-1.75cm}

\begin{IEEEbiography}[{\includegraphics[width=1in,height=1.1in,clip,keepaspectratio]{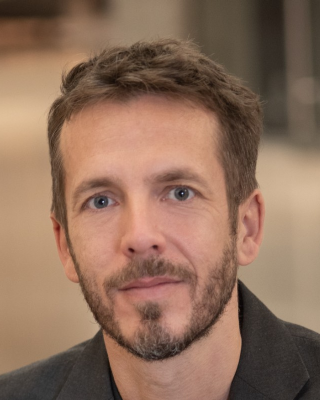}}]{Gwenael Poitau} has 20 years of experience
developing wireless communication systems for
the telecom and defense markets. He has worked for Cavium/Marvell, and has served as Head of Engineering at Ultra Electronics. He co-founded the Resilient Machine Learning Institute in collaboration with École de Technologie Supérieure. He is a Wireless AI Technology Director at Dell Technologies. 
\end{IEEEbiography}

\vspace{-1.75cm}

\begin{IEEEbiography}[{\includegraphics[width=1in,height=1.1in,clip,keepaspectratio]{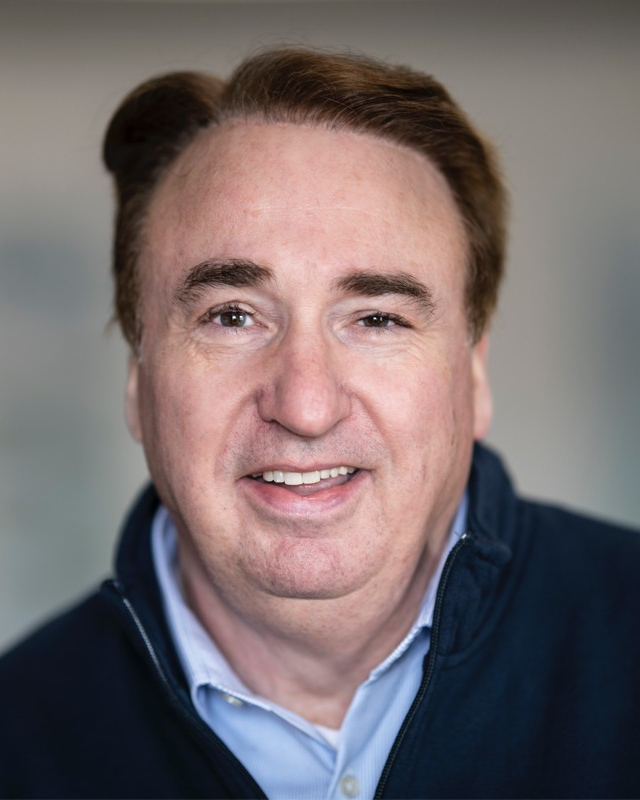}}]{Michael Healy} is a Technical Leader in Dell Technologies. He has 14 issued patents and has contributed to revenue streams such as the Cisco DOCSIS 3.0, SMPTE, and PCI-SIG. 
At IBM he led an IBM POWER-ISA CPU Vector Processor silicon project.  At AMD he designed CPU Processors and GPUs for the computing industry.  At Cisco, he delivered products to the DOCSIS Cable Telecom industry.
\end{IEEEbiography}

\vspace{-1.75cm}

\begin{IEEEbiography}[{\includegraphics[width=1in,height=1.1in,clip,keepaspectratio]{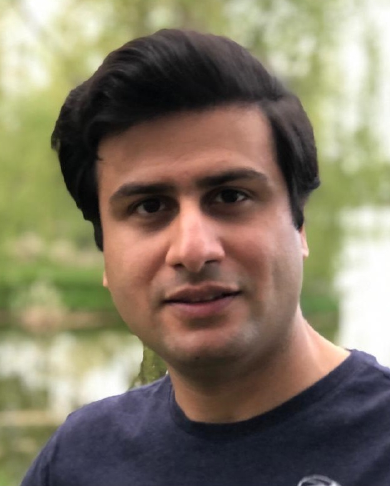}}]{Mohammad Alavirad} is a Principal Technical Staff at Dell Technologies. He has worked for Ranovus, Viavi, and Ciena and has served as a guest editor for IEEE Photonics Technology Letters, Quantum Electronics, and Applied Physics Letters. His research lies on Open RAN, 
and the development of B5G/6G cellular architectures for mmWave Communication systems.
\end{IEEEbiography}

\vspace{-1.75cm}

\begin{IEEEbiography}[{\includegraphics[width=1in,height=1.1in,clip,keepaspectratio]{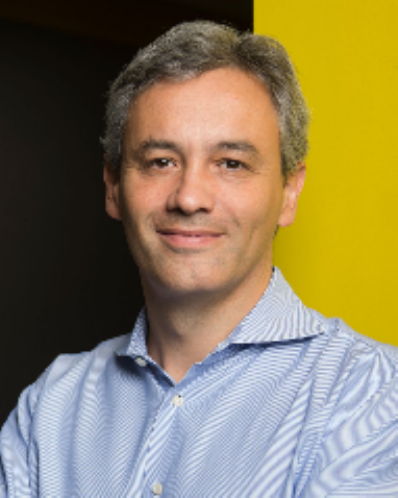}}]{Tommaso Melodia}
is the Director of the Institute for the Wireless Internet 
of Things and Research Director for the Platforms for Advanced Wireless Research. He has served as
Associate Editor of IEEE Transactions on Wireless Communications, Transactions on Mobile Computing and Elsevier Computer Networks, and as TPC Chair
for IEEE INFOCOM, General Chair for IEEE SECON, ACM
Nanocom, and WUWnet.  He is an IEEE Fellow and ACM Senior Member.
\end{IEEEbiography}

\end{document}

%% file: acronyms.tex
\newacronym{3gpp}{3GPP}{3rd Generation Partnership Project}
\newacronym{4g}{4G}{4th generation}
\newacronym{5g}{5G}{5th generation}
\newacronym{6g}{6G}{6th generation}
\newacronym{5gc}{5GC}{5G Core}
\newacronym{adc}{ADC}{Analog to Digital Converter}
\newacronym{aerpaw}{AERPAW}{Aerial Experimentation and Research Platform for Advanced Wireless}
\newacronym{ai}{AI}{Artificial Intelligence}
\newacronym{aimd}{AIMD}{Additive Increase Multiplicative Decrease}
\newacronym{am}{AM}{Acknowledged Mode}
\newacronym{amc}{AMC}{Adaptive Modulation and Coding}
\newacronym{amf}{AMF}{Access and Mobility Management Function}
\newacronym{aops}{AOPS}{Adaptive Order Prediction Scheduling}
\newacronym{api}{API}{Application Programming Interface}
\newacronym{xapp}{xApp}{Intelligent Application}
\newacronym{apn}{APN}{Access Point Name}
\newacronym{ap}{AP}{Application Protocol}
\newacronym{aqm}{AQM}{Active Queue Management}
\newacronym{ausf}{AUSF}{Authentication Server Function}
\newacronym{avc}{AVC}{Advanced Video Coding}
\newacronym{awgn}{AGWN}{Additive White Gaussian Noise}
\newacronym{balia}{BALIA}{Balanced Link Adaptation Algorithm}
\newacronym{bbu}{BBU}{Base Band Unit}
\newacronym{bdp}{BDP}{Bandwidth-Delay Product}
\newacronym{ber}{BER}{Bit Error Rate}
\newacronym{bf}{BF}{Beamforming}
\newacronym{bler}{BLER}{Block Error Rate}
\newacronym{brr}{BRR}{Bayesian Ridge Regressor}
\newacronym{bs}{BS}{Base Station}
\newacronym{bsr}{BSR}{Buffer Status Report}
\newacronym{bss}{BSS}{Business Support System}
\newacronym{ca}{CA}{Carrier Aggregation}
\newacronym{caas}{CaaS}{Connectivity-as-a-Service}
\newacronym{cb}{CB}{Code Block}
\newacronym{cc}{CC}{Congestion Control}
\newacronym{ccid}{CCID}{Congestion Control ID}
\newacronym{cco}{CC}{Carrier Component}
\newacronym{cdd}{CDD}{Cyclic Delay Diversity}
\newacronym{cdf}{CDF}{Cumulative Distribution Function}
\newacronym{cdn}{CDN}{Content Distribution Network}
\newacronym{cn}{CN}{Core Network}
\newacronym{codel}{CoDel}{Controlled Delay Management}
\newacronym{comac}{COMAC}{Converged Multi-Access and Core}
\newacronym{cord}{CORD}{Central Office Re-architected as a Datacenter}
\newacronym{cornet}{CORNET}{COgnitive Radio NETwork}
\newacronym{cosmos}{COSMOS}{Cloud Enhanced Open Software Defined Mobile Wireless Testbed for City-Scale Deployment}
\newacronym{cots}{COTS}{Commercial Off-the-Shelf}
\newacronym{cp}{CP}{Control Plane}
\newacronym{cpu}{CPU}{Central Processing Unit}
\newacronym{cqi}{CQI}{Channel Quality Information}
\newacronym{cr}{CR}{Cognitive Radio}
\newacronym{cran}{CRAN}{Cloud \gls{ran}}
\newacronym{crs}{CRS}{Cell Reference Signal}
\newacronym{csi}{CSI}{Channel State Information}
\newacronym{csirs}{CSI-RS}{Channel State Information - Reference Signal}
\newacronym{cu}{CU}{Central Unit}
\newacronym{d2tcp}{D$^2$TCP}{Deadline-aware Data center TCP}
\newacronym{d3}{D$^3$}{Deadline-Driven Delivery}
\newacronym{dac}{DAC}{Digital to Analog Converter}
\newacronym{dag}{DAG}{Directed Acyclic Graph}
\newacronym{das}{DAS}{Distributed Antenna System}
\newacronym{dash}{DASH}{Dynamic Adaptive Streaming over HTTP}
\newacronym{dc}{DC}{Dual Connectivity}
\newacronym{dccp}{DCCP}{Datagram Congestion Control Protocol}
\newacronym{dce}{DCE}{Direct Code Execution}
\newacronym{dci}{DCI}{Downlink Control Information}
\newacronym{dctcp}{DCTCP}{Data Center TCP}
\newacronym{dl}{DL}{Downlink}
\newacronym{dmr}{DMR}{Deadline Miss Ratio}
\newacronym{dmrs}{DMRS}{DeModulation Reference Signal}
\newacronym{drlcc}{DRL-CC}{Deep Reinforcement Learning Congestion Control}
\newacronym{drs}{DRS}{Discovery Reference Signal}
\newacronym{du}{DU}{Distributed Unit}
\newacronym{e2e}{E2E}{end-to-end}
\newacronym{ecaas}{ECaaS}{Edge-Cloud-as-a-Service}
\newacronym{ecn}{ECN}{Explicit Congestion Notification}
\newacronym{edf}{EDF}{Earliest Deadline First}
\newacronym{embb}{eMBB}{Enhanced Mobile Broadband}
\newacronym{empower}{EMPOWER}{EMpowering transatlantic PlatfOrms for advanced WirEless Research}
\newacronym{enb}{eNB}{evolved Node Base}
\newacronym{endc}{EN-DC}{E-UTRAN-\gls{nr} \gls{dc}}
\newacronym{epc}{EPC}{Evolved Packet Core}
\newacronym{eps}{EPS}{Evolved Packet System}
\newacronym{es}{ES}{Edge Server}
\newacronym{etsi}{ETSI}{European Telecommunications Standards Institute}
\newacronym[firstplural=Estimated Times of Arrival (ETAs)]{eta}{ETA}{Estimated Time of Arrival}
\newacronym{eutran}{E-UTRAN}{Evolved Universal Terrestrial Access Network}
\newacronym{faas}{FaaS}{Function-as-a-Service}
\newacronym{fapi}{FAPI}{Functional Application Platform Interface}
\newacronym{fdd}{FDD}{Frequency Division Duplexing}
\newacronym{fdm}{FDM}{Frequency Division Multiplexing}
\newacronym{fdma}{FDMA}{Frequency Division Multiple Access}
\newacronym{fed4fire}{FED4FIRE+}{Federation 4 Future Internet Research and Experimentation Plus}
\newacronym{fir}{FIR}{Finite Impulse Response}
\newacronym{fit}{FIT}{Future \acrlong{iot}}
\newacronym{fpga}{FPGA}{Field Programmable Gate Array}
\newacronym{fr2}{FR2}{Frequency Range 2}
\newacronym{fs}{FS}{Fast Switching}
\newacronym{fscc}{FSCC}{Flow Sharing Congestion Control}
\newacronym{ftp}{FTP}{File Transfer Protocol}
\newacronym{fw}{FW}{Flow Window}
\newacronym{ge}{GE}{Gaussian Elimination}
\newacronym{gnb}{gNB}{Next Generation Node Base}
\newacronym{nextg}{NextG}{Next Generation}
\newacronym{gop}{GOP}{Group of Pictures}
\newacronym{gpr}{GPR}{Gaussian Process Regressor}
\newacronym{gpu}{GPU}{Graphics Processing Unit}
\newacronym{gtp}{GTP}{GPRS Tunneling Protocol}
\newacronym{gtpc}{GTP-C}{GPRS Tunnelling Protocol Control Plane}
\newacronym{gtpu}{GTP-U}{GPRS Tunnelling Protocol User Plane}
\newacronym{gtpv2c}{GTPv2-C}{\gls{gtp} v2 - Control}
\newacronym{gw}{GW}{Gateway}
\newacronym{harq}{HARQ}{Hybrid Automatic Repeat reQuest}
\newacronym{hetnet}{HetNet}{Heterogeneous Network}
\newacronym{hh}{HH}{Hard Handover}
\newacronym{hol}{HOL}{Head-of-Line}
\newacronym{hqf}{HQF}{Highest-quality-first}
\newacronym{hss}{HSS}{Home Subscription Server}
\newacronym{http}{HTTP}{HyperText Transfer Protocol}
\newacronym{ia}{IA}{Initial Access}
\newacronym{iab}{IAB}{Integrated Access and Backhaul}
\newacronym{ic}{IC}{Incident Command}
\newacronym{ietf}{IETF}{Internet Engineering Task Force}
\newacronym{imsi}{IMSI}{International Mobile Subscriber Identity}
\newacronym{imt}{IMT}{International Mobile Telecommunication}
\newacronym{iot}{IoT}{Internet of Things}
\newacronym{ip}{IP}{Internet Protocol}
\newacronym{itu}{ITU}{International Telecommunication Union}
\newacronym{kpi}{KPI}{Key Performance Indicator}
\newacronym{kpm}{KPM}{Key Performance Measurement}
\newacronym{kvm}{KVM}{Kernel-based Virtual Machine}
\newacronym{los}{LOS}{Line-of-Sight}
\newacronym{lsm}{LSM}{Link-to-System Mapping}
\newacronym{lstm}{LSTM}{Long Short Term Memory}
\newacronym{lte}{LTE}{Long Term Evolution}
\newacronym{lxc}{LXC}{Linux Container}
\newacronym{m2m}{M2M}{Machine to Machine}
\newacronym{mac}{MAC}{Medium Access Control}
\newacronym{manet}{MANET}{Mobile Ad Hoc Network}
\newacronym{mano}{MANO}{Management and Orchestration}
\newacronym{mc}{MC}{Multi-Connectivity}
\newacronym{mcc}{MCC}{Mobile Cloud Computing}
\newacronym{mchem}{MCHEM}{Massive Channel Emulator}
\newacronym{mcs}{MCS}{Modulation and Coding Scheme}
\newacronym{mec}{MEC}{Multi-access Edge Computing}
\newacronym{mec2}{MEC}{Mobile Edge Cloud}
\newacronym{mfc}{MFC}{Mobile Fog Computing}
\newacronym{mgen}{MGEN}{Multi-Generator}
\newacronym{mi}{MI}{Mutual Information}
\newacronym{mib}{MIB}{Master Information Block}
\newacronym{miesm}{MIESM}{Mutual Information Based Effective SINR}
\newacronym{mimo}{MIMO}{Multiple Input, Multiple Output}
\newacronym{ml}{ML}{Machine Learning}
\newacronym{mlr}{MLR}{Maximum-local-rate}
\newacronym[plural=\gls{mme}s,firstplural=Mobility Management Entities (MMEs)]{mme}{MME}{Mobility Management Entity}
\newacronym{mmtc}{mMTC}{Massive Machine-Type Communications}
\newacronym{mmwave}{mmWave}{millimeter wave}
\newacronym{mpdccp}{MP-DCCP}{Multipath Datagram Congestion Control Protocol}
\newacronym{mptcp}{MPTCP}{Multipath TCP}
\newacronym{mr}{MR}{Maximum Rate}
\newacronym{mrdc}{MR-DC}{Multi \gls{rat} \gls{dc}}
\newacronym{mse}{MSE}{Mean Square Error}
\newacronym{mss}{MSS}{Maximum Segment Size}
\newacronym{mt}{MT}{Mobile Termination}
\newacronym{mtd}{MTD}{Machine-Type Device}
\newacronym{mtu}{MTU}{Maximum Transmission Unit}
\newacronym{mumimo}{MU-MIMO}{Multi-user \gls{mimo}}
\newacronym{mvno}{MVNO}{Mobile Virtual Network Operator}
\newacronym{nalu}{NALU}{Network Abstraction Layer Unit}
\newacronym{nas}{NAS}{Non-Access Stratum}
\newacronym{nbiot}{NB-IoT}{Narrow Band IoT}
\newacronym{nfv}{NFV}{Network Function Virtualization}
\newacronym{nfvi}{NFVI}{Network Function Virtualization Infrastructure}
\newacronym{nic}{NIC}{Network Interface Card}
\newacronym{nlos}{NLOS}{Non-Line-of-Sight}
\newacronym{now}{NOW}{Non Overlapping Window}
\newacronym{nsm}{NSM}{Network Service Mesh}
\newacronym[type=hidden]{nr}{NR}{New Radio}
\newacronym{nrf}{NRF}{Network Repository Function}
\newacronym{nsa}{NSA}{Non Stand Alone}
\newacronym{nse}{NSE}{Network Slicing Engine}
\newacronym{nssf}{NSSF}{Network Slice Selection Function}
\newacronym{o2i}{O2I}{Outdoor to Indoor}
\newacronym{oai}{OAI}{OpenAirInterface}
\newacronym{oaicn}{OAI-CN}{\gls{oai} \acrlong{cn}}
\newacronym{oairan}{OAI-RAN}{\acrlong{oai} \acrlong{ran}}
\newacronym{oam}{OAM}{Operations, Administration and Maintenance}
\newacronym{ofdm}{OFDM}{Orthogonal Frequency Division Multiplexing}
\newacronym{olia}{OLIA}{Opportunistic Linked Increase Algorithm}
\newacronym{omec}{OMEC}{Open Mobile Evolved Core}
\newacronym{onap}{ONAP}{Open Network Automation Platform}
\newacronym{onf}{ONF}{Open Networking Foundation}
\newacronym{onos}{ONOS}{Open Networking Operating System}
\newacronym{oom}{OOM}{\gls{onap} Operations Manager}
\newacronym{opnfv}{OPNFV}{Open Platform for \gls{nfv}}
\newacronym{orbit}{ORBIT}{Open-Access Research Testbed for Next-Generation Wireless Networks}
\newacronym{os}{OS}{Operating System}
\newacronym{oss}{OSS}{Operations Support System}
\newacronym{pa}{PA}{Position-aware}
\newacronym{pase}{PASE}{Prioritization, Arbitration, and Self-adjusting Endpoints}
\newacronym{pawr}{PAWR}{Platforms for Advanced Wireless Research}
\newacronym{pbch}{PBCH}{Physical Broadcast Channel}
\newacronym{pcef}{PCEF}{Policy and Charging Enforcement Function}
\newacronym{pcfich}{PCFICH}{Physical Control Format Indicator Channel}
\newacronym{pcrf}{PCRF}{Policy and Charging Rules Function}
\newacronym{pdcch}{PDCCH}{Physical Downlink Control Channel}
\newacronym{pdcp}{PDCP}{Packet Data Convergence Protocol}
\newacronym{pdsch}{PDSCH}{Physical Downlink Shared Channel}
\newacronym{pdu}{PDU}{Packet Data Unit}
\newacronym{pf}{PF}{Proportionally Fair}
\newacronym{pgw}{PGW}{Packet Gateway}
\newacronym{phich}{PHICH}{Physical Hybrid ARQ Indicator Channel}
\newacronym{phy}{PHY}{Physical}
\newacronym{pmch}{PMCH}{Physical Multicast Channel}
\newacronym{pmi}{PMI}{Precoding Matrix Indicators}
\newacronym{powder}{POWDER}{Platform for Open Wireless Data-driven Experimental Research}
\newacronym{ppo}{PPO}{Proximal Policy Optimization}
\newacronym{ppp}{PPP}{Poisson Point Process}
\newacronym{prach}{PRACH}{Physical Random Access Channel}
\newacronym{prb}{PRB}{Physical Resource Block}
\newacronym{psnr}{PSNR}{Peak Signal to Noise Ratio}
\newacronym{pss}{PSS}{Primary Synchronization Signal}
\newacronym{pucch}{PUCCH}{Physical Uplink Control Channel}
\newacronym{pusch}{PUSCH}{Physical Uplink Shared Channel}
\newacronym{qam}{QAM}{Quadrature Amplitude Modulation}
\newacronym{qci}{QCI}{\gls{qos} Class Identifier}
\newacronym{qoe}{QoE}{Quality of Experience}
\newacronym{qos}{QoS}{Quality of Service}
\newacronym{quic}{QUIC}{Quick UDP Internet Connections}
\newacronym{rach}{RACH}{Random Access Channel}
\newacronym{ran}{RAN}{Radio Access Network}
\newacronym[firstplural=end to endcess Technologies (RATs)]{rat}{RAT}{end to endcess Technology}
\newacronym{rcn}{RCN}{Research Coordination Network}
\newacronym{rec}{REC}{Radio Edge Cloud}
\newacronym{ra}{RA}{Resource Allocation}
\newacronym{red}{RED}{Random Early Detection}
\newacronym{renew}{RENEW}{Reconfigurable Eco-system for Next-generation End-to-end Wireless}
\newacronym{rf}{RF}{Radio Frequency}
\newacronym{rfc}{RFC}{Request for Comments}
\newacronym{rfr}{RFR}{Random Forest Regressor}
\newacronym{ric}{RIC}{RAN Intelligent Controller}
\newacronym{rlc}{RLC}{Radio Link Control}
\newacronym{rlf}{RLF}{Radio Link Failure}
\newacronym{rlnc}{RLNC}{Random Linear Network Coding}
\newacronym{rmr}{RMR}{RIC Message Router}
\newacronym{rmse}{RMSE}{Root Mean Squared Error}
\newacronym{rnis}{RNIS}{Radio Network Information Service}
\newacronym{rr}{RR}{Round Robin}
\newacronym{rrc}{RRC}{Radio Resource Control}
\newacronym{rrm}{RRM}{Radio Resource Management}
\newacronym{rru}{RRU}{Remote Radio Unit}
\newacronym{rs}{RS}{Remote Server}
\newacronym{rsrp}{RSRP}{Reference Signal Received Power}
\newacronym{rsrq}{RSRQ}{Reference Signal Received Quality}
\newacronym{rss}{RSS}{Received Signal Strength}
\newacronym{rssi}{RSSI}{Received Signal Strength Indicator}
\newacronym{rtt}{RTT}{Round Trip Time}
\newacronym{ru}{RU}{Radio Unit}
\newacronym{rw}{RW}{Receive Window}
\newacronym{rx}{RX}{Receiver}
\newacronym{s1ap}{S1AP}{S1 Application Protocol}
\newacronym{sa}{SA}{standalone}
\newacronym{sack}{SACK}{Selective Acknowledgment}
\newacronym{sap}{SAP}{Service Access Point}
\newacronym{sc2}{SC2}{Spectrum Collaboration Challenge}
\newacronym{scef}{SCEF}{Service Capability Exposure Function}
\newacronym{sch}{SCH}{Secondary Cell Handover}
\newacronym{scoot}{SCOOT}{Split Cycle Offset Optimization Technique}
\newacronym{sctp}{SCTP}{Stream Control Transmission Protocol}
\newacronym{sdap}{SDAP}{Service Data Adaptation Protocol}
\newacronym{sdk}{SDK}{Software Development Kit}
\newacronym{sdm}{SDM}{Space Division Multiplexing}
\newacronym{sdma}{SDMA}{Spatial Division Multiple Access}
\newacronym{sdn}{SDN}{Software-defined Networking}
\newacronym{sdr}{SDR}{Software-defined Radio}
\newacronym{seba}{SEBA}{SDN-Enabled Broadband Access}
\newacronym{sgsn}{SGSN}{Serving GPRS Support Node}
\newacronym{sgw}{SGW}{Service Gateway}
\newacronym{si}{SI}{Study Item}
\newacronym{sib}{SIB}{Secondary Information Block}
\newacronym{sinr}{SINR}{Signal to Interference plus Noise Ratio}
\newacronym{sip}{SIP}{Session Initiation Protocol}
\newacronym{siso}{SISO}{Single Input, Single Output}
\newacronym{sla}{SLA}{Service Level Agreement}
\newacronym{sm}{SM}{Service Model}
\newacronym{smf}{SMF}{Session Management Function}
\newacronym{smo}{SMO}{Service Management and Orchestration}
\newacronym{sms}{SMS}{Short Message Service}
\newacronym{smsgmsc}{SMS-GMSC}{\gls{sms}-Gateway}
\newacronym{snr}{SNR}{Signal-to-Noise-Ratio}
\newacronym{son}{SON}{Self-Organizing Network}
\newacronym{sptcp}{SPTCP}{Single Path TCP}
\newacronym{srb}{SRB}{Service Radio Bearer}
\newacronym{srn}{SRN}{Standard Radio Node}
\newacronym{srs}{SRS}{Sounding Reference Signal}
\newacronym{ss}{SS}{Synchronization Signal}
\newacronym{sss}{SSS}{Secondary Synchronization Signal}
\newacronym{st}{ST}{Spanning Tree}
\newacronym{svc}{SVC}{Scalable Video Coding}
\newacronym{tb}{TB}{Transport Block}
\newacronym{tcp}{TCP}{Transmission Control Protocol}
\newacronym{tdd}{TDD}{Time Division Duplexing}
\newacronym{tdm}{TDM}{Time Division Multiplexing}
\newacronym{tdma}{TDMA}{Time Division Multiple Access}
\newacronym{tfl}{TfL}{Transport for London}
\newacronym{tfrc}{TFRC}{TCP-Friendly Rate Control}
\newacronym{tft}{TFT}{Traffic Flow Template}
\newacronym{tgen}{TGEN}{Traffic Generator}
\newacronym{tip}{TIP}{Telecom Infra Project}
\newacronym{tm}{TM}{Transparent Mode}
\newacronym{to}{TO}{Telco Operator}
\newacronym{tr}{TR}{Technical Report}
\newacronym{trp}{TRP}{Transmitter Receiver Pair}
\newacronym{ts}{TS}{Technical Specification}
\newacronym{tti}{TTI}{Transmission Time Interval}
\newacronym{ttt}{TTT}{Time-to-Trigger}
\newacronym{tx}{TX}{Transmitter}
\newacronym{uas}{UAS}{Unmanned Aerial System}
\newacronym{uav}{UAV}{Unmanned Aerial Vehicle}
\newacronym{udm}{UDM}{Unified Data Management}
\newacronym{udp}{UDP}{User Datagram Protocol}
\newacronym{udr}{UDR}{Unified Data Repository}
\newacronym{ue}{UE}{User Equipment}
\newacronym{uhd}{UHD}{\gls{usrp} Hardware Driver}
\newacronym{ul}{UL}{Uplink}
\newacronym{um}{UM}{Unacknowledged Mode}
\newacronym{uml}{UML}{Unified Modeling Language}
\newacronym{upa}{UPA}{Uniform Planar Array}
\newacronym{upf}{UPF}{User Plane Function}
\newacronym{urllc}{URLLC}{Ultra Reliable and Low Latency Communications}
\newacronym{usa}{U.S.}{United States}
\newacronym{usim}{USIM}{Universal Subscriber Identity Module}
\newacronym{usrp}{USRP}{Universal Software Radio Peripheral}
\newacronym{utc}{UTC}{Urban Traffic Control}
\newacronym{vim}{VIM}{Virtualization Infrastructure Manager}
\newacronym{vm}{VM}{Virtual Machine}
\newacronym{vnf}{VNF}{Virtual Network Function}
\newacronym{volte}{VoLTE}{Voice over \gls{lte}}
\newacronym{voltha}{VOLTHA}{Virtual OLT HArdware Abstraction}
\newacronym{vr}{VR}{Virtual Reality}
\newacronym{vran}{vRAN}{Virtualized \gls{ran}}
\newacronym{vss}{VSS}{Video Streaming Server}
\newacronym{wbf}{WBF}{Wired Bias Function}
\newacronym{wf}{WF}{Waterfilling}
\newacronym{wlan}{WLAN}{Wireless Local Area Network}
\newacronym{osm}{OSM}{Open Source \gls{nfv} Management and Orchestration}
\newacronym{pnf}{PNF}{Physical Network Function}
\newacronym{drl}{DRL}{Deep Reinforcement Learning}
\newacronym{rl}{RL}{Reinforcement Learning}
\newacronym{mtc}{MTC}{Machine-type Communications}
\newacronym{osc}{OSC}{O-RAN Software Community}
\newacronym{rc}{RC}{RAN Control}
\newacronym{dqn}{DQN}{Deep Q-Network}
\newacronym{v2x}{V2X}{Vehicle-to-everything}
\newacronym{gbsm}{GBSM}{Geometry-Based Stochastic Model}
\newacronym{gbs}{GBSM}{Geometry-Based Stochastic}
\newacronym{quadriga}{QuaDRiGa}{QUAsi Deterministic RadIo channel GenerAtor}
\newacronym{relu}{ReLU}{Rectified Linear Unit} 
\newacronym{mpc}{MPC}{Multipath Component}
\newacronym{nn}{NN}{Neural Network}
\newacronym{sgd}{SGD}{Stochastic Gradient Descent}
\newacronym{cpi}{CPI}{Conservative Policy Iteration}
\newacronym{trpo}{TRPO}{Trust Region Policy Optimization}
\newacronym{mrat}{multi-RAT}{Multi-Radio Access Technology}
\newacronym{se}{SE}{Spectrum Efficiency}
\newacronym{marl}{MARL}{Multi-Agent \gls{drl}}
\newacronym{noma}{NOMA}{Non-Orthogonal Multiple Access}
\newacronym{td3}{TD3}{Twin Delayed DDPG}
\newacronym{ddpg}{DDPG}{Deep Deterministic Policy Gradient}

%% file: main.bbl
\begin{thebibliography}{10}
\providecommand{\url}[1]{#1}
\csname url@samestyle\endcsname
\providecommand{\newblock}{\relax}
\providecommand{\bibinfo}[2]{#2}
\providecommand{\BIBentrySTDinterwordspacing}{\spaceskip=0pt\relax}
\providecommand{\BIBentryALTinterwordstretchfactor}{4}
\providecommand{\BIBentryALTinterwordspacing}{\spaceskip=\fontdimen2\font plus
\BIBentryALTinterwordstretchfactor\fontdimen3\font minus \fontdimen4\font\relax}
\providecommand{\BIBforeignlanguage}[2]{{%
\expandafter\ifx\csname l@#1\endcsname\relax
\typeout{** WARNING: IEEEtran.bst: No hyphenation pattern has been}%
\typeout{** loaded for the language `#1'. Using the pattern for}%
\typeout{** the default language instead.}%
\else
\language=\csname l@#1\endcsname
\fi
#2}}
\providecommand{\BIBdecl}{\relax}
\BIBdecl

\bibitem{polese2023understanding}
M.~Polese, L.~Bonati, S.~D’Oro, S.~Basagni, and T.~Melodia, ``{Understanding O-RAN: Architecture, interfaces, algorithms, security, and research challenges},'' \emph{IEEE Communications Surveys \& Tutorials}, 2023.

\bibitem{sutton2018reinforcement}
R.~S. Sutton and A.~G. Barto, \emph{{Reinforcement learning: An introduction}}.\hskip 1em plus 0.5em minus 0.4em\relax MIT press, 2018.

\bibitem{wang2022self}
X.~Wang, J.~D. Thomas, R.~J. Piechocki, S.~Kapoor, R.~Santos-Rodr{\'\i}guez, and A.~Parekh, ``{Self-play learning strategies for resource assignment in Open-RAN networks},'' \emph{Computer Networks}, vol. 206, p. 108682, 2022.

\bibitem{polese2022colo}
M.~Polese, L.~Bonati, S.~D’Oro, S.~Basagni, and T.~Melodia, ``{{ColO-RAN:} Developing machine learning-based xApps for open RAN closed-loop control on programmable experimental platforms},'' \emph{IEEE Transactions on Mobile Computing}, 2022.

\bibitem{d2022orchestran}
S.~D’Oro, L.~Bonati, M.~Polese, and T.~Melodia, ``{OrchestRAN: Network automation through orchestrated intelligence in the open RAN},'' in \emph{IEEE Conference on Computer Communications}, 2022, pp. 270--279.

\bibitem{johnson2022nexran}
D.~Johnson, D.~Maas, and J.~Van Der~Merwe, ``{NexRAN: Closed-loop RAN slicing in POWDER-A top-to-bottom open-source open-RAN use case},'' in \emph{Proceedings of the 15th ACM Workshop on Wireless Network Testbeds, Experimental evaluation \& CHaracterization}, 2022, pp. 17--23.

\bibitem{kak2023hexran}
A.~Kak, V.-Q. Pham, H.-T. Thieu, and N.~Choi, ``{{HexRAN}: A Programmable Multi-RAT Platform for Network Slicing in the Open RAN Ecosystem},'' \emph{arXiv preprint arXiv:2304.12560}, 2023.

\bibitem{EURECOM+7416}
C.-C. Chen, C.-Y. Chang, and N.~Nikaein, ``Flexslice: Flexible and real-time programmable ran slicing framework,'' in \emph{IEEE Global Communications Conference}, Kuala Lumpur, 2023.

\bibitem{wiebusch2023towards}
R.~Wiebusch, N.~A. Wagner, D.~Overbeck, F.~Kurtz, and C.~Wietfeld, ``{Towards open 6g: Experimental o-ran framework for predictive uplink slicing},'' in \emph{IEEE International Conference on Communications}.\hskip 1em plus 0.5em minus 0.4em\relax IEEE, 2023, pp. 4834--4839.

\bibitem{yeh2023deep}
S.-p. Yeh, S.~Bhattacharya, R.~Sharma, and H.~Moustafa, ``{Deep Learning for Intelligent and Automated Network Slicing in 5G Open RAN (ORAN) Deployment},'' \emph{Open Journal of the Communications Society}, 2023.

\bibitem{kouchaki2022actor}
M.~Kouchaki and V.~Marojevic, ``{Actor-Critic Network for O-RAN Resource Allocation: xApp Design, Deployment, and Analysis},'' in \emph{IEEE Globecom Workshops (GC Wkshps)}, 2022, pp. 968--973.

\bibitem{filali2023communication}
A.~Filali, B.~Nour, S.~Cherkaoui, and A.~Kobbane, ``{Communication and computation O-RAN resource slicing for URLLC services using deep reinforcement learning},'' \emph{IEEE Communications Standards Magazine}, vol.~7, no.~1, pp. 66--73, 2023.

\bibitem{10329913}
M.~Zangooei, M.~Golkarifard, M.~Rouili, N.~Saha, and R.~Boutaba, ``{Flexible RAN Slicing in Open RAN With Constrained Multi-Agent Reinforcement Learning},'' \emph{IEEE Journal on Selected Areas in Communications}, vol.~42, no.~2, pp. 280--294, 2024.

\bibitem{iturria2022multi}
P.~Iturria~Rivera, H.~Zhang, H.~Zhou, S.~Mollahasani, and M.~Erol~Kantarci, ``{Multi-Agent Team Learning in Virtualized Open Radio Access Networks (O-RAN)},'' \emph{Sensors}, vol.~22, p. 5375, 07 2022.

\bibitem{zhang2022team}
H.~Zhang, H.~Zhou, and M.~Erol-Kantarci, ``{Team learning-based resource allocation for open radio access network (O-RAN)},'' in \emph{IEEE International Conference on Communications}.\hskip 1em plus 0.5em minus 0.4em\relax IEEE, 2022, pp. 4938--4943.

\bibitem{zhang2022federated}
H.~Zhang, H.~Zhou, and M.~Erol~Kantarci, ``{Federated deep reinforcement learning for resource allocation in O-RAN slicing},'' in \emph{IEEE Global Communications Conference (GLOBECOM)}, 2022, pp. 958--963.

\bibitem{9933014}
F.~Rezazadeh, L.~Zanzi, F.~Devoti, H.~Chergui, X.~Costa-Pérez, and C.~Verikoukis, ``{On the Specialization of FDRL Agents for Scalable and Distributed 6G RAN Slicing Orchestration},'' \emph{IEEE Transactions on Vehicular Technology}, vol.~72, no.~3, pp. 3473--3487, 2023.

\bibitem{luong2019applications}
N.~C. Luong, D.~T. Hoang, S.~Gong, D.~Niyato, P.~Wang, Y.-C. Liang, and D.~I. Kim, ``{Applications of deep reinforcement learning in communications and networking: A survey},'' \emph{IEEE Communications Surveys \& Tutorials}, vol.~21, no.~4, pp. 3133--3174, 2019.

\bibitem{alwarafy2021deep}
A.~Alwarafy, M.~Abdallah, B.~S. Ciftler, A.~Al-Fuqaha, and M.~Hamdi, ``{Deep reinforcement learning for radio resource allocation and management in next generation heterogeneous wireless networks: A survey},'' \emph{arXiv preprint arXiv:2106.00574}, 2021.

\bibitem{suh2022deep}
K.~Suh, S.~Kim, Y.~Ahn, S.~Kim, H.~Ju, and B.~Shim, ``{Deep reinforcement learning-based network slicing for beyond 5G},'' \emph{IEEE Access}, vol.~10, pp. 7384--7395, 2022.

\bibitem{liu2020deepslicing}
Q.~Liu, T.~Han, N.~Zhang, and Y.~Wang, ``{DeepSlicing: Deep reinforcement learning assisted resource allocation for network slicing},'' in \emph{GLOBECOM 2020-2020 IEEE Global Communications Conference}.\hskip 1em plus 0.5em minus 0.4em\relax IEEE, 2020, pp. 1--6.

\bibitem{sun2019dynamic}
G.~Sun, Z.~T. Gebrekidan, G.~O. Boateng, D.~Ayepah-Mensah, and W.~Jiang, ``{Dynamic reservation and deep reinforcement learning based autonomous resource slicing for virtualized radio access networks},'' \emph{Ieee Access}, vol.~7, pp. 45\,758--45\,772, 2019.

\bibitem{yan2023deep}
D.~Yan, B.~K. Ng, W.~Ke, and C.-T. Lam, ``{Deep Reinforcement Learning Based Resource Allocation for Network Slicing with Massive MIMO},'' \emph{IEEE Access}, 2023.

\bibitem{naderializadeh2021resource}
N.~Naderializadeh, J.~J. Sydir, M.~Simsek, and H.~Nikopour, ``{Resource management in wireless networks via multi-agent deep reinforcement learning},'' \emph{IEEE Transactions on Wireless Communications}, vol.~20, no.~6, pp. 3507--3523, 2021.

\bibitem{azimi2021energy}
Y.~Azimi, S.~Yousefi, H.~Kalbkhani, and T.~Kunz, ``{Energy-efficient deep reinforcement learning assisted resource allocation for 5G-RAN slicing},'' \emph{IEEE Transactions on Vehicular Technology}, vol.~71, no.~1, pp. 856--871, 2021.

\bibitem{rahimi2022novel}
A.~M. Rahimi, A.~Ziaeddini, and S.~Gonglee, ``{A novel approach to efficient resource allocation in load-balanced cellular networks using hierarchical DRL},'' \emph{Journal of Ambient Intelligence and Humanized Computing}, vol.~13, no.~5, pp. 2887--2901, 2022.

\bibitem{bonati2023openran}
L.~Bonati, M.~Polese, S.~D’Oro, S.~Basagni, and T.~Melodia, ``{{OpenRAN Gym}: AI/ML development, data collection, and testing for O-RAN on PAWR platforms},'' \emph{Computer Networks}, vol. 220, p. 109502, 2023.

\bibitem{bonati2021colosseum}
L.~Bonati, P.~Johari, M.~Polese, S.~D’Oro, S.~Mohanti, M.~Tehrani-Moayyed, D.~Villa, S.~Shrivastava, C.~Tassie, K.~Yoder \emph{et~al.}, ``{C}olosseum: {L}arge-scale {W}ireless {E}xperimentation through {H}ardware-in-the-{L}oop {N}etwork {E}mulation,'' in \emph{IEEE International Symposium on Dynamic Spectrum Access Networks (DySPAN)}, 2021, pp. 105--113.

\bibitem{tsampazi2023comparative}
M.~Tsampazi, S.~D'Oro, M.~Polese, L.~Bonati, G.~Poitau, M.~Healy, and T.~Melodia, ``{A Comparative Analysis of Deep Reinforcement Learning-based xApps in O-RAN},'' \emph{IEEE Global Communications Conference (GLOBECOM)}, 2023.

\bibitem{tensorflow2015-whitepaper}
\BIBentryALTinterwordspacing
M.~Abadi, A.~Agarwal, P.~Barham, E.~Brevdo, Z.~Chen, C.~Citro, G.~S. Corrado, A.~Davis, J.~Dean, M.~Devin, S.~Ghemawat, I.~Goodfellow, A.~Harp, G.~Irving, M.~Isard, Y.~Jia, R.~Jozefowicz, L.~Kaiser, M.~Kudlur, J.~Levenberg, D.~Man\'{e}, R.~Monga, S.~Moore, D.~Murray, C.~Olah, M.~Schuster, J.~Shlens, B.~Steiner, I.~Sutskever, K.~Talwar, P.~Tucker, V.~Vanhoucke, V.~Vasudevan, F.~Vi\'{e}gas, O.~Vinyals, P.~Warden, M.~Wattenberg, M.~Wicke, Y.~Yu, and X.~Zheng, ``{{TensorFlow}: Large-Scale Machine Learning on Heterogeneous Systems},'' 2015, software available from tensorflow.org. [Online]. Available: \url{https://www.tensorflow.org/}
\BIBentrySTDinterwordspacing

\bibitem{wang2023resource}
Q.~Wang, Y.~Liu, Y.~Wang, X.~Xiong, J.~Zong, J.~Wang, and P.~Chen, ``{Resource allocation based on Radio Intelligence Controller for Open RAN towards 6G},'' \emph{IEEE Access}, 2023.

\bibitem{fiandrino2023explora}
C.~Fiandrino, L.~Bonati, S.~D'Oro, M.~Polese, T.~Melodia, and J.~Widmer, ``{EXPLORA: AI/ML EXPLainability for the Open RAN},'' \emph{Proceedings of the ACM on Networking}, vol.~1, no. CoNEXT3, pp. 1--26, 2023.

\bibitem{brik2022deep}
B.~Brik, K.~Boutiba, and A.~Ksentini, ``{Deep learning for B5G open radio access network: Evolution, survey, case studies, and challenges},'' \emph{IEEE Open Journal of the Communications Society}, vol.~3, pp. 228--250, 2022.

\bibitem{TFAgents}
\BIBentryALTinterwordspacing
S.~Guadarrama, A.~Korattikara, O.~Ramirez, P.~Castro, E.~Holly, S.~Fishman, K.~Wang, E.~Gonina, N.~Wu, E.~Kokiopoulou, L.~Sbaiz, J.~Smith, G.~Bartók, J.~Berent, C.~Harris, V.~Vanhoucke, and E.~Brevdo, ``{{TF-Agents}: A library for Reinforcement Learning in TensorFlow},'' \url{https://github.com/tensorflow/agents}, 2018, [Online; accessed 25-June-2019]. [Online]. Available: \url{https://github.com/tensorflow/agents}
\BIBentrySTDinterwordspacing

\bibitem{mnih2015human}
V.~Mnih, K.~Kavukcuoglu, D.~Silver, A.~A. Rusu, J.~Veness, M.~G. Bellemare, A.~Graves, M.~Riedmiller, A.~K. Fidjeland, G.~Ostrovski \emph{et~al.}, ``{Human-level control through deep reinforcement learning},'' \emph{Nature}, vol. 518, no. 7540, pp. 529--533, 2015.

\bibitem{kaloxylos2021ai}
\BIBentryALTinterwordspacing
A.~Kaloxylos, A.~Gavras, D.~Camps, M.~Ghoraishi, and H.~Hrasnica, ``{AI and ML--Enablers for beyond 5G Networks},'' 2021. [Online]. Available: \url{https://5g-ppp.eu/wp-content/uploads/2021/05/AI-MLforNetworks-v1-0.pdf}
\BIBentrySTDinterwordspacing

\bibitem{yang2019application}
J.~Yang, X.~You, G.~Wu, M.~M. Hassan, A.~Almogren, and J.~Guna, ``{Application of reinforcement learning in UAV cluster task scheduling},'' \emph{Future generation computer systems}, vol.~95, pp. 140--148, 2019.

\bibitem{schulman2017proximal}
J.~Schulman, F.~Wolski, P.~Dhariwal, A.~Radford, and O.~Klimov, ``{Proximal policy optimization algorithms},'' \emph{arXiv preprint arXiv:1707.06347}, 2017.

\bibitem{mnih2013playing}
V.~Mnih, K.~Kavukcuoglu, D.~Silver, A.~Graves, I.~Antonoglou, D.~Wierstra, and M.~Riedmiller, ``{Playing atari with deep reinforcement learning},'' \emph{arXiv preprint arXiv:1312.5602}, 2013.

\bibitem{kakade2002approximately}
S.~Kakade and J.~Langford, ``{Approximately optimal approximate reinforcement learning},'' in \emph{Proceedings of the Nineteenth International Conference on Machine Learning}, 2002, pp. 267--274.

\bibitem{gomez2016srslte}
I.~Gomez-Miguelez, A.~Garcia-Saavedra, P.~D. Sutton, P.~Serrano, C.~Cano, and D.~J. Leith, ``{{srsLTE}: An Open-Source Platform for {LTE} Evolution and Experimentation},'' in \emph{Proceedings of the Tenth ACM International Workshop on Wireless Network Testbeds, Experimental Evaluation, and Characterization}, 2016, pp. 25--32.

\bibitem{bonati2021scope}
L.~Bonati, S.~D'Oro, S.~Basagni, and T.~Melodia, ``{{SCOPE}: An Open and Softwarized Prototyping Platform for NextG Systems},'' in \emph{Proceedings of the 19th Annual International Conference on Mobile Systems, Applications, and Services}, 2021, pp. 415--426.

\bibitem{mgen}
{U.S. Naval Research Laboratory, "Multi-Generator ({MGEN}) Network Test Tool"}. \url{https://www.nrl.navy.mil/itd/ncs/products/mgen}. 2019.

\bibitem{alliance2021ran}
O.~Alliance, ``{O-RAN Working Group 2:“O-RAN AI/ML workflow description and requirements 1.03},'' \emph{O-RAN. WG2. AIML-v01. 03 Technical Specification}, 2021.

\bibitem{kim2022adaptive}
M.~Kim, J.-S. Kim, M.-S. Choi, and J.-H. Park, ``{Adaptive Discount Factor for Deep Reinforcement Learning in Continuing Tasks with Uncertainty},'' \emph{Sensors}, vol.~22, no.~19, p. 7266, 2022.

\bibitem{zhang2021minibatch}
C.~Zhang, Q.~Song, Z.~Meng \emph{et~al.}, ``{Minibatch Recursive Least Squares Q-Learning},'' \emph{Computational Intelligence and Neuroscience}, 2021.

\bibitem{lillicrap2015continuous}
T.~P. Lillicrap, J.~J. Hunt, A.~Pritzel, N.~Heess, T.~Erez, Y.~Tassa, D.~Silver, and D.~Wierstra, ``{Continuous control with deep reinforcement learning},'' \emph{arXiv preprint arXiv:1509.02971}, 2015.

\bibitem{hessel2018rainbow}
M.~Hessel, J.~Modayil, H.~Van~Hasselt, T.~Schaul, G.~Ostrovski, W.~Dabney, D.~Horgan, B.~Piot, M.~Azar, and D.~Silver, ``{Rainbow: Combining improvements in deep reinforcement learning},'' in \emph{Proceedings of the AAAI conference on artificial intelligence}, vol.~32, no.~1, 2018.

\bibitem{fortunato2017noisy}
M.~Fortunato, M.~G. Azar, B.~Piot, J.~Menick, I.~Osband, A.~Graves, V.~Mnih, R.~Munos, D.~Hassabis, O.~Pietquin \emph{et~al.}, ``{Noisy networks for exploration},'' \emph{arXiv preprint arXiv:1706.10295}, 2017.

\bibitem{schulman2015trust}
J.~Schulman, S.~Levine, P.~Abbeel, M.~Jordan, and P.~Moritz, ``{Trust region policy optimization},'' in \emph{International conference on machine learning}.\hskip 1em plus 0.5em minus 0.4em\relax PMLR, 2015, pp. 1889--1897.

\bibitem{fujimoto2019off}
S.~Fujimoto, D.~Meger, and D.~Precup, ``{Off-policy deep reinforcement learning without exploration},'' in \emph{International conference on machine learning}.\hskip 1em plus 0.5em minus 0.4em\relax PMLR, 2019, pp. 2052--2062.

\bibitem{9068634}
Y.~Xiao, Q.~Zhang, F.~Liu, J.~Wang, M.~Zhao, Z.~Zhang, and J.~Zhang, ``{NFVdeep: Adaptive Online Service Function Chain Deployment with Deep Reinforcement Learning},'' in \emph{IEEE/ACM International Symposium on Quality of Service}, 2019, pp. 1--10.

\bibitem{kozlica2023deep}
R.~Kozlica, S.~Wegenkittl, and S.~Hir{\"a}nder, ``{Deep q-learning versus proximal policy optimization: Performance comparison in a material sorting task},'' in \emph{IEEE 32nd International Symposium on Industrial Electronics (ISIE)}, 2023, pp. 1--6.

\end{thebibliography}
